\documentclass[12pt]{article}
\usepackage{epsf}
\def\lQ{\Lambda_{\rm QCD}}
\newcommand{\nn}{\nonumber}
\newcommand{\be}{\begin{equation}}
\newcommand{\ee}{\end{equation}}
\newcommand{\bea}{\begin{eqnarray}}
\newcommand{\eea}{\end{eqnarray}}

\def\als{\alpha_{\rm s}}
\def\siml{{\
\lower-1.2pt\vbox{\hbox{\rlap{$<$}\lower6pt\vbox{\hbox{$\sim$}}}}\ }}  
\def\simg{{\
\lower-1.2pt\vbox{\hbox{\rlap{$>$}\lower6pt\vbox{\hbox{$\sim$}}}}\ }}

\newcommand{\MS}{\overline{\rm MS}}
\newcommand{\RS}{\rm RS}

\newcommand{\PV}{\rm PV}
\newcommand{\OS}{\rm OS}

\begin{document}
\begin{titlepage}
\begin{flushright}
\tt{UB-ECM-PF 03/22\\
GUTPA 03/10/01}
\end{flushright}

\vspace{1cm}
\begin{center}
\begin{Large}
{\bf QCD phenomenology of static sources
and gluonic excitations at short distances}\\[2cm]
\end{Large} 
{\large Gunnar S.\ Bali}\footnote{g.bali@physics.gla.ac.uk}\\
{\it Department of Physics \& Astronomy, The University of Glasgow,
Glasgow G12 8QQ, Scotland \\}
\vspace{0.5cm}
{\large Antonio Pineda}\footnote{pineda@ecm.ub.es}\\
{\it Dept. d'Estructura i Constituents de la Mat\`eria,
  U. Barcelona \\ Diagonal 647, E-08028 Barcelona, Catalonia, Spain
        \\}
\end{center}

\vspace{1cm}

\begin{abstract}
New lattice data for the $\Pi_u$ and $\Sigma_u^-$ potentials at short
distances are presented. We compare perturbation theory to the lower
static hybrid potentials and find good agreement at short distances,
once the renormalon ambiguities are accounted for.  We use the
non-perturbatively determined continuum-limit static hybrid and ground
state potentials at short distances to determine the gluelump
energies.  The result is consistent with an estimate obtained from the
gluelump data at finite lattice spacings.  For the
lightest gluelump, we obtain
$\Lambda_{B}^{\RS}(\nu_f=2.5\,r_0^{-1})=[2.25\pm 0.10({\rm
latt.})\pm 0.21 ({\rm th.})\pm 0.08 (\Lambda_{\MS})]\,r_0^{-1}$
in the quenched
approximation with $r_0^{-1}\approx 400$~MeV. 
We show that, to quote sensible numbers for the
absolute values of the gluelump energies, it is necessary to handle
the singularities of the singlet and octet potentials in the Borel
plane. We propose to subtract the renormalons of the short-distance
matching coefficients, the potentials in this case. For the singlet
potential the leading renormalon is already known and related to that
of the pole mass, for the octet potential a new renormalon appears,
which we approximately evaluate. We also apply our methods to heavy
light mesons in the static limit and from the lattice simulations 
available in the literature we obtain the quenched result 
$
\overline{\Lambda}^{\RS}(\nu_f=2.5\,r_0^{-1})=
	       \left[1.17\pm 0.08({\rm latt.})\pm 0.13
({\rm th.}) \pm 0.09 (\Lambda_{\MS}) \right]\,r_0^{-1}
$.
We calculate $m_{b,\MS}(m_{b,\MS})$ and apply our methods to
gluinonia whose dynamics are governed by the singlet potential
between adjoint sources. We can exclude
non-standard linear short-distance contributions
to the static potentials, with good accuracy.
\vspace{5mm} \\ 
PACS numbers: 12.38.Bx, 12.38.Gc, 12.38.Cy, 12.39.Hg, 14.80.Ly, 14.65.Fy
\end{abstract}

\end{titlepage}
\vfill
\setcounter{footnote}{0} 
\vspace{1cm}

\section{Introduction}

In recent years, we have witnessed growing interest in the physics of
gluelumps and static hybrid potentials. In many cases this has been
driven by increasingly reliable lattice simulations of their
properties
\cite{latticeshort2,detar,Bali:2000vr,Collins:1997cb,Perantonis:1990dy,Ford:as,Campbell:fe}.
These results expose models of low energy QCD to stringent tests and
therefore enhance our understanding of the underlying dynamics. The
short distance physics of the static hybrid potentials is of
particular importance.  In this region, hybrids and
gluelumps are intimately related and well suited to investigate
the interplay between perturbative and non-perturbative physics.
At short distances, $r$,
one is faced with widely separated
scales: $1/r \gg \lQ$. In such situations, effective field theories
(EFTs) are particularly useful since they enable the
physics associated with the different scales to be factorized
in a very efficient and model
independent way. One EFT designed to deal with the kinematical case of
interest to us corresponds to potential non-relativistic QCD
(pNRQCD)~\cite{Mont}, in the static
limit~\cite{pNRQCD}.

In Ref.~\cite{pNRQCD} the gluelumps and the short distance regime of
the static hybrids were studied within this EFT
framework and general features identified. Some results known from the
past \cite{Hasenfratz:1980jv,Perantonis:1990dy,BBV,FM,Bali:2000gf}
were recovered within a unified framework and in some cases extended.

One can go beyond this analysis and use lattice data plus the
knowledge of the (perturbative) octet potential to obtain numerical
values for gluelump masses in a particular scheme.  However,
analogously to the situation with the static singlet potential, the
convergence of the perturbative series of the octet potential does not
appear very promising. This is a general problem when different scales
are factorized, and in particular perturbative from non-perturbative
ones.  The bad convergence is also related to the problem of
factorizing non-perturbative quantities, without defining their
perturbative counterparts~\cite{Bigi:1994em},
and is usually believed to be due to the existence
of singularities in the Borel transform of the perturbative
quantity.
These singularities appear to be
due to scales of order $e^{-n} \times$-(the typical scale of the
perturbative quantity) in an $n$-loop calculation.
In Ref.~\cite{RS} one of the present authors proposed
that, since these singularities are related to energy scales much lower than
the ones that are supposedly included in the perturbative object, they
should be subtracted from it and introduced in the matrix elements of
the effective theory.  This programme has been worked out for the pole mass
and the static singlet potential~\cite{RS,staticpot}. Here we
apply the same approach to the static octet potential. This will allow
us to determine absolute values for the gluelump masses from
the spectrum of the static hybrids, as well as to study up to which
scale one can use perturbation theory to describe hybrid potentials.

This paper is organized as follows. In Sec.~\ref{pNRQCD} we will
work out the
r\^ole of gluelumps in pNRQCD, and how gluelumps and hybrid potentials
are interrelated.
In Sec.~\ref{lattice} we will then
sketch how our lattice data have been obtained,
before discussing and classifying
renormalons and power corrections
in the continuum $\MS$ scheme as well as in a lattice scheme
in Sec.~\ref{statoct}.
In the same section we will also generalise
the renormalon subtracted ($\RS$) scheme of Ref.~\cite{RS} 
to the case of the octet potential
and discuss the scale dependence.
In Sec.~\ref{pheno} we will obtain the gluelump masses
in the $\RS$ scheme and relate these results to the lattice scheme.
We will compare to previous literature and
predict the gluelump spectrum.
In Sec.~\ref{statlight} we will determine the binding energy 
of static-light mesons as well as the bottom mass,
before we discuss generalisations to and relations with
adjoint potentials, gluinonium and other objects with
relevance to short-distance QCD in Sec.~\ref{gluino}.

\section{Hybrid potentials and gluelumps}
\label{pNRQCD}
We discuss the relationship between hybrid potentials and
gluelumps at short distances. First we consider the EFT picture,
before we discuss the symmetries that are relevant in the
non-perturbative case. Finally we compare these expectations
to lattice data.

\subsection{pNRQCD and gluelumps}
The pNRQCD Lagrangian at leading order in $1/m$ and in the multipole expansion reads~\cite{Mont,pNRQCD},
\begin{eqnarray}
L_{\rm pNRQCD} =
\int d^3\!{\bf r}\,d^3{\bf R}\,{\rm Tr} \,\Biggl\{ {\rm S}^\dagger \left( i\partial_0 - V_s  \right) {\rm S} 
+ {\rm O}^\dagger \left( iD_0 - V_o \right) {\rm O} \Biggr\}
-\int d^3{\bf R} {1\over 4} F_{\mu \nu}^{a} F^{\mu \nu \, a}+O(r).
\label{pnrqcd0}
\end{eqnarray}
All the gauge fields in Eq.\ (\ref {pnrqcd0}) are evaluated 
in ${\bf R}$ and $t$, in particular $F^{\mu \nu \, a} \equiv F^{\mu \nu \, a}
({\bf R},t)$ and $iD_0 {\rm O} \equiv i \partial_0 {\rm O} - g [A_0({\bf
  R},t),{\rm O}]$. 
The singlet and octet potentials
$V_{i}$, $i=s,o$ are to be regarded as matching coefficients, which depend
on the scale $\nu_{us}$ separating soft gluons from ultrasoft ones. In the
static limit ``soft'' energies are of $O(1/r)$ and ``ultrasoft'' energies
are of $O(\als/r)$. 
Notice that the hard scale, $m$, plays no r\^ole in this limit.
The only assumption 
made so far 
concerns the size of $r$, i.e.\ $1/r \gg \lQ$, such that the potentials can be 
computed in perturbation theory.  Also note that throughout this paper
we will adopt a Minkowski space-time notation.

The spectrum of the singlet state reads,
\be
\label{Es}
E_s(r)=2m_{\OS}+V_s(r)+O(r^2)
\,,
\ee
where $m_{\OS}$ denotes an on-shell (OS) mass.
One would normally apply pNRQCD to quarkonia and in this
case $m_{\OS}$ represents the heavy quark pole mass.
For the static hybrids, the spectrum reads
\be
\label{EH}
E_H(r)=2m_{\OS}+V_o(r)+\Lambda_H^{\OS}+O(r^2)
\,,
\ee
where
\be
\Lambda_H^{\OS}\equiv \lim_{T\to\infty} i\frac{\partial}{\partial T}\ln 
\langle H^a(T/2)\phi^{adj}_{ab}(T/2,-T/2)H^b(-T/2) \rangle
\label{LH}
\,,
\ee 
\be
\phi(T/2,-T/2) \equiv \phi(T/2,{\bf R},-T/2,{\bf R}) 
= {\rm P} \exp \left\{ - ig \displaystyle \int_{-T/2}^{T/2} \!\! dt \, A_0({\bf R},t) \right\}
\,,
\ee
and $H$ represents some gluonic field, for examples see
Table~\ref{tablegluelumps} in Sec.~\ref{higherg}.

Eq.~(\ref{EH}) allows us to relate the energies
of the static hybrids $E_H$ to the energies of the gluelumps,
 \be
\label{eqgl}
\Lambda_H^{\OS}=\left[E_H(r)-E_s(r)\right]-\left[V_o(r)-V_s(r)\right]+O(r^2).
\ee
This equation encapsulates one of the central ideas of this paper.
The combination $E_H-E_s$ is renormalon-free in perturbation theory
[up to possible $O(r^2)$ effects], and can be calculated unambiguously
non-perturbatively: the ultraviolet (UV) renormalons related to the infrared
(IR) renormalons
of twice the pole mass cancel each other. However, $\Lambda_H$
contains an UV renormalon that corresponds to the leading IR renormalon of
$V_o$.

The shapes (of some) of the $E_H(r)$ have been computed on the lattice
for instance in
Refs.~\cite{latticeshort2,detar,Bali:2000vr,Collins:1997cb,Perantonis:1990dy,Ford:as}.
On the other hand the values of (some) $\Lambda_H$ have also been
computed within a variety of models as well as in lattice
simulations~\cite{FM,Campbell:1985kp}.  Consistency would require that
the values of $\Lambda_H$ obtained from $E_H-E_s$ and the values of
$\Lambda_H$ directly obtained from gluelump
computations should
agree. This will be checked in Sec.~\ref{seclaa} below.

Gluelump states are created by a static source in
the octet (adjoint) representation attached to some gluonic content
($H$) such that the state becomes a singlet under gauge
transformations. This is what would happen to heavy gluinos in the
static approximation. Hence sometimes gluelumps are also referred to
as gluinoballs or glueballinos
in the literature.  Without further information, their
energy is only fixed up to a global constant. Only the
energy splittings between different states have well defined
continuum limits in lattice simulations.  In lattice regularization at
a lattice spacing $a$ the normalization ambiguity is reflected in a
linear divergence $\propto a^{-1}$ while in dimensional regularization
one encounters an UV renormalon.  In the HQET picture of a heavy-light
meson one faces a similar problem. In this situation one also has a
static source (in the fundamental representation in this case), which
has to be attached to some light-quark (and gluonic) content to become
a singlet under gauge transformations. The binding energy
${\overline{\Lambda}}$ is again only defined up to a global
constant~\cite{Martinelli:1998vt} and only its sum with the pole mass is 
unambiguous:
\be
\label{MB}
M_B=m_{b,\OS}+{\overline{\Lambda}}^{\OS}+O(1/m_b)
\,.
\ee 
We will investigate this situation in Sec.~\ref{statlight} below.

\subsection{Symmetries of hybrid potentials and gluelumps}
The spectrum of open QCD string states can be completely classified by
the quantum numbers associated with the underlying symmetry group, up
to radial excitations.  In this case, these are the distance between the
endpoints, the gauge group representation under which these endpoints
transform (in what follows we consider the fundamental
representation), and the symmetry group of cylindrical rotations with
reflections $D_{\infty h}$. The irreducible representations of the
latter group are conventionally labelled by the spin along the axis,
$\Lambda$, where $\Sigma,\Pi,\Delta$ refer to $\Lambda=0,1,2$,
respectively, with a subscript $\eta=g$ for gerade (even) $PC=+$ or $\eta=u$
for ungerade (odd) $PC=-$ transformation properties.  All
$\Lambda\geq 1$ representations are two-dimensional. The
one-dimensional $\Sigma$ representations have, in addition to the
$\eta$ quantum number, a $\sigma_v$ parity with respect to
reflections on a plane that includes the two endpoints. This is
reflected in an additional $\pm$ superscript.  The state associated
with the static singlet potential transforms according to the
representation $\Sigma_g^+$ while the two lowest lying hybrid potentials
are within the $\Pi_u$ and $\Sigma_u^-$ representations, respectively.

In contrast, point-like QCD states are characterised by the $J^{PC}$
of the usual $O(3)\otimes {\mathcal C}$ rotation group as well as by
the gauge group representation of the source. In the pure gauge
sector, gauge invariance requires this representation to have
vanishing triality, such that the source can be screened to a singlet
by the glue. States created by operators in the singlet representation
are known as glueballs, octet states as gluelumps. In contrast to
gluelump states, where the octet source propagates through the gluonic
background, the normalization of glueball states with respect to the
vacuum energy is unambiguous.

Since $D_{\infty h}\subset
O(3)\otimes{\mathcal C}$, in the limit $r\rightarrow 0$ certain hybrid
levels must become degenerate. For instance, in this limit, the
$\Sigma_u^-$ state corresponds to a $J^{PC}=1^{+-}$ state with $J_z=0$ while
the $\Pi_u$ doublet corresponds to its $J_z=\pm 1$ partners.  The
gauge transformation property of the hybrid potential creation
operator will also change in this limit, ${\mathbf 3}\otimes{\mathbf
3^*}={\mathbf 1}\oplus{\mathbf 8}$, such that hybrids will either
approach gluelumps [cf.~Eq.~(\ref{EH})] or glueballs, in an
appropriate normalization. In the case of glueballs the correct
normalization can be obtained by considering the difference
$E_H(r)-E_s(r)$ from which the pole mass cancels. We will discuss the
situation with respect to gluelumps in detail in Sec.~\ref{statoct}
below.

In perturbation theory, the ground state potential corresponds to
the singlet potential while hybrid potentials 
will have the perturbative expansion of the octet
potential.

Recently, Philipsen~\cite{Philipsen:2002az} suggested to
non-perturbatively generalise the octet potential, employing a
definition that resembles the perturbative one,
after gauge fixing to Laplacian Coulomb gauge.
He proved that this construction is equivalent to a gauge invariant
correlation function whose eigenvalues will resemble masses of
physical states. In the limit $r\rightarrow 0$ the suggested
operator will be an adjoint temporal Schwinger line, dressed with a
non-local but symmetric gluon cloud, with the $J^{PC}$ quantum numbers
of the vacuum. A similar construction is mentioned in paragraph 2
of Sec.~\ref{statlight} below, as a possible non-perturbative normalization
point for gluelump energies. The static ``octet'' potential suggested in
Ref.~\cite{Philipsen:2002az} will have the $\Sigma_g^+$ symmetry and,
up to a different non-perturbative off-set, the same perturbative
expansion and power term/renormalon structure as the hybrid potentials
discussed below. Due to the nature of its creation operator
which is non-local, even in the $r=0$ limit,
at present it is not obvious to us how
this non-perturbative state can be interpreted in terms of the local states
we are considering in this paper, certainly an open question that
should be addressed in the future.

\subsection{Hybrid and gluelump mass splittings}
\label{sechsplit}
\begin{figure}[h]
\hspace{-0.1in}
\epsfxsize=4.8in
\centerline{\epsffile{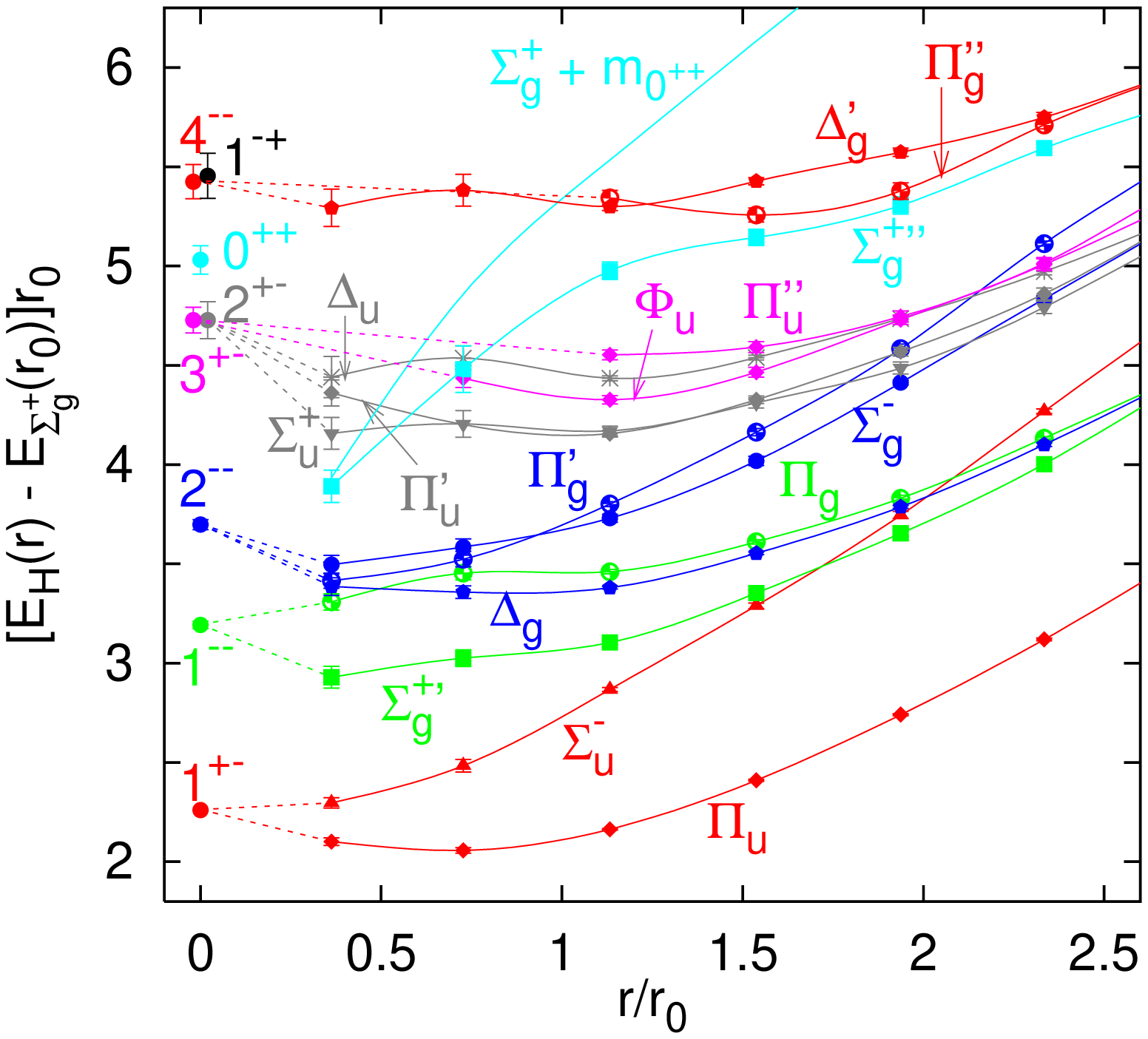}}
\caption {{\it Different hybrid potentials~\cite{latticeshort2} at a
lattice spacing $a_{\sigma}\approx 0.2$~fm~$\approx 0.4\,r_0$,
where $r_0\approx 0.5$~fm, in
comparison with the gluelump spectrum, extrapolated to the continuum
limit~\cite{FM} (circles, left-most data points). The gluelump spectrum
has been shifted by an arbitrary constant to adjust the $1^{+-}$ state
with the $\Pi_u$ and $\Sigma_u^-$ potentials at short distance.  In
addition, we include the sum of the ground state ($\Sigma_g^+$)
potential and the scalar glueball mass $m_{0^{++}}$~\cite{glueball,glueball2}.
The lines are
drawn to guide the eye.}}
\label{splitting}
\end{figure}

We would like to establish if lattice data on hybrid potentials
reproduces the degeneracies expected from the above discussion in the
short distance region. In the limit $r\rightarrow 0$, any given
$\Lambda\geq 1$ hybrid potential can be subduced from any $J^{PC}$ state with
$J\geq \Lambda$ and $PC=+$ for $\eta=g$ or $PC=-$ for $\eta=u$ representations.
For instance the $\Pi_u$ is embedded in
$1^{+-},1^{-+},2^{+-},2^{-+},\cdots$.  The situation is somewhat different
for $\Lambda=0$ states, which have the additional $\sigma_v$ parity:
the $\Sigma_g^+$ representation
can be obtained from $0^{++},1^{--},2^{++},\cdots$, $\Sigma_g^-$ from
$0^{--},1^{++},\cdots$, $\Sigma_u^+$ from $0^{+-},1^{-+},\cdots$ and
$\Sigma_u^-$ from $0^{-+},1^{+-},\cdots$.  We list all combinations of
interest to us in Table~\ref{gluedegen}.  The ordering of low lying
gluelumps has been established in Ref.~\cite{FM} and reads with
increasing mass: $1^{+-}, 1^{--}, 2^{--}, 2^{+-}, 3^{+-}, 0^{++},
4^{--}, 1^{-+}$,
with a $3^{--}$ state in the region of the $4^{--}$ and $1^{-+}$.
The $2^{+-}$ and $3^{+-}$ as well as the $4^{--}$ and
$1^{-+}$ states are degenerate within present statistical
uncertainties\footnote{The splittings between all states with respect
to the $1^{+-}$ ground state have been extrapolated 
to the continuum limit in
Ref.~\cite{FM} and we add our own
extrapolations for the $4^{--}$ and $1^{-+}$ states to these, based on
the tables of this reference.}.  The continuum limit gluelump masses
are displayed as circles at the left of Fig.~\ref{splitting}, where we
have added the (arbitrary) overall constant $2.26/r_0$ to the gluelump
splittings to match the hybrid potentials.  The similarity of this
value to our estimate of the gluelump energy in Sec.~\ref{RSglump}
below is purely accidental.

\begin{table}[h]
\begin{center}\begin{tabular}{|c||c|}
\hline
point particle $J^{PC}$&open string $\Lambda^{\sigma_v}_{\eta}$\\\hline\hline
$1^{+-}$&$\Sigma_u^-, \Pi_u$\\
$1^{--}$&$\Sigma_g^{+\prime},\Pi_g$\\
$2^{--}$&$\Sigma_g^-,\Pi_g',\Delta_g$\\
$2^{+-}$&$\Sigma_u^+,\Pi_u',\Delta_u$\\
$3^{+-}$&$\Sigma_u^{-'},\Pi_u'',\Delta_u',\Phi_u$\\
$0^{++}$&$\Sigma_g^{+\prime\prime}$\\
$4^{--}$&$\Sigma_g^{-\prime},\Pi_g'',\Delta_g',\Phi_g,\Gamma_g$\\
$1^{-+}$&$\Sigma_u^{+\prime},\Pi_u'''$\\\hline
\end{tabular}\end{center}
\caption{{\it Expected degeneracies of hybrid potentials
at short distance, based on the level ordering of the gluelump spectrum.
Note that if the $3^{+-}$ gluelump turned out to be lighter than the $2^{+-}$
then the $\Sigma_u^{-'},\Pi_u',\Delta_u,\Phi_u$ potentials would approach the
$3^{+-}$ state while the $\Sigma_u^+,\Pi_u'',\Delta_u'$ potentials
would approach the $2^{+-}$ instead.}}
\label{gluedegen}
\end{table}

Juge, Kuti and Morningstar~\cite{latticeshort2} have, 
for the first time, comprehensively determined the spectrum of hybrid
potentials. We convert their data,
computed at their smallest lattice spacing $a_{\sigma}\approx 0.2$~fm, 
into units of $r_0\approx 0.5$~fm~\cite{Sommer:1993ce}.  Since the
results have been obtained with an improved action and on anisotropic
lattices with $a_{\tau}\approx a_{\sigma}/4$, one might expect lattice
artifacts to be small\footnote{On the lattice the relevant symmetry
group is $D_{4h}$ rather than $D_{\infty h}$ (see
e.g.~Ref.~\cite{hamermesh}). In the continuum limit the $A_{1\eta}$
potentials will correspond to $\Sigma^{+}_{\eta}$, the $A_{2\eta}$
potentials to $\Sigma^-_{\eta}$ and the $E_{\eta}$ potentials to
$\Pi_{\eta}$, where $\eta=u, g$.  The radial excitations could in
principal correspond to higher spin potentials and in fact one of the
three observed excitations of $E_u$ will correspond to the $\Phi_u$
ground state. In all other cases, associating the lowest possible
continuum spin to a given lattice potential seems to agree with the
ordering suggested by the gluelump spectrum (as well as in the large
distance string limit~\cite{latticeshort2}).  $B_{1\eta}$ and
$B_{2\eta}$ both correspond to $\Delta_{\eta}$.  In either case (as
well as for $\Delta_g'$), at the short distances displayed in the
figure, the two lattice representations agree with each other,
supporting the view that violations of rotational symmetry are small.
In this case we only display the lattice representations with better
statistical accuracy, i.e.\ the $B_{1\eta}^{(\prime)}$s.},
at least for the lower
lying potentials. Hence we compare these data, normalized to
$E_{\Sigma_g^+}(r_0)$, with the continuum expectations of the
gluelumps~\cite{FM}.  The full lines are cubic splines to guide the eye
while the dashed lines indicate the gluelumps towards which we would
expect the respective potentials to converge.

The first 7 hybrid potentials are compatible with the degeneracies
suggested by Table~\ref{gluedegen}.  The next state is trickier
since it is not clear whether $2^{+-}$ or $3^{+-}$ is lighter. In the
figure we depict the case for a light $2^{+-}$. This would mean that
$(\Sigma_u^+, \Pi_u',\Delta_u)$ approach the $2^{+-}$ while
$(\Sigma_u^{-\prime},\Pi_u'',\Delta_u',\Phi_u)$ approach the $3^{+-}$.
Note that of the latter four potentials only data for $\Pi_u''$ and
$\Phi_u$ are available. Also note that the continuum states
$\Pi_u',\Pi_u''$ and $\Phi_u$ are all obtained from the same $E_u$
lattice representation.  For the purpose of the figure we make an
arbitrary choice to distribute the former three
states among the $E_u',E_u''$ and
$E_u'''$ lattice potentials.  To firmly establish their ordering one
would have to investigate radial excitations in additional lattice
hybrid channels and/or clarify the gluelump spectrum in more detail.
Should the $2^{+-}$ and $3^{+-}$ hybrid levels be inverted then
$(\Sigma_u^{-\prime},\Pi_u',\Delta_u,\Phi_u)$ will converge to the
$3^{+-}$ while $(\Sigma_u^+,\Pi_u'',\Delta_u')$ will approach the
$2^{+-}$.  We note that the ordering of the hybrid potentials, with a
low $\Sigma_u^+$, makes the first interpretation more suggestive.

Finally the $\Sigma_g^{+\prime\prime}$ potential seems to head towards
the $0^{++}$ gluelump but suddenly turns downward, approaching the
(lighter) sum of ground state potential and scalar glueball~\cite{glueball,glueball2}
instead.
The latter type of decay will eventually happen for all lattice
potentials but only at extremely short distances.  We also remark that
all potentials will diverge as $r\rightarrow 0$. This does not affect
our comparison with the gluelump results, since we have normalized
them to the $\Pi_u/\Sigma_u^-$ potentials at the shortest distance
available. (The gluelump values are plotted at $r=0$ to
simplify the figure.)

On a qualitative level the short-distance data are very consistent
with the expected degeneracies.  {}From the figure we see that at
$r\approx 2\,r_0\approx 1\,$fm the spectrum of hybrid potentials
displays the equi-distant band structure one would qualitatively expect
from a string picture. Clearly this region, as well as the cross-over
region to the short-distance behaviour $r_0<r< 2\, r_0$, cannot be
expected to be within the perturbative domain: at best one can possibly
imagine perturbation theory to be valid for the left-most 2 data
points. With the exception of the $\Pi_u$, $\Pi_u'$ and $\Phi_u$
potentials there are also no clear
signs for the onset of the short distance $1/r$ behaviour with a
positive coefficient as expected from perturbation
theory. Furthermore, most of the gaps within multiplets of hybrid
potentials, that are to leading order indicative of the size of the
non-perturbative $r^2$ term, are still quite significant, even at
$r=0.4\,r_0$; for instance the difference between the $\Sigma_u^-$ and
$\Pi_u$ potentials at this smallest distance is about 0.28~$r_0^{-1}\approx
110$~MeV.

\subsection{The difference between the $\Pi_u$ and $\Sigma_u^-$ hybrids}
\label{pisisplit}
\begin{figure}[ht]
\hspace{-0.1in}
\epsfxsize=4.8in
\centerline{
\centerline{\epsffile{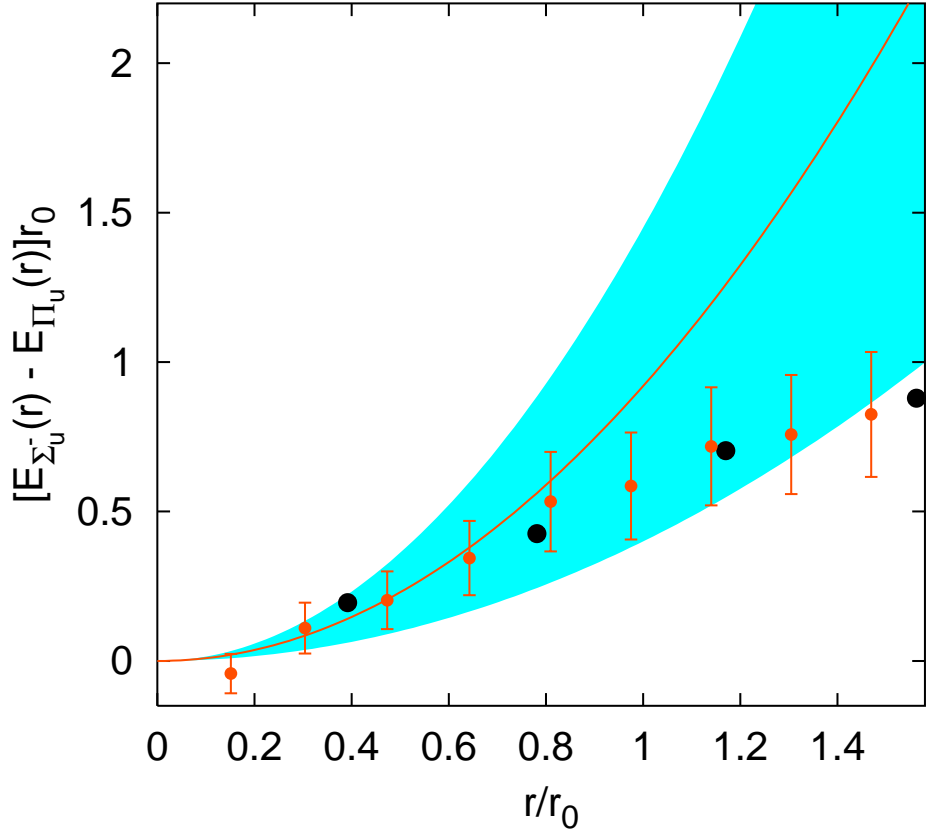}}
}
\caption {{\it Splitting between the $\Sigma_u^-$ and the $\Pi_u$ 
potentials, extrapolated to the continuum limit,
and the comparison with a quadratic fit to the
$r\siml 0.5\,r_0$ data points ($r_0^{-1}\approx 0.4$~GeV). The big
circles correspond to the data of Juge et al.~\cite{latticeshort2}, 
obtained at finite lattice spacing $a_{\sigma}\approx 0.39\,r_0$. 
The errors in this case are smaller than the symbols.}}
\label{r2}
\end{figure}

{}From the above considerations it is clear that
for a more quantitative study we need lattice data at
shorter distances. In this paper we have obtained these for the
lowest two gluonic excitations, $\Pi_u$ and $\Sigma_u^-$ (see
Sec.~\ref{lattice}). We display their
differences in the continuum limit in Fig.~\ref{r2}.
We see how these approach zero at small $r$, as expected
from the short distance expansion. pNRQCD predicts that the next effects should
be of $O(r^2)$ (and renormalon-free). In fact, we can fit
the lattice
data rather well with a $\Delta E_{\Pi_u-\Sigma_g^+} =A_{\Pi_u-\Sigma_u^-}r^2$ ansatz for
short distances, with slope
(see Fig. \ref{r2}),
\be
\label{eqa}
A_{\Pi_u-\Sigma_u^-}=0.92^{+0.53}_{-0.52}\,r_0^{-3}
\,,
\ee
where the error is purely statistical (lattice). This fit has been done using
points $r \siml 0.5\, r_0$. By increasing the fit range to $r \siml 0.8\, r_0$
the following result is obtained,
\be
A_{\Pi_u-\Sigma_u^-}=(0.83\pm 0.29)\,r_0^{-3}
\,,
\ee
indicating stability of the result Eq.~(\ref{eqa}) above.

In order to estimate systematic errors one can add a quartic term: $b\,r^4$
(only even powers of $r$ appear in the multipole expansion of this quantity).
If the result is stable, our determination 
of $A_{\Pi_u-\Sigma_u^-}$ should not change much. Actually this is what
happens. If we fit up to $r\siml 0.5\, r_0$, we obtain
the central value $A_{\Pi_u-\Sigma_u}r_0^3=0.93$ with a very small quartic
coefficient,
$b\,r_0^5=-0.05$. If we increase the range to $r\siml 0.8\, r_0$,
we obtain the same central value,
$A_{\Pi_u-\Sigma_u^-}r_0^3=0.93$, but with a slightly bigger quartic term,
$b\,r_0^5=-0.18$. Introducing the quartic term enhances the stability
of $A_{\Pi_u-\Sigma_u^-}$ under variations of the fit range.
{}From this
discussion we conclude that the systematic error is negligible, in comparison
to the error displayed in our result Eq.~(\ref{eqa}).

We remark that within the
framework of static pNRQCD and to second order in the
multipole expansion, one can relate
the slope $A_{\Pi_u-\Sigma_u^-}$ to gluonic correlators of QCD.   
 
\section{Lattice determination of hybrid potentials}
\label{lattice}
We extract the hybrid potentials in two sets of simulations,
using the Wilson gauge action on an isotropic lattice with volume
$24^3\times 48$ at $\beta=6.2$ ($a\approx 0.14\,r_0$)
as well as on three anisotropic lattices
with spatial lattice spacings $a_{\sigma}\approx 0.33,0.23,0.16\, r_0$,
respectively, with anisotropy $a_{\sigma}\approx 4\,a_{\tau}$. The former
result has been obtained in the context of the study of
Ref.~\cite{Collins:1997cb} (and has been published in
Ref.~\cite{Bali:2000vr})
while the simulation parameters, statistics and smearing
of the latter runs are identical
to those of Ref.~\cite{Bali:2000un}:
$(\beta,\xi_0)=(5.8,3.1),(6.0,3.2),(6.2,3.25)$.
The isotropic data are used as a consistency check and in
Sec.~\ref{seclatt} below, while we
extrapolate the data obtained on the anisotropic lattices to the continuum
limit.

Some time was spent on improving the shape of the hybrid creation
operators to optimize the overlap with the ground
state~\cite{Collins:1997cb}.
The $\Pi_u$ potential has been determined on-axis as well as
along a plane-diagonal, ${\mathbf r}/a_{\sigma}\propto (1,1,0)$, while the
$\Sigma_u^-$ potential has only been obtained on-axis.
Typically we achieved ground state overlaps of around 65~\%
for both potentials at $\beta=5.8$ and between 85~\% and 90~\%
at the larger two $\beta$ values. Typical fit ranges for one-exponential
fits to correlation functions for the
$\Pi_u (\Sigma_u^-)$ potential were $8\leq t/a_{\tau}\leq 18$, 
($9\leq t/a_{\tau}\leq 14$) at $\beta=5.8$,
$9\leq t/a_{\tau}\leq 24$ ($11\leq t/a_{\tau}\leq 21$) at $\beta=6.0$ and
$13\leq t/a_{\tau}\leq 30$ ($15\leq t/a_{\tau}\leq 25$) at $\beta=6.2$.
For all further details of the analysis we refer to Ref.~\cite{Bali:2000un}
where potentials between sources in non-fundamental representations of
$SU(3)$ were extracted using exactly the same methods.

\begin{figure}[h]
\hspace{-0.1in}
\epsfxsize=4.8in
\centerline{\epsffile{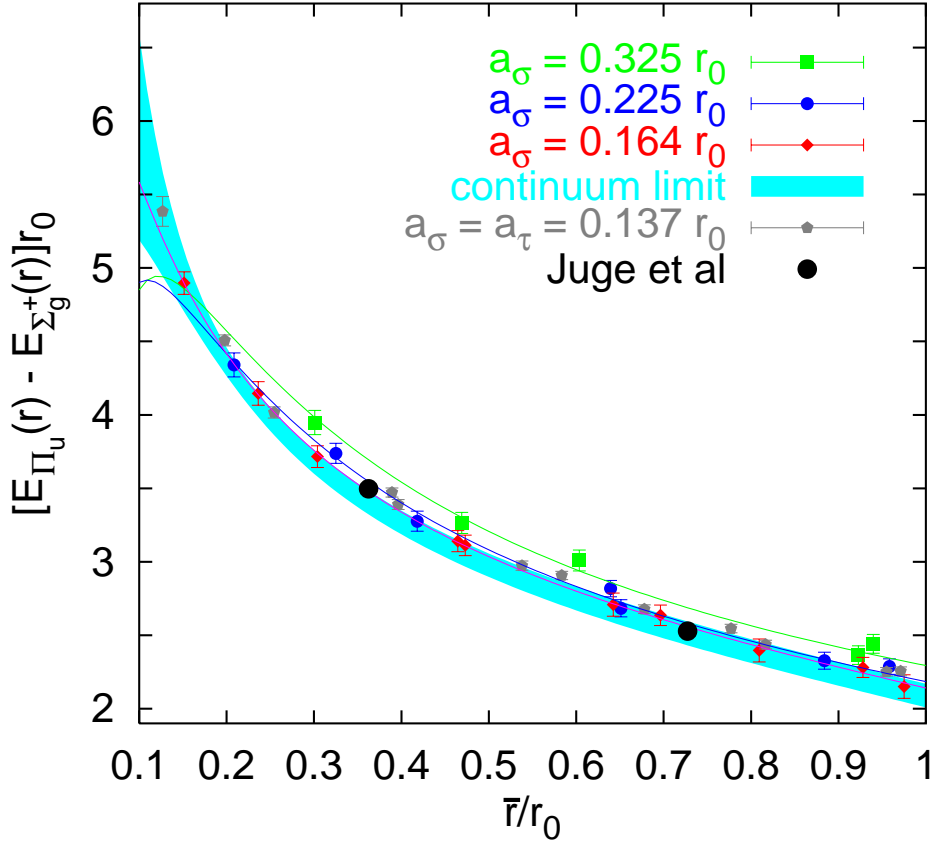}}
\caption {{\it Continuum limit extrapolation of the
difference between the $\Pi_u$ and the $\Sigma_g^+$
potentials vs.\ $\overline{r}=r[1+O(a^2/r^2)]$ as described in
the text [Eqs.~(\ref{overl}) and (\ref{latr})].
The Juge et al.\ data are from Ref.~\cite{latticeshort2}.}}
\label{continuum}
\end{figure}

Subsequently, the potentials as well as differences between potentials
have been extrapolated to the continuum limit.
As one such example we display the difference between the $\Pi_u$ and
the singlet potential in Fig.~\ref{continuum} for distances $r\leq r_0$.
In this extrapolation we somewhat deviate from Ref.~\cite{Bali:2000un}:
we follow Ref.~\cite{NS} in removing the lattice artifacts
to leading order in $\alpha_s$, by 
plotting the data as a function of the inverse
lattice Coulomb propagator,
\begin{equation}
\label{overl}
\overline{r}=a_{\sigma}\left[\frac{1}{{\mathbf R}}\right]_L^{-1}\,,
\end{equation}
rather than of $r$. The lattice Coulomb propagator for the Wilson
gauge action is given by,
\begin{equation}
\label{latr}
\left[\frac{1}{{\mathbf R}}\right]_L=4\pi\int_{-\pi}^{\pi}\!\frac{d^3Q}{\left(2\pi\right)^3}\,\frac{\cos({\mathbf Q}{\mathbf R})}{4\sum_i\sin\left(\frac{Q_i}{2}\right)}\,,
\end{equation}
and agrees with the continuum $1/R$-function up to $O(a^2/r^2)$
lattice artifacts.
${\mathbf R}={\mathbf r}/a$ denotes an integer valued three-vector and
the $Q_i=q_ia_{\sigma}$ are dimensionless.
For the $\Pi_u$ potential
this procedure
removes violations of rotational symmetry within the statistical errors
and brings the plane-diagonal
points in-line with the on-axis data. Unfortunately, we cannot perform
a similar internal test for the $\Sigma_u^-$ potential which we only determined
for on-axis separations.

The next step involved fitting differences between hybrid potentials
and $\Sigma_g^+$, $\Delta E_H=E_H-E_{\Sigma_g^+}$,
for $r\geq 2a$ to the phenomenological interpolation,
\begin{equation}
\Delta E_H(r)=c_1+\frac{c_2+c_3\ln(r)}{r}+c_4r^2\,,
\end{equation}
with parameters $c_i$.
We then extrapolated these interpolating curves
to the continuum limit,
assuming the leading order $a_{\sigma}^2$ dependence. This
was done
separately for different
pairs of two lattice spacings. The central value
of the extrapolation is given by the result obtained
from  the
$a_{\sigma}\approx 0.33\,r_0$ and $a_{\sigma}\approx 0.16\,r_0$
data sets. The error is estimated by the squared sum of
the statistical error of the fine lattice data set and the difference
between the above extrapolation and an extrapolation obtained
from the $a_{\sigma}\approx 0.23\,r_0$ and $a_{\sigma}\approx 0.16\,r_0$
data sets. With decreasing $r$ the interpolating fits become less well
constrained and hence the latter systematic uncertainty
increases. The resulting error band is depicted in Fig.~\ref{continuum}.
Reassuringly, the $a_{\sigma}\approx 0.16\,r_0$ data are already in agreement
with the continuum limit
and the $a_{\sigma}\approx 0.23\,r_0$ data agree within
errors: the fine lattice
data set effectively already corresponds to the continuum
limit. The more precise isotropic  reference data ($a\approx 0.14\,r_0$)
are also close to the continuum limit. We also notice that the
first three data points of the
coarse lattice data by Juge et al.~\cite{latticeshort2}
($a_{\sigma}\approx 0.39\,r_0$) are compatible with our extrapolation.
The same observations hold true for the $\Sigma_u^-$ potentials.

Rather than representing the continuum limit extrapolated potentials
by error bands, in the remaining parts of this paper we add the difference
between (finite $a$) interpolation and (continuum limit)
extrapolation to the fine lattice data points
and increase their errors by the systematic uncertainty involved in
the extrapolation.

\section{Static octet potential}
\label{statoct}
We will discuss the octet potential in the OS ($=$ ``pole mass'')
scheme, compute the normalization constant of the
renormalon and generalise the RS
renormalon subtracted scheme~\cite{RS} to this case. We will also discuss
the structure of power divergences on the lattice and the analogous
lattice scheme. Finally we discuss the running of the gluelump
mass from one scale to another.

\subsection{OS scheme for the octet potential}
\label{OSscheme}

The octet potential in the case $1/r \gg \lQ$ can be computed order by
order in perturbation theory. Nevertheless, it is not an IR
safe object \cite{short}. Its perturbative expansion reads, 
\be
\label{Vo}
V_o(r;\nu_{us}) \simeq \sum_{n=0}^\infty V_{o,n} \als^{n+1},
\ee
where we have made explicit its dependence on the IR cutoff
$\nu_{us}$ and $\als=\als(\nu)$, where we define 
$$
\nu {d \als \over d
\nu}=-2\als\left\{\beta_0{\als \over 4 \pi}+\beta_1\left({\als \over 4
\pi}\right)^2 + \cdots\right\}
.$$
In what follows we will always identify $\alpha_s$ with $\alpha_{\MS}$.
The first two coefficients $V_{o,0}$, $V_{o,1}$ are known,
as well as the leading-log terms of
$V_{o,3}$ \cite{short} (for the renormalization-group improved expression see
Ref.~\cite{RG}). Note, however, that these leading logs are not
associated to the first IR renormalon. For $V_{o,2}$ there 
exists a preliminary computation~\cite{Schroeder2},
\be
V_{o,2}= -{1 \over N_c^2-1}V_{s,2}+\delta V_{o,2}\,,
\ee
\be
\delta V_{o,2} \approx -{1 \over 2N_c}{1 \over (4\pi)^2}21 C_A^2{1
\over r}
\,,
\ee
which we will use in what follows. $V_{s,2}$ has been computed
in Ref.~\cite{2loop}. For 
$V_{o,3}$, we will use the renormalon-based estimate that we
obtain in Sec.~\ref{secpot} below (Table~\ref{tabv}). 

\begin{figure}[ht]
\makebox[1.0cm]{\phantom b}
\put(20,137){$r_0V_{o}(r)$}
\put(248,137){$r_0V_{o}(r)$}
\put(-30,10){\epsfxsize=8truecm \epsfbox{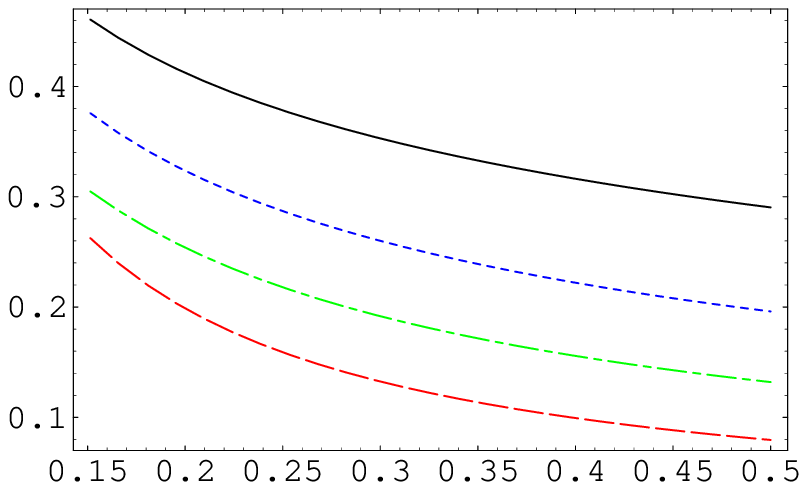}}
\put(210,10){\epsfxsize=8truecm \epsfbox{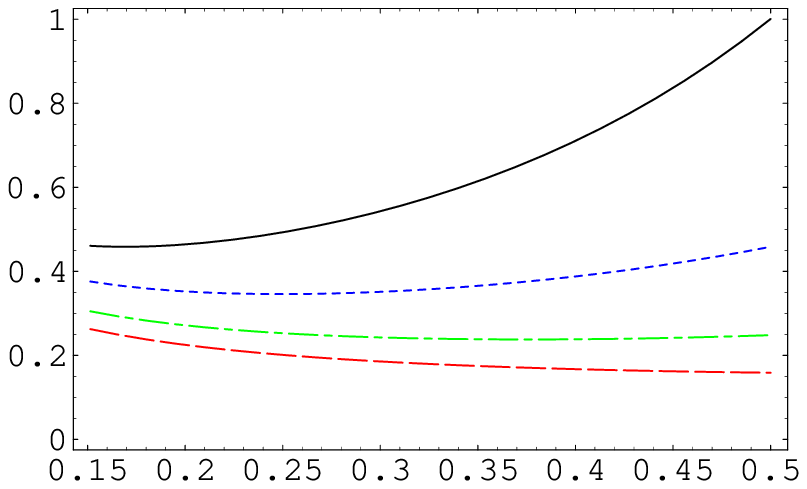}}
\put(100,1){$r/r_0$}
\put(300,1){$r/r_0$}
\put(-15,1){$a)$}
\put(220,1){$b)$}
\caption {{\it $r_0V_{o}(r)$ (the octet
potential in the OS scheme) at tree level (dashed lines), one-loop
(dashed-dotted lines), two loops (dotted lines) and three loops (estimate)
plus the leading single ultrasoft log (solid lines).
Fig.\ a) corresponds to the scale $\nu=\nu_i$ [cf.~Eq.~(\ref{eqnui})]
and Fig.\ b) to $\nu=1/r$. In both
cases, $\nu_{us}=2.5\,r_0^{-1}$. Only the solid curves depend on
this choice.}} 
\label{combinedOS}
\vspace{1mm}
\end{figure}

Studying the convergence of perturbation theory of the octet
potential in the OS scheme, conclusions similar to those in
Ref.~\cite{staticpot} are obtained. The poor convergence is
demonstrated in Fig.~\ref{combinedOS}, where we try two choices of
the scale $\nu$. In part (a) we use $\nu=\nu_i$ where,
\be
\label{eqnui}
\nu_i={r'}^{-1}=6.604\,r_0^{-1}\approx 2.6\,\mbox{GeV}\,,
\ee
corresponds to
the shortest distance $r'$ for which the continuum limit extrapolated
lattice potentials are available. In part (b) we vary $\nu=1/r$.
Obviously the curves depicted in
the two parts of the figure agree with each other at $r=r'\approx 0.15\,r_0$.
Note the difference in the vertical scale.

\subsection{Static octet potential normalization constant}
\label{secpot}
We define the Borel transform of the octet potential as follows,
\be\label{borelb}
V_o = \int\limits_0^\infty\mbox{d} t \,e^{-t/\als}\,B[V_o](t)
\,,
\qquad 
B[V_o](t)\equiv \sum_{n=0}^\infty 
V_{o,n} \frac{t^n}{n!} . 
\ee
The behaviour of the perturbative expansion
Eq.\ (\ref{Vo}) at large
orders is dictated by the closest singularity to the origin of its
Borel transform, which happens to be located at
$t=2\pi/\beta_0$. This singularity has two
sources: one is an UV renormalon which cancels with the
renormalon of twice the pole mass, the other is an IR renormalon that 
cancels with the UV renormalon of the gluelump energy. 
This result follows from the structure of the effective theory and 
the consequent factorization of the different scales in Eq.\ (\ref{EH}). 
Being more precise, the behaviour of the Borel transform of the static 
octet potential near the
closest singularity
to the origin [$u=1/2$ where we define $u=\beta_0 t/(4\pi)$] reads,
\be
\label{BVo}
B[V_{o}](t(u))=N_{V_o}\nu {1 \over
(1-2u)^{1+b}}\left(1+c_1(1-2u)+c_2(1-2u)^2+\cdots \right)+({\rm
analytic\; term})\,,
\ee
where by {\it analytic term}, we mean a function that is analytic up to
 the next IR renormalon at $u=3/2$. 
This 
dictates the behaviour of the perturbative expansion at large orders to be,
\be\label{generalV}
V_{o,n} \stackrel{n\rightarrow\infty}{=} N_{V_o}\,\nu\,\left({\beta_0 \over 2\pi}\right)^n
 \,{\Gamma(n+1+b) \over
 \Gamma(1+b)}
\left(
1+\frac{b}{(n+b)}c_1+\frac{b(b-1)}{(n+b)(n+b-1)}c_2+ \cdots
\right)
.
\ee
The 
structure of the renormalon is equal to the singlet one.
This is due to the fact that the number of octet fields is conserved at leading order 
in the multipole expansion and that the mass (potential) does not renormalize at 
this order. Therefore the values of the coefficients $b, c_1, c_2, \ldots$
above are the 
same as for the case of the static potential and the pole mass and can be found
in Refs.~\cite{Beneke2,renormalons,RS}. We display them here for ease of reference:
\be
b={\beta_1 \over 2\beta_0^2}\,,
\ee
\be
c_1={1 \over 4\,b\beta_0^3}\left({\beta_1^2 \over \beta_0}-\beta_2\right)
\,,
\ee
and 
\be
c_2={1 \over b(b-1)}
{\beta_1^4 + 4 \beta_0^3 \beta_1 \beta_2 - 2 \beta_0 \beta_1^2 \beta_2 + 
   \beta_0^2 (-2 \beta_1^3 + \beta_2^2) - 2 \beta_0^4 \beta_3 
\over 32 \beta_0^8}
\,.
\ee
The only difference with respect to
the static singlet potential is the value of $N_{V_o}$.
The cancellation of the renormalon  in 
Eq.~(\ref{EH}) requires,
\be
\label{eqnm}
2N_m+N_{V_o}+N_{\Lambda}=0
\,,\ee
where $N_{\Lambda}$ is the normalization constant of the renormalon
of the gluelump mass ($B[\Lambda]$ reads the same as
Eq. (\ref{BVo}), with the replacement $N_{V_o} \mapsto
N_{\Lambda}$).  Therefore, unlike in the static singlet potential
case, we cannot fix $N_{V_o}$ from the knowledge of $N_m$ alone. Yet we will
(approximately) determine $N_{V_o}$ from low orders in perturbation theory
of the octet potential.  Note also that $N_{\Lambda}$ is independent
of $H$, the specific gluonic content of the gluelump, since it only
depends on the high energy behaviour, which is universal.
To leading non-trivial order one obtains, $N_{V_o}=C_A/2-C_f,
N_{\Lambda}=C_A/2$.

\begin{figure}[h]
\hspace{-0.1in}
\epsfxsize=4.8in
\centerline{
\put(60,190){{\Large $N_{V_o}(x)$}}
\epsffile{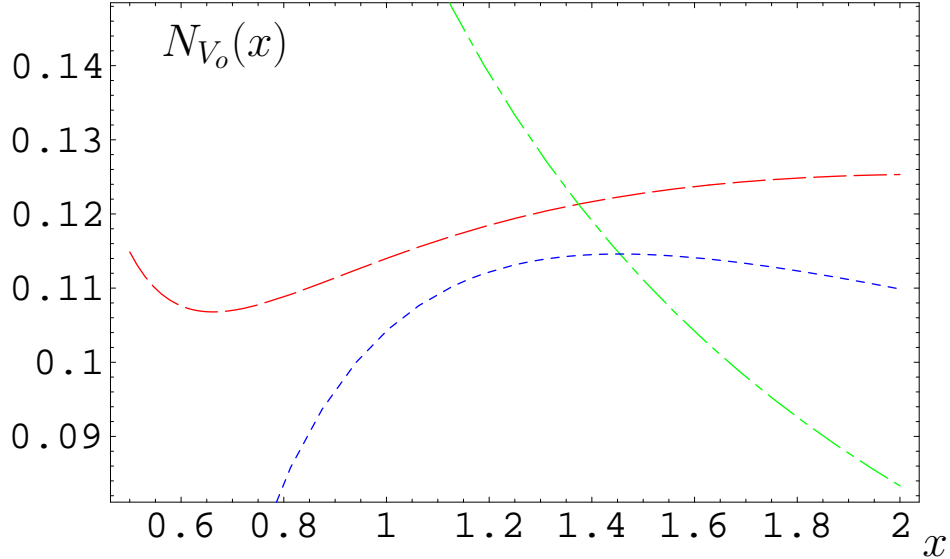}
\put(1,1){{\Large $x$}}
}
\caption {{\it $x \equiv {\nu r}$ dependence of
  $N_{V_o}$ for $n_f=0$ at LO (dashed-dotted line), NLO (dotted line)
and NNLO (dashed line).}} 
\label{figNVo}
\end{figure}

In analogy to  Refs.~\cite{Lee,RS,staticpot} we define the new function,
\bea
D_{V_o}(u)&=&\sum_{n=0}^{\infty}D_{V_o}^{(n)} u^n = (1-2u)^{1+b}B[V_{o}^{(0)}](t(u))
\\
\nn
&
=&N_{V_o}\nu\left(1+c_1(1-2u)+c_2(1-2u)^2+\cdots
\right)+(1-2u)^{1+b}({\rm analytic\; term})\,,
\eea
and try to approximately determine $N_{V_o}$ by using the
 first three coefficients of this series. In analogy to
Refs.~\cite{RS,staticpot}, we fix $\nu=1/r$ and
obtain [up to $O(u^3)|_{u=1/2}$],
\be
\label{nvo}
N_{V_o}=0.166667-0.0624292+0.00976333 = 0.114001 \,.  
\ee
The
convergence is rather good and, moreover, we have a sign
alternating series. In fact, the scale dependence is becoming milder 
when we go to higher orders (see
Fig.~\ref{figNVo}). Note that if the two loop coefficient $V_{o,2}$ had
been equal to that of the singlet case \cite{2loop} (with 
colour factor $C_f \mapsto C_A/2-C_f$), we would have
obtained $N_{V_o}=0.146542$.


We can now compute estimates for $V_{o,n}$ by using Eq.
(\ref{generalV}). These, as well as estimates for $V_{s,n}$,
are displayed in Table \ref{tabv} for $n_f=0$.
We can see
that the exact results are reasonably well reproduced.
Hence we feel confident
that we are near the asymptotic regime dominated by the first IR
renormalon and that for higher $n$ our predictions will accurately
approximate the exact results.

\begin{table}[h]
\addtolength{\arraycolsep}{0.2cm}
$$
\begin{array}{|l||c|c|c|c|c|}
\hline
{\tilde V}_{o,n}= r V_{o,n}  & {\tilde V}_{o,0} & {\tilde V}_{o,1} 
& {\tilde V}_{o,2} & {\tilde V}_{o,3} & {\tilde V}_{o,4} 
\\ \hline\hline
{\rm exact}\; (n_f=0) & 0.166667 & 0.305472 & 1.27419 & --- &
 ---  \\
{\rm Eq.}\; (\ref{generalV})\; (n_f=0) & 0.110552 & 0.244266 & 1.14193
  &  6.97413 &  54.4562 \\
 \hline
\hline
{\tilde V}_{s,n}= r V_{s,n}  & {\tilde V}_{s,0} & {\tilde V}_{s,1} 
& {\tilde V}_{s,2} & {\tilde V}_{s,3} & {\tilde V}_{s,4} 
\\ \hline\hline
{\rm exact}\; (n_f=0) & -1.33333 & -2.44378& -11.7893 & --- &
 ---  \\
{\rm estimate} \; (n_f=0) & -1.20643 & -2.66564 & -12.4616 & -76.1075 & -594.2718\\
 \hline
\end{array}
$$
\caption{{\it Values of $V_{o,n}$ with $\nu=1/r$: 
exact result (where available) and the
  estimate using Eq.\ (\ref{generalV}). We also display the estimates of
$V_{s,n}$ with $\nu=1/r$ (extracted from Ref.~\cite{RS}).}}
\label{tabv}
\end{table}

In order to avoid large corrections from terms depending on $\nu_{us}$,
the predictions should be understood with $\nu_{us}=1/r$ and later-on
one can use the renormalization group equations for the static potential
\cite{RG} to keep track of the $\nu_{us}$ dependence.

\subsection{RS scheme for the octet potential}
\label{secrs}
In Sec.~\ref{OSscheme} we have demonstrated the poor convergence of
the perturbative expansion of the octet potential in the OS
scheme. This bad behaviour is usually believed to be due to the
singularities in the Borel transform of the perturbative expansion.
Nevertheless, these singularities are fake since they cancel with
singularities in the matrix elements.  On the other hand this lack of
convergence of perturbation theory arises because at higher orders in
perturbation theory smaller and smaller momenta contribute to
the short-distance matching coefficients of the effective theory. This
clashes with the logic of scale separation in the EFT
formalism.  The solution advocated in Ref.~\cite{RS} was
to subtract this behaviour from the matching coefficients. At the
practical level this was implemented by subtracting the
Borel plane singularities
of the matching coefficients.  In
Refs.~\cite{RS,staticpot} this has been worked out for the pole mass
and the  static singlet potential and we refer to these references for
the definitions and further details. In particular Eq. (\ref{Es}) 
reads,
\be
\label{EsRS}
E_s(r)=2m_{\RS}(\nu_f)+V_{s,\RS}(r;\nu_f)+O(r^2)
\,,
\ee
where 
\be
m_{\RS}(\nu_f)= m_{\OS}-\delta m_{\RS}(\nu_f)
\,,
\ee
\be
V_{s,\RS}(r;\nu_f)=V_{s}(r)+2\delta m_{\RS}(\nu_f)\,,
\ee
and (in the above equation we have already used the fact that the renormalon 
of the singlet potential cancels with the one of minus twice the pole
mass)\footnote{Actually, throughout this paper we
use the RS' scheme as defined in Ref.~\cite{RS}
instead of the RS scheme, since we believe this to have a more
physical interpretation. For simplicity of notation we will however
refer to this modified scheme as ``RS scheme'', omitting the
``prime''.},
\be
\label{deltamRS}
\delta m_{\RS}(\nu_f)= \sum_{n=1}^\infty N_m\,\nu_f\,\left({\beta_0 \over
2\pi}\right )^n \als^{n+1}(\nu_f)\,\sum_{k=0}^\infty
c_k{\Gamma(n+1+b-k) \over \Gamma(1+b-k)} \,.  
\ee

\begin{figure}[h]
\hspace{-0.1in}
\epsfxsize=5in
\centerline{
\put(-30,190){{\Large $r_0V_{o,\RS}$}}
\epsffile{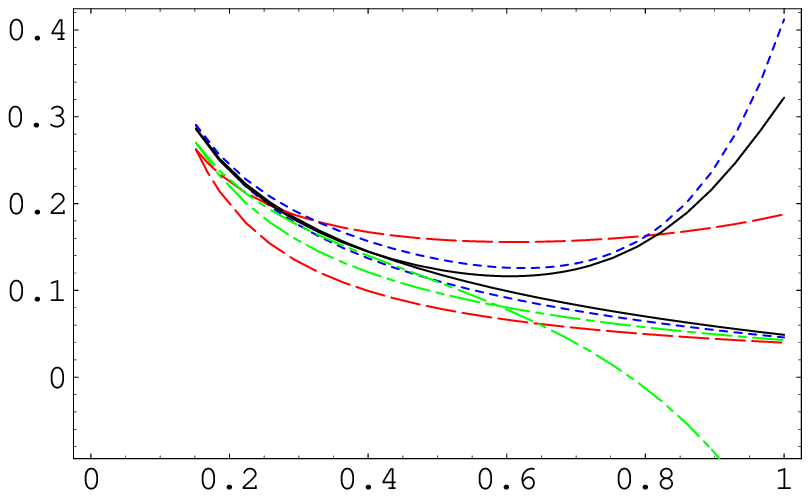}
\put(1,1){{\Large $r/r_0$}}}
\caption {{\it $r_0V_{o,\RS}$ at tree level (dashed lines), one-loop
(dashed-dotted lines), two loops (dotted lines) and three loops (estimate)
plus the leading single ultrasoft log (solid lines). For the scale of
$\als(\nu)$, we set $\nu=\nu_i$ (stable behaviour at large
distances) or $\nu=1/r$ (diverging at large distances).
We kept
$\nu_f=2.5\,r_0^{-1}$ fixed.}} 
\label{VRSnur}
\end{figure}

\begin{figure}[h]
\hspace{-0.1in}
\epsfxsize=5in
\centerline{
\put(150,190){$r_0(V_{o,\RS}(r)-V_{o,\RS}(r'))+C$}
\centerline{\epsffile{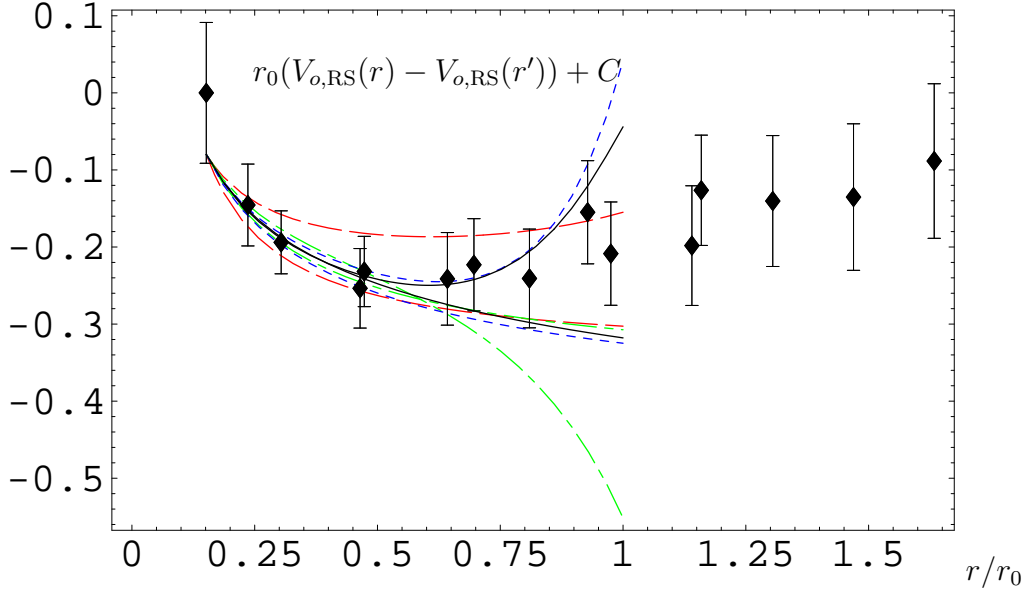}}
\put(-50,1){$r/r_0$}
}
\caption {{\it $r_0[V_{o,\RS}(r)-V_{o,\RS}(r')]+C$ at tree level
(dashed lines), one-loop
(dashed-dotted lines), two loops (dotted lines) and three loops (estimate)
plus the leading single ultrasoft log (solid lines) compared with the
non-perturbative continuum-limit results for
$E_{\Pi_u}(r)-E_{\Pi_u}(r')$ (symbols with error bars). For the scale of
$\als(\nu)$, we set $\nu=\nu_i=1/r'$ (stable behaviour at large
distances) or $\nu=1/r$ (diverging at large distances).
A (small) constant $C$ is arbitrarily
adjusted to show agreement with the lattice data.}}
\label{VRSlattnur}
\end{figure}

For the static hybrids, the spectrum reads,
\be
\label{EHRS}
E_H(r)=2m_{\RS}(\nu_f)+V_{o,\RS}(r;\nu_f)+\Lambda_H^{\RS}(\nu_f)+O(r^2)
\,.
\ee
Obviously, we have to define the octet
potential and the gluelump mass above. In the RS scheme the 
octet potential reads,
\be 
V_{o,\RS}(\nu_f)=V_o-\delta V_{o,\RS}
=\sum_{n=0}^\infty V^{\RS}_{o,n}\als^{n+1} \,, 
\ee 
where 
\be \delta
V_{o,\RS}\equiv \sum_{n=1}^\infty N_{V_o}\,\nu_f\,\left({\beta_0 \over
2\pi}\right )^n \als^{n+1}(\nu_f)\,\sum_{k=0}^\infty
c_k{\Gamma(n+1+b-k) \over \Gamma(1+b-k)} \,.  
\ee 
This specifies the gluelump mass which reads,
\be
\label{lamrs}
\Lambda_H^{\RS}(\nu_f)=\Lambda_H -\delta\Lambda_{\RS}(\nu_f),
\ee
where
\be
\delta\Lambda_{\RS}(\nu_f)
=
\sum_{n=1}^\infty
N_{\Lambda_H}\,\nu_f\,\left({\beta_0 \over 2\pi}\right )^n
\als^{n+1}(\nu_f)\,\sum_{k=0}^\infty c_k{\Gamma(n+1+b-k) \over
\Gamma(1+b-k)} \,.  
\ee 
Note that the potentials and $\Lambda_H^{\RS}$
depend on $\nu_f$ which, in the context of pNRQCD,
can be thought of as a matching scale between ultrasoft
and soft physics. In what follows, we will set
$\nu_f=2.5\,r_0^{-1}$. Results for different values of $\nu_f$
can be obtained using the running on $\nu_f$, which is renormalon
independent.

Analogously to the discussion of Ref.~\cite{staticpot}, we can study the
convergence of the perturbative expansion in the RS scheme. 
In Fig.~\ref{VRSnur} we can see
that the stability is greatly improved, compared to the OS scheme discussed
in the previous section.
No matter whether we choose to work with
$\als(\nu_i)$ or $\als(1/r)$, the expansions converge towards the same
curve.  In Fig.~\ref{VRSlattnur} we can also see that they agree
with the continuum
limit lattice data (we have
to subtract an unknown constant for this comparison).
In this figure the errors of $E_{\Pi_u}(r)-E_{\Pi_u}(r')$ for $r>r'$
are  purely statistical while the (strongly correlated) systematic error
of the continuum limit extrapolation is only displayed for the first
data point [$E_{\Pi_u}(r')-E_{\Pi_u}(r')=0$], where it is largest.

The price we pay to obtain convergent expansions in $\als$ for the
potentials is the introduction of power-like terms
(proportional to $\nu_f$, with logarithmic corrections).
This behaviour very much resembles that of
lattice regularization with a hard cut-off which we discuss below.

\subsection{Lattice scheme for the octet potential}
\label{seclatt}
It is conceptionally illuminating also to consider the situation in
lattice regularization. In this case, the inverse lattice spacing
$a^{-1}$ results in a hard UV cut-off of the gluon
momenta. Feynman diagrams are UV finite and EFT matrix
elements are manifestly renormalon-free as long as they are obtained
in non-perturbative numerical simulations.  The price paid is the
existence of power divergences $\propto a^{-1}$, which cannot be
eliminated in the continuum limit.

The analogy with the previous sections can be made quite evident. In
particular, all the quantities that we have defined in the OS and RS
schemes can also be defined in a lattice scheme. There are some
differences however. The lattice gluelump $\Lambda_H^L(a)$ has a
power divergence to start with (which can be traded in for a
renormalon ambiguity when subtracted in perturbation theory). In this
sense it is similar to $\Lambda_H^{\RS}(\nu_f)$.  While formally many
expressions resemble those of the RS case, $a^{-1}$ plays a slightly
different r\^ole than $\nu_f$ that separates soft from ultrasoft
scales since $a\ll r\ll \nu_f^{-1}$.  Another difference is that at
finite lattice spacings the potentials remain finite as $r\rightarrow
0$. In particular, we will see that gluelumps are the $r\rightarrow 0$
limits of hybrid potentials (at finite lattice spacing), in
perturbation theory as well as non-perturbatively. This should not be
surprising since the $r\rightarrow 0$ limit at finite lattice spacing
corresponds to the situation $r \ll a$. This means that the
ultraviolet cutoff $\sim a^{-1}$ is much smaller than $r^{-1}$ and that the
dynamical degrees of freedom are only the ultrasoft ones. Actually, in
this situation, $\nu_f$ and $1/a$ play an analogous r\^ole. 

Let us illustrate the above by first considering perturbation
theory, before discussing the scale separation and how the lattice
scheme translates into other schemes, at finite lattice spacings as
well as in the continuum limit.

For simplicity we will consider the Wilson discretization of the
continuum action.  In this case the ``lattice Coulomb term''
$[1/{\mathbf R}]_L$ takes the form Eq.~(\ref{latr}).  For instance one
can calculate the finite value, $[1/{0}]_L=3.17\ldots$.  Using this
notation, one finds the lattice results,
\bea
\label{vsl}
V_{s,L}({\mathbf r};a)&=&-C_f\alpha_La^{-1}\left[\frac{1}{\mathbf
R}\right]_L
\left[1+O(\alpha_L)\right]
+
2\delta m_{\rm stat}^L(a)\,,\\
\label{vol}
V_{o,L}({\mathbf r};a)&=&\left(\frac{C_A}{2}-C_f\right)\alpha_La^{-1}\left[\frac{1}{\mathbf R}\right]_L\left[1+O(\alpha_L)\right]+
2\delta m_{\rm stat}^L(a)\,,
\eea
where the ``self energy'' is given by,
\bea\nn
a\delta m_{\rm stat}^L(a)&=&\frac{C_f}{2}\alpha_L\left[\frac{1}{{0}}\right]_L+\cdots\\
&=&\frac{C_f}{2}\alpha_L\left(v_1+v_2\frac{\alpha_L}{4\pi}+v_3\frac{\alpha_L^2}{(4\pi)^2}+\cdots\right)\,.\label{v0}
\eea
Note that unlike in dimensional regularization,
by using a hard cut-off, such power divergencies
appear naturally as part of the perturbative expansion.
Eqs.~(\ref{vsl}) and (\ref{vol}) are both known to $O(\als^2)$ and
Eq.~(\ref{vsl}) (as well as the difference $V_o-V_s$) is
also known approximately
to $O(\als^3)$, up to $O(\als^3a^2/r^2)$
lattice corrections~\cite{Bali:2002wf}.
In pure gauge theory with Wilson action, the coefficients
of the expansion of $\delta m^L_{\rm stat}$
read~\cite{Duncan:1994uq,Martinelli:1998vt,Bali:2002wf,DiRenzo:2000nd},
\bea
\label{v1}
v_1&=&3.1759115\ldots,\\
\label{v2}
v_2&=&0.21003(5)\times 10^3\,,\\
\label{v3}
v_3&=&20.4(3)\times 10^3\,.
\eea
$\alpha_L=3/(2\pi\beta)$ denotes the lattice coupling 
at a scale $a^{-1}$ which can be
translated into other schemes such as $\overline{MS}$ by means of
a perturbative computation,
\begin{equation}
\alpha_L=\als(a^{-1})\left[1-b_1\frac{\als(a^{-1})}{4\pi}-
\left(b_2-2b_1^2\right)\frac{\als^2(a^{-1})}{(4\pi)^2}+\cdots\right]\,
\end{equation}
with~\cite{Christou:1998ws},
\bea
\label{eqb1}
b_1&\approx&73.93539066\ldots\,,\\
\label{eqb2}
b_2&\approx&b_1^2+1388.1645\,.
\eea

Let us now consider the singlet case. We have,
\be
E^L_{\Sigma_g^+}({\mathbf r};a)=V_{s,L}({\mathbf r};a)+\lQ \left[O(\lQ^2 r^2)
+O(\lQ^2a^2)+O(a^2/r^2)\right]\,,
\ee
where $\lQ$ represents a generic non-perturbative scale like $r_0^{-1}$.
The last two terms account for possible non-perturbative lattice artifacts,
which vanish as $a\rightarrow 0$.
{}From the quarkonium energy $E_s(r)$ at $r\gg a$,
we can non-perturbatively obtain
the heavy quark mass in a lattice scheme,
\be
\label{mLL}
m_L(a)=\frac{1}{2}\left[E_s(r)-E_{\Sigma^+_g}^L({\mathbf r};a)\right]
+O(a^2/r^2)\,.\\
\ee
By redefining,
\be
\overline{V}_{s,L}({\mathbf r};a)=V_{s,L}({\mathbf r};a)-2\delta m_{\rm stat}^L(a)\,,
\ee
we can then achieve formal correspondence to Eqs.\ (\ref{EsRS}) and (\ref{Es}),
respectively:
\bea
\label{EsL1}
E_s(r)&=&2m_L(a)+V_{s,L}({\mathbf r};a)+O(r^2)\\
&=&2m_{OS}+\overline{V}_{s,L}({\mathbf r};a)+O(r^2)
\label{EsL}
\,,
\eea
where the above two equations are correct up to $O(\lQ^2a^2)$ and $O(a^2/r^2)$
lattice corrections.

We can relate the heavy quark mass in the lattice scheme to the OS scheme,
\be
\label{mL}
m_{L}(a)=m_{\OS}-\delta m_{\rm stat}^L(a)\,.
\ee
$\delta m_{\rm stat}^L(a)$ contains the same renormalon
as $m_{\OS}$, such that Eq.\ (\ref{mL}) has good convergence
properties when expanded in terms of $\als$.
$m_L(a)$ is proportional to $a^{-1}$, with logarithmic, as well as
$O(a^2)$ lattice corrections.
One can convert
$m_L(a)$ order by order in perturbation theory into say
$m_{\MS}(\nu)$, without renormalon ambiguity.

In the lattice scheme we also have
$E^L_{\Sigma_g^+}(0;a)=V_{s,L}(0;a)=0$: the sources are ``smeared
out'' on a scale $a$ since the gluon, due to the UV cut-off, cannot
resolve structures smaller than the lattice spacing.  Consequently,
the Coulomb term does not diverge as $r\rightarrow 0$ but approaches a
finite value in units of $a$.  In perturbation theory, in the limit
$r\rightarrow 0$, the lattice $[1/{\mathbf R}]_L$ term exactly cancels
with $2\delta m_{\rm stat}^L$: the perturbative expansion of
$V_{s,L}({\mathbf r};a)$, Eq.~(\ref{vsl}) above, does not contain the
renormalon associated with the pole mass. Non-perturbatively, in the
limit $r\rightarrow 0$ the Wilson loop becomes a time independent
constant, such that $E^L_{\Sigma_g^+}(0;a)=0$ too.
As $r>0$ the perturbative $V_{s,L}$ acquires a power term.

\medskip

Next we consider the hybrid case. We can calculate the gluelump
mass in perturbation theory\footnote{In the context of perturbation theory we do not distinguish between different gluelumps since
the mass splittings have an entirely non-perturbative origin.},
\be
a\delta\Lambda_L(a)=\frac{C_A}{2}\alpha_L\left[\frac{1}{{0}}\right]_L+\cdots
=\frac{C_A}{2}\alpha_L\left(v_1+v_2\frac{\alpha_L}{4\pi}\right)+
\cdots\,,
\ee
where $v_1$ and $v_2$ are the same as for the case of $\delta m_{\rm stat}^L$
and can be found in Eqs.~(\ref{v1}) and (\ref{v2}) above. Note that
the $O(\alpha_s^3)$ term is expected to be different and is not
known at present. However, $\Lambda_H$ is related to the difference
between $V_o$ and $V_s$, such that any difference
with respect to the $v_3$
of Eq.~(\ref{v3}) above will be suppressed by a colour factor $1/N_c^2$.

\begin{figure}[h]
\hspace{-0.1in}
\epsfxsize=4.8in
\centerline{
\epsffile{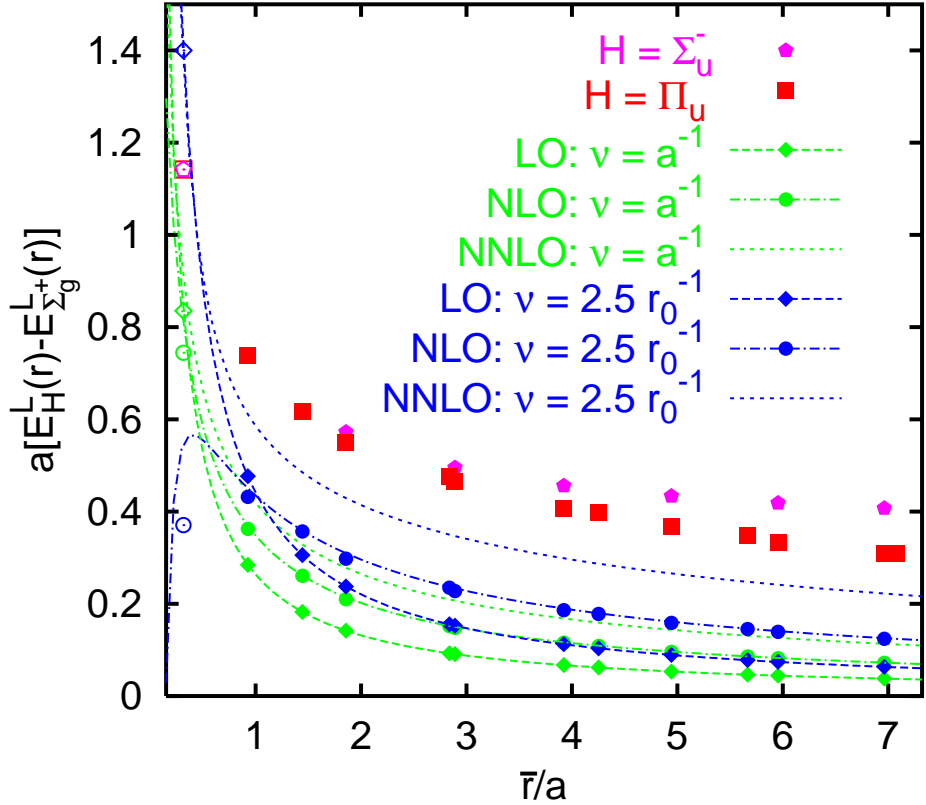}}
\caption {{\it Splitting between the lowest two hybrids 
and the $\Sigma_g^+$
potentials (pentagons and squares)
as a function of $\overline{r}/a$ [see Eq.~(\ref{overl})]
at a fixed lattice spacing, $a\approx 0.137\,r_0$, in comparison to
$V_{o,L}({\mathbf r};a)-V_{s,L}({\mathbf r};a)=\overline{V}_{o,L}({\mathbf r};a)-\overline{V}_{o,L}({\mathbf r};a)+\delta\Lambda_L(a)$
at tree level (dashed lines, diamonds), one loop (dashed-dotted lines, circles)
and two loops (dotted lines, $r\gg a$ estimates). The open symbols correspond to
the respective gluelumps, non-perturbatively (square with pentagon) and
in lattice perturbation theory (diamonds and circles).}}
\label{antonio}
\end{figure}

The tree level expression for $V_{o,L}$ is displayed in Eq.~(\ref{vol}).
While the perturbative
expansion of $V_{s,L}$ was unaffected by the renormalon of the
pole mass, the one of $V_{o,L}$ contains
the same renormalon as the expansion of $\delta\Lambda_L$.
For $r\gg a$ the
renormalon-free combination $V_{o,L}({\mathbf r};a)-\delta\Lambda_L(a)$
plays the r\^ole of $V_{o,\RS}(r;\nu_f)$ in Eq.~(\ref{EHRS}).
At $r=0$ we have,
$V_{o,L}(0;a)=\delta\Lambda_L(a)$ as well as 
the non-perturbative equality,
\begin{equation}
\label{footeq}
E^L_{\Pi_u}(0;a)=E^L_{\Sigma_u^-}(0;a)=\Lambda_B^L(a)\,.
\end{equation}
We redefine,
\be
\overline{V}_{o,L}({\mathbf r};a)=V_{o,L}({\mathbf r};a)
-\delta\Lambda_L(a)-2\delta m_{\rm stat}^L(a)\,,
\ee
to achieve formal correspondence with Eqs.~(\ref{EH}) and (\ref{eqgl}):
\be
\label{uros}
E_H^L({\mathbf r};a)-E^L_{\Sigma_g^+}({\mathbf r};a)=\Lambda_H^L(a)
+\left[\overline{V}_{o,L}({\mathbf r};a)-\overline{V}_{s,L}({\mathbf r};a)\right]+O(r^2)\,.
\ee
Note that $E^L_{H}({\mathbf r};a)= E_{H}(r)-2m_L(a)+O(a^2/r^2)$,
in analogy
to Eq.~(\ref{mLL}).
The combination,
\be
\overline{V}_{o,L}({\mathbf r};a)-\overline{V}_{s,L}({\mathbf r};a)=
\frac{C_A}{2}\alpha_La^{-1}\left(\left[\frac{1}{\mathbf R}\right]_L-
\left[\frac{1}{{0}}\right]_L\right)+O(\alpha^2)\,,
\ee
vanishes for $r=0$ and is renormalon-free. The same holds true for
$E_H^L(0;a)-E_{\Sigma_g^+}^L(0;a)-\Lambda_H^L(a)=0$: Eq.~(\ref{uros})
is not only valid for
$r> a$ but also for\footnote{Based
on the results of Sec.~\ref{seclaa} below as
well as of Ref.~\cite{glueball}, we know that
the $1^{+-}$ glueball will become
lighter than the gluelump $\Lambda_B^L(a)$ around
$a< r_c \approx r_0/7$, when using the Wilson action.
In fact we discussed a similar situation in
Sec.~\ref{sechsplit} above, for the $\Sigma_g^{+\prime}$ potential.
This limit is not yet relevant for the $\Pi_u$ and $\Sigma_u^-$ potentials
at the lattice spacings covered
in this paper. In the case $a<r_c$,
Eq.~(\ref{uros}) will still apply for $r_c^{-1}\gg r^{-1}\gg \lQ$, however,
Eq.~(\ref{footeq}) will become modified;
it would apply to the first radial excitations
in the hybrid channels rather than to the ground states,
until finally around $a\approx r_0/12$ a
continuum of two-glueball states is encountered.}
$r=0$.
We have,
\be
\Lambda_H^L(a)=\Lambda_H^{\OS}+\delta\Lambda_L(a)\,.
\ee
Note that the above equation is only correct up to
non-perturbative $O(\lQ^2a^2)$ contributions to $\Lambda^L_Hr_0$.
Again $\Lambda_H^L$ is renormalon-free but has a power divergence.
By subtracting $\delta\Lambda_L(a)$ order by order in perturbation
theory one can
obtain an on shell $\Lambda_H^{\OS}$, but at the price of a renormalon ambiguity.
Note the similarity between the above equation and Eq.~(\ref{lamrs}).

\begin{figure}[th]
\hspace{-0.1in}
\epsfxsize=4.8in
\centerline{
\epsffile{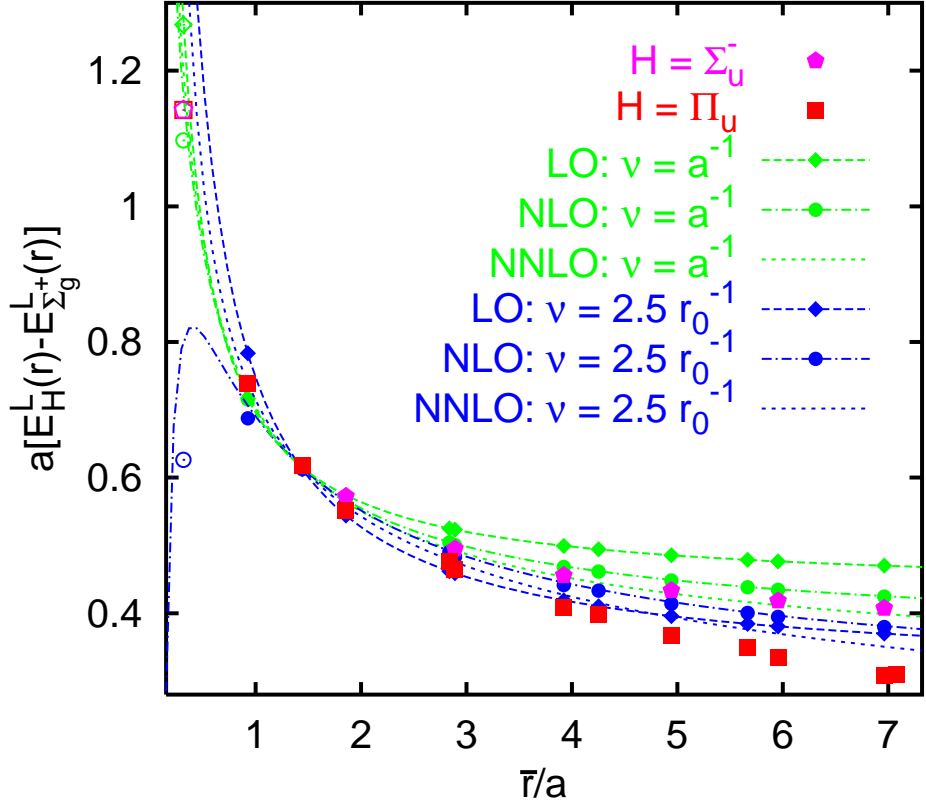}}
\caption {{\it Splitting between the lowest two hybrids 
and the $\Sigma_g^+$
potentials (pentagons and squares)
as a function of $\overline{r}/a$ [see Eq.~(\ref{overl})]
at a fixed lattice spacing, $a\approx0.137\,r_0$, in comparison with
$\overline{V}_{o,L}({\mathbf r};a)-\overline{V}_{o,L}({\mathbf r};a)+C$
at tree level (dashed lines, diamonds), one loop (dashed-dotted lines, circles)
and two loops (dotted lines, $r\gg a$ estimates).
The vertical normalization $C$ has been
adjusted to produce agreement at $r/a=\sqrt{2}$.
The open symbols correspond to
the respective gluelumps, non-perturbatively (square with pentagon) and
in lattice perturbation theory (diamond and circles).}}
\label{antonio2}
\end{figure}

In Fig.~\ref{antonio} we compare non-perturbative data on the splitting
between hybrid potentials with respect to the ground state potential
with the perturbative expectation. The data have been obtained
by us on an isotropic lattice at $\beta=6.2$ with lattice spacing
$a\approx 0.137\,r_0$. Both gaps, $E^L_{\Pi_u}-E^L_{\Sigma_g^+}$ (squares)
and $E^L_{\Sigma_u^-}-E^L_{\Sigma_g^+}$ (pentagons)
are plotted as a function of $\overline{r}/a$ [see Eq.~(\ref{overl})].
The differences are indicative of the size of the expected non-perturbative
$O(r^2)$ contributions.
We compare the non-perturbative data
to the perturbative expectation
for $V_{o,L}({\mathbf r};a)-V_{s,L}({\mathbf r};a)$. The latter perturbation theory will
suffer from the same renormalon ambiguity as $\delta\Lambda_L(a)$
and the difference between perturbation theory and
non-perturbative data corresponds to $\Lambda_B^{\OS}$.
The left-most points (open symbols) correspond to
the $\Lambda_B$ gluelump, plotted at $\overline{r}/a=[1/0]_L\approx
0.315$. 

The evaluation was done both in terms of $\alpha_s(a^{-1})$ and
$\alpha_s(\nu_f)$ where $\nu_f=2.5\,r_0^{-1}\approx 0.34\,a^{-1}$.
To simplify the figure we disregard the uncertainty in the
determination of $\Lambda_{\MS}=0.602(48)\,r_0^{-1}$~\cite{Lambda}.
At LO and NLO lattice perturbation theory 
results are available~\cite{Bali:2002wf}
(diamonds and squares). Since everything is plotted
as a function of $\overline{r}/a=[1/{\mathbf R}]_L^{-1}$ all diamonds
lie exactly on top of the dashed continuous $r\gg a$ curves while at small
distances there are
differences between the dashed-dotted NLO curves and the exact NLO
results (circles). In addition we plot the $r\gg a$ limits
to NNLO (dotted curves).
The shapes of the perturbative curves remain qualitatively stable
while the normalization is not convergent as the order of the expansion
is increased and is also strongly affected by the scale of $\alpha_s(\nu)$.
This behaviour reflects the presence of the renormalon of $\Lambda_B^{\OS}$, 
quite similar to what we can see in Fig. \ref{combinedOS}a.

By comparing with the renormalon-free right hand side
({\em rhs}) of Eq.~(\ref{uros}) a
better convergence can be achieved. However, such a comparison is
only possible up to $O(\alpha_s^2)$ as we do not exactly know the
$O(\alpha_s^3)$ contribution to the counterterm $\delta\Lambda_L(a)$ in the
lattice scheme.
Instead we choose to demonstrate the quality of the perturbative expansion
in Fig.~\ref{antonio2} by adding global normalization constants to all curves
in such a
way that agreement is produced at $r/a=\sqrt{2}$. (We shall return to
the question of renormalon cancellation in $\Lambda_H^L(a)$ in
Sec.~\ref{seclaa} below.) Indeed the differences between NNLO and
NLO are smaller than those between NLO and LO. Moreover, at higher
orders the scale dependence is reduced. The $\nu=a^{-1}$ curves seem
to describe the data better at small $r$ while the $\nu=\nu_f$
curves seem to work better at larger $r$. Up to distances as big as
$\overline{r}=r_0\approx 7.3\,a$ the perturbative curves seem to have
an accuracy better than the non-perturbative uncertainties, estimated
by the difference $E_{\Sigma_u^-}(r)-E_{\Pi_u}(r)$.

We mentioned above that while formally the lattice spacing $a^{-1}$
appears in the same places in the lattice scheme as the scale
$\nu_f$ did in the RS scheme of Sec.~\ref{secrs},
these two scales should not be confused
with each other as $a^{-1}> r^{-1}> \nu_f> \lQ$.
Conceptionally we have been discussing the
situation in which the potentials are evaluated in perturbation theory
at scales $\nu>\nu_f$ while $\Lambda_H^{\RS}$ is an
ultrasoft matrix elements, associated to physics at scales
smaller than $\nu_f$. The lattice encapsulates the same physical picture.
For instance, to each finite order in 
perturbation theory\footnote{In fact this is one way to define
$2\delta m^L_{\rm stat}$ in perturbation theory: the $r$-independent
part of the Fourier-transform of the momentum space lattice
potential~\cite{Bali:2002wf}.}, 
$V_{s/o,L}({\mathbf r};a)\stackrel{r\rightarrow\infty}{\longrightarrow}
2 \delta m_{\rm stat}^L(a)$ and $V_{o,L}(0,a)=\delta\Lambda_L(a)$:
the power contribution to the lattice mass $\delta m_{\rm stat}^L$
(whose perturbation theory
is affected by the IR renormalon of the on shell mass)
corresponds to the UV behaviour of the potentials
while the power contribution to $\Lambda_H^L$
(whose perturbative expansion has the UV renormalon of $\Lambda_H^{\OS}$)
is associated
with the low energy behaviour of $V_{o,L}$.
This is the same renormalon/power term
structure as in the continuum OS/RS schemes.

For $a\ll r<\lQ^{-1}$ lattice effects become invisible
and the formulae elaborated above apply under the replacement
$\RS\mapsto L$. To illustrate this quasi-continuum limit,
we eliminate the $a^{-1}$ dependence from the
expressions altogether, which is straight forward:
\bea
E_s(r)&=&2m_L(a)+E^L_{\Sigma_g^+}(r;a)\,,\\
\label{EsLtilde}
&=&2m_{L}(\nu_f^{-1})+V_{s,L}(r;\nu_f^{-1})+O(r^2)
\,,
\eea
where
\bea
\label{tildeV}
V_{s,L}(r;\nu_f^{-1})&=&V_{s,L}(r;a)-2\delta m_{\rm stat}^L(a)+2\delta m_{\rm stat}^L(\nu_f^{-1}),\\
m_L(\nu_f^{-1})&=&m_{L}(a)+\delta m_{\rm stat}^L(a)-\delta m_{\rm stat}^L(\nu_f^{-1})\,.
\eea
Note that the running from one scale to another is renormalon-free.
For the hybrid case we can directly write,
\be
E_H(r)=2m_L(\nu_f^{-1})+[V_{o,L}(r;\nu_f^{-1})-\delta\Lambda_L(\nu_f^{-1})]+\Lambda_H(\nu_f^{-1})+O(r^2)\,,
\ee
where the combination $V_{o,L}-\delta\Lambda_L$ replaces the $V_{o,RS}$
of Eq.~(\ref{EHRS}).

Finally, we mention that the situation $r=0$ on the lattice resembles
the $r\ll \nu_f^{-1}$ continuum situation. Unlike in the continuum, however,
on the lattice, even at $r=0$, all observables remain finite as
$a^{-1}$ provides us with a hard UV cut-off.

\subsection{Scale dependence}
\label{scadep}
As we have mentioned in the previous sections, the running of pole mass
and gluelump energies with $\nu_f$,
in the RS scheme, and with $a$, in the lattice scheme, is renormalon-free.
Therefore, the functional dependence
can be described by a convergent expansion in
perturbation theory. Nevertheless, in order to achieve the renormalon
cancellation, the same scale $\nu$ has to be used in the perturbative
expansion. This produces large logs if the scales $\nu_f$ and
$\nu_f'$ are widely separated and, eventually, some errors, if one
works to finite order in perturbation theory. In the RS scheme, there
exists a solution to this problem. Even though $\delta m_{\RS}(\nu_f)$
suffers from the renormalon ambiguity, the difference $\delta
m_{\RS}(\nu_f)-\delta m_{\RS}(\nu_f')$ is renormalon-free.
We can perform a resummation of $\delta m_{\RS}(\nu_f)$
with any prescription to avoid the singularity in the Borel
plane since it will cancel in the difference. We will take here the
Principal Value (PV) prescription, which yields,
\be
\label{PV}
\delta m_{\RS}^{\rm PV}(\nu_f)= N_m\nu_f\als(\nu_f)\sum_{s=0}^\infty
c_s 
\left[
D_{b-s}\left( -{2\pi \over \beta_0\als}\right) 
-1
\right]
\,,  
\ee 
where
\be
\label{DPV}
D_{b}(-x)=x
\left\{
e^{-x}(-x)^b\left[\Gamma(-b)-\Gamma(-b,-x)\right]-\cos(\pi b)e^{-x}\Gamma(-b)
\right\}
\,,
\ee
and,
\be
\Gamma(b,x)=\int_x^\infty\! dt\, t^{b - 1} e^{-t}\,,
\ee
denotes the incomplete $\Gamma$ function.

\begin{figure}[h]
\hspace{-0.1in}
\epsfxsize=4.8in
\centerline{
\put(100,190){{\Large$-\delta m_{\RS}^{\rm PV}(\nu_f)+\delta m_{\RS}^{\rm PV}(\nu_f')$}}
\epsffile{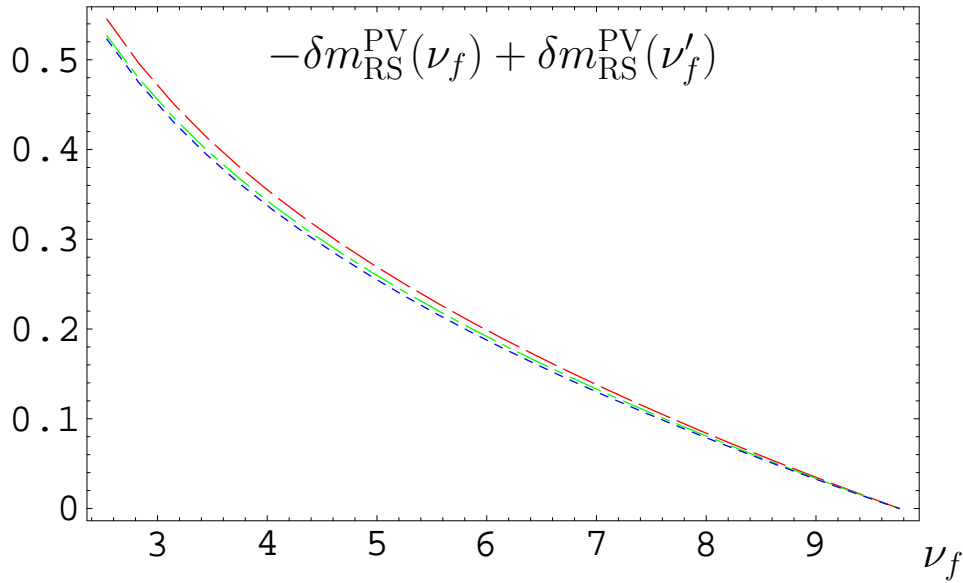}\put(1,1){{\Large $\nu_f$}}}
\caption {{\it $-\delta m_{\RS}^{\rm PV}(\nu_f)+\delta m_{\RS}^{\rm PV}(\nu_f')$
at LO (dashed line), NLO (dashed-dotted line) and NNLO (dotted line) according to the sum in Eq.\ (\ref{PV}). We take
$\nu_f'=9.76\, r_0^{-1}$. }}
\label{PVsums}
\end{figure}

The second term in Eq. (\ref{DPV}) corresponds to $\Lambda_{\MS}$,
once introduced in the sum of Eq. (\ref{PV}). It cancels from
the combination\footnote{
One may wonder if this cancellation
materializes itself in practice since we only
know the first three terms of
the series. However, we checked this numerically
and the results turned out to be virtually indistinguishable.},
$\delta m_{\RS}^{\rm PV}(\nu_f)-\delta m_{\RS}^{\PV}(\nu_f')$,
and we will not consider it any longer. The sum of Eq.\ (\ref{PV})
represents softer and softer singularities in the Borel
plane. Therefore, we expect at least the difference $\delta
m_{\RS}^{\PV}(\nu_f)-\delta m_{\RS}^{\PV}(\nu_f')$ to
converge (although, obviously, we have
no mathematical proof of this). Since the first three terms are
known we can check if this actually happens. We can see that this is
so with a high degree of confidence in Fig.\ \ref{PVsums}.

\begin{figure}[h]
\hspace{-0.1in}
\epsfxsize=4.8in
\centerline{
\put(150,190){{\Large$-\delta m_{\RS}(\nu_f)+\delta m_{\RS}(\nu_f')$}}
\put(170,140){{\Large LO}}
\put(80,80){{\Large NLO}}
\put(80,150){{\Large NNLO}}
\put(60,120){{\Large PV}}
\epsffile{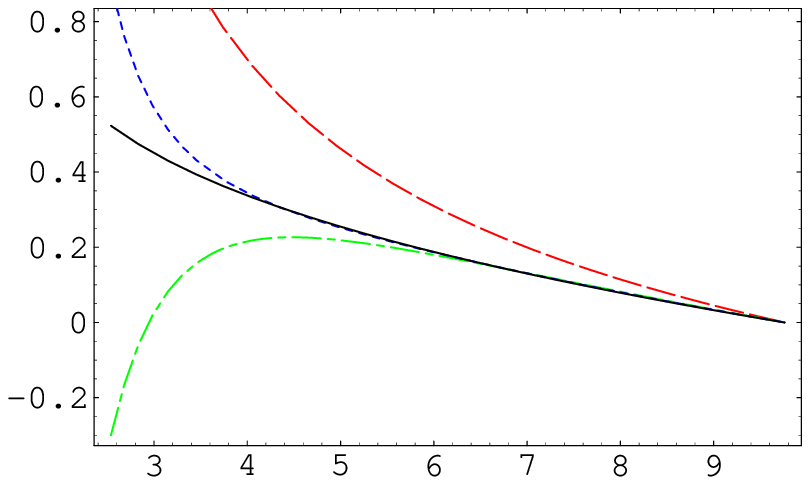}\put(1,1){{\Large $\nu_f$}}}
\caption {{\it $-\delta m_{\RS}(\nu_f)+\delta m_{\RS}(\nu_f')$
at LO (dashed line), NLO (dashed-dotted line) and NNLO (dotted line)
in perturbation
theory [see Eq.\ (\ref{FO}) with $\nu=\nu_f$) versus the Principal Value result (solid line). We take
$\nu_f'=9.76\, r_0^{-1}$. }}
\label{PVvsFO}
\end{figure}

We can also compare $-\delta m_{\RS}^{\rm PV}(\nu_f)+\delta
m_{\RS}^{\rm PV}(\nu_f')$ with the corresponding difference, calculated at
finite order in perturbation theory:
\bea
\label{FO}
&&-\delta m_{\RS}(\nu_f)+\delta m_{\RS}(\nu_f')
=
-
{\nu_f'-\nu_f \over2}{\tilde V}_{s,1} \als^2(\nu)
\\
\nn
&&
\qquad
\qquad
-
\left\{
{\nu_f'-\nu_f \over2}{\tilde V}_{s,2} +
\left[{\nu_f\over 2}{\beta_0 \over \pi}\ln\left({\nu_f \over \nu}\right)-{\nu_f'\over
2}{\beta_0 \over \pi}\ln\left({\nu_f' \over \nu}\right)\right]{\tilde V}_{s,1}\right\}\als^3(\nu)+\cdots
\,. 
\eea
We depict this comparison in Fig.\ \ref{PVvsFO}, where we take
$\nu=\nu_f$ to minimise one of the logs. We see how the finite order
results approach the PV curve\footnote{For finite
order computations we take $\als$ with one, two, three etc.\ loop
running according to the order in
$\als$ at which we work.
If instead, we  use $\als$ with 4-loop running (the
highest accuracy known until now) the convergence to the PV
result is accelerated.}, which we will use in what follows
wherever we need the running.

A similar behaviour holds if, instead of $\delta m_{\RS}$, we study
$\delta \Lambda_{\RS}$. 

For the lattice scheme we cannot perform an analytical resummation
as higher order terms are unknown. On the other hand, there exist
non-perturbative lattice determinations of the static masses
[$\Lambda_H^L(a)$ and ${\mathcal E}(a)=\overline{\Lambda}^L(a)$]
for different lattice
spacings. They provide us with non-perturbative measurements of the
running against which the finite order results can be tested.
It is also possible to relate results in both schemes by 
perturbative renormalon-free expressions.
We will investigate both, the running within the lattice scheme
and the translation between both schemes
in Secs.~\ref{seclaa} \ref{ssssrs} and \ref{v02} below.

\section{Phenomenological analysis of the gluelump spectrum}
\label{pheno}
We will determine the lowest gluelump energy, $\Lambda_B$, from two different
observables in two different schemes:
from the non-perturbative difference
$E_{\Pi_u}(r)-E_{\Sigma_g^+}(r)$ in the continuum limit in
the RS scheme as well as from gluelump energies $\Lambda_B^L(a)$
obtained at finite lattice spacings in a lattice scheme. Lattice and RS scheme
can be translated into each other and we find internal consistency.
We finally present results on the whole gluelump spectrum and compare our findings
to previous literature. 

The situation discussed here is similar to the one encountered
in the ``binding energy'' in static-light systems which we will
address in Sec.~\ref{statlight} below. These mesons
very much resemble
gluelumps, with the only difference that the source is in the
fundamental representation and screened by a light quark rather than
by a gluonic operator.

\subsection{Determination of $\Lambda_B^{\RS}$
from the static potentials}
\label{RSglump}
We intend to determine $\Lambda_{B}$ from the hybrid potentials.  For
this purpose we will use our $n_f=0$ lattice continuum limit data on
$\Delta E_{\Pi_u}(r)=E_{\Pi_u}(r)-E_{\Sigma_g^+}(r)$ as obtained in
Sec.~\ref{lattice}.  Using this difference allows us to eliminate the
power divergence that appears in lattice simulations of
the potentials (or, in the continuum OS scheme,
the renormalon associated with the
pole mass). Note that the difference has a well defined continuum
limit. It is also interesting to see that the large distance linear term
is cancelled as well.  At the same time, $\Lambda_B$ will still
additively contribute to this combination, see Eq.~(\ref{eqgl}).  In
order to extract this non-perturbative constant, the perturbative
difference between octet and singlet potentials has to be subtracted.
For a reliable determination, the perturbative series has to be well
defined and show convergence. However, this is complicated by the
contribution from the renormalon discussed above and can only
be achieved in a scheme where such renormalon singularities are taken
into account. We have worked out the RS scheme in Sec.~\ref{secrs},
which is well suited for this purpose.

\begin{figure}[h]
\hspace{-0.1in}
\epsfxsize=4.8in
\centerline{
\put(100,190){{\Large$r_0((V_{o,\RS}-V_{s,\RS})(r)+\Lambda_B^{\RS})$}}
\epsffile{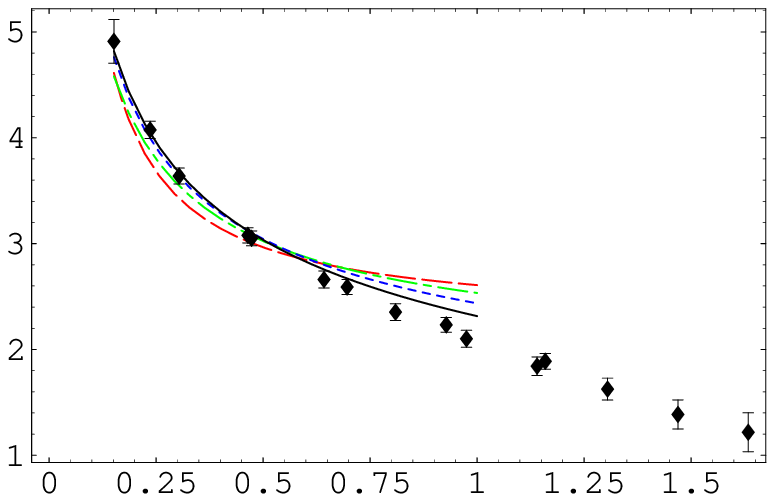}\put(1,1){{\Large $r/r_0$}}}
\caption {{\it Splitting between the $\Pi_u$ and the $\Sigma_g^+$
potentials and the comparison with Eq. (\ref{fit}) for $\nu=\nu_i$ [see 
Eq.~(\ref{eqnui})] at
$\nu_f=2.5\,r_0^{-1}$. $r_0[(V_{o,\RS}-V_{s,\RS})(r)+\Lambda_B^{\RS}]$
is plotted  at tree level (dashed line), one-loop
(dashed-dotted line), two loops (dotted line) and three loops (estimate)
plus the leading single ultrasoft log (solid line).}}
\label{VRSlattnu}
\end{figure}


\begin{figure}[h]
\hspace{-0.1in}
\epsfxsize=4.8in
\centerline{\put(150,190){{\Large 
$r_0((V_{o,\RS}-V_{s,\RS})(r)+\Lambda_B^{\RS})$}}
\centerline{\epsffile{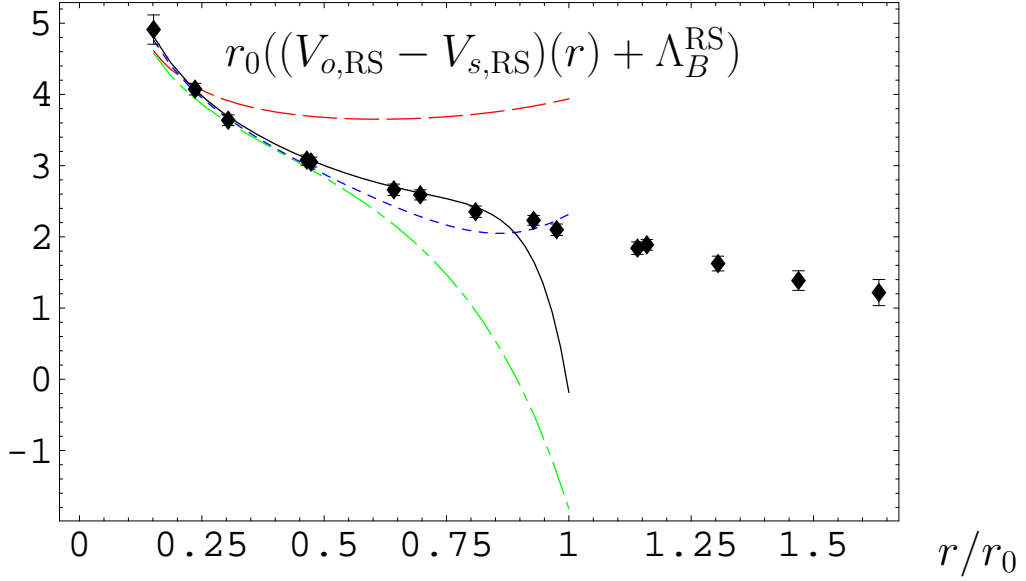}}
\put(-50,1){{\Large $r/r_0$}}}
\caption {{\it Splitting between the $\Pi_u$ and the $\Sigma_g^+$
potentials and the comparison with Eq. (\ref{fit}) with $\nu=1/r$
for $\nu_f=2.5\,r_0^{-1}$. $r_0[(V_{o,\RS}-V_{s,\RS})(r)+\Lambda_B^{\RS}]$
is plotted versus r
at tree level (dashed line), one-loop (dashed-dotted line), two-loops (dotted
line) and three loops (estimate) plus the RG expression for the
ultrasoft logs (solid line).}}
\label{VRSlattr}
\end{figure}
 
We fit
$\Lambda_{B}$ using the following equality (see Figs.~\ref{VRSlattnu}
and \ref{VRSlattr} for the quality of the fit):
\be
\label{fit}
E_{\Pi_u}(r)-E_{\Sigma_g^+}(r)=\Lambda_{B}^{\RS}(\nu_f)+V_{o,\RS}(r;\nu_f)
-V_{s,\RS}(r;\nu_f)
\,,
\ee
where the non-perturbatively obtained left hand side ({\em lhs})
is renormalon-free
but on the {\em rhs}
the renormalon can be shifted between the
two contributions, the ultrasoft matrix element $\Lambda_B$ and 
the soft Wilson coefficient $V_o-V_s$, at a given order of perturbation theory.
This is why we have to specify the scheme, the RS scheme in our case, which
we use to eliminate (or to reduce) this ambiguity.

We fix $\nu_f=2.5\,r_0$ and the final result reads, 
\be
\label{fitLambdaRSB}
\Lambda_{B}^{\RS}(\nu_f=2.5\,r_0^{-1})=\left[2.25\pm 0.10({\rm latt.})\pm 0.21
({\rm th.})\pm 0.08 (\Lambda_{\MS}) \right]\,r_0^{-1}
\,.
\ee
Note that $\Lambda_B$ is the only fit parameter. Also note that the above
value corresponds to the $n_f=0$ case.
The errors of this determination stem from several sources (for the above 
fit we use lattice data up to distances of around 0.5 $r_0$):

{\bf 1)} ``latt." denotes the statistical error of the fit: $\pm 0.10$.

{\bf 2)} ``th." stands for the theoretical errors. We first consider
the error due to the truncation of the perturbative series (higher
orders in perturbation theory/scale dependence). We obtain a first
estimate by performing the perturbative expansion in $\als(\nu_i)$
or in $\als(1/r)$. This provides us with an estimate of neglected subleading
logarithms.  Actually, in both cases one and the same number,
$\Lambda_{B}^{\RS}\approx 2.25\, r_0^{-1}$, is obtained, which we
take as our central
value. The effects of higher orders in perturbation theory are
estimated by considering the convergence of the determination of
$\Lambda_{B}^{\RS}$ at each order in perturbation theory. Working with
$\als(\nu_i)$, the series $\{2.43,2.37,2.28,2.25\}$ is
obtained. This series seems to show convergence for the last terms. In
any case, the corrections are small.  Working with $\als(1/r)$, the
series $\{2.00,2.40,2.31,2.25\}$ is obtained. This series is clearly
convergent although the corrections are larger than when using
$\als(\nu_i)$ as the expansion parameter. To be
conservative we will take the difference between the last two terms as
the error made by truncating the perturbative series: $\pm
0.06$. There is also some source of error from the normalization
constant of the renormalon of the singlet and octet potential. For the
singlet potential (following Ref.~\cite{RS}) we estimate a $10 \%$ error in
$N_{V_s}$, which produces a $\pm 0.10$ error. For the octet potential,
the error is very small compared with other sources of error. Even if,
conservatively, we consider the general shift produced by setting
$\delta V_{o,2}=0$ (note that this also accounts for the error in
perturbation theory of the octet static potential) our result only
changes by $\approx 0.01/0.02$. We will neglect this error to avoid
double counting. In the above analysis we have neglected 
non-perturbative effects. On general grounds they have the structure
\be
\label{deltanp}
\delta_{np}(E_{\Pi_u}-E_{\Sigma_g^+}) \simeq r^2F\left( V_o^{\RS}-V_s^{\RS}
\over \lQ \right)+B\,r^2
\,.
\ee
The $B\,r^2$ term is due to ${\bf r} \cdot O^\dagger {\bf E} O$ type
contributions in the pNRQCD Lagrangian (see Ref.\ \cite{pNRQCD} for
details). The other term in Eq.\ (\ref{deltanp}) is due to ${\bf r}
\cdot O^\dagger {\bf E} S$ type contributions.
This produces a perturbative mass gap.
$F$ is the convolution of a short distance and a long distance
piece, depending on the ratio of $V_o^{\RS}-V_s^{\RS}$ over the
masses of the gluelumps. For the purpose of estimating
the uncertainty
it seems reasonable to keep the lowest order in this
expansion. This is equivalent
to having a quadratic contribution,
\be
\delta_{np}(E_{\Pi_u}-E_{\Sigma_g^+}) \simeq A_{\Pi_u-\Sigma_g^+}r^2
\,.
\ee
If we introduce this term into the fit, we obtain $r_0\Lambda_B\approx 2.30$
[working with $\als(1/r)$] with $A_{\Pi_u-\Sigma_g^+} \simeq -0.4\,r_0^{-3}$.
We take the difference
as an indication of the error due to non-perturbative effects.
By summing linearly all the above errors we obtain $\pm 0.21$.

{\bf 3)} ``$\Lambda_{\MS}$'': this error is due to the uncertainty in
$\Lambda_{\MS} =[0.602 \pm 0.48]\,r_0^{-1}$~\cite{Lambda}: $\pm 0.08$.

We have performed the fit using lattice data within a window of
inverse distances ranging from about $\nu_i\approx 2.6$~GeV down to
$\nu_f\approx~1$~GeV.  From the plots (see Figs.\ \ref{VRSlattnu} and
\ref{VRSlattr}) one can actually see that the curves follow the
lattice data up to values $r \siml r_0$. This corresponds to
very low energies ($< 500$ MeV).  Being conservative, we will not
use data determined at these low energies without a better understanding of
the dynamics. Nonetheless, such a fit would actually produce very
similar numbers to the ones quoted above. This is even more so if a
quadratic term is included. In general, introducing more lattice
points reduces the statistical errors (``latt.''). Including a
quadratic term will reduce the theoretical error on $\Lambda_B$ since
some of the changes that occur when altering the order of perturbation
theory can be absorbed into a variation of
$A_{\Pi_u-\Sigma_g^+}$. However, the addition of a second fit
parameter increases the statistical error and also the uncertainty due
to $\Lambda_{\MS}$.  We conclude that while the individual errors
depend on the precise fitting details the total error remains
remarkably stable.

One might ask whether, in addition to $\Lambda_B$, a reliable value of
$A_{\Pi_u-\Sigma_g^+}$ can be obtained. This however would require
more lattice data at short distances as well as a more 
detailed understanding of the $r^2$ renormalon of the static singlet
potential.

We do not consider the $\Sigma_u^-$ data in this section as we have
already established in Sec.~\ref{pisisplit}
above that the difference with respect to the
$\Pi_u$ potential is proportional to $r^2$ to leading order. Hence we
cannot obtain any independent new information on $\Lambda_B$ from
these data, that have larger statistical errors.

\subsection{Determination of $\Lambda_B^{\RS}$ from $\Lambda_B^{L}$}
\label{seclaa}

There exists a direct determination of $\Lambda_B^L(a)$ (the
$1^{+-}$ or $B$ gluelump) by Foster and
Michael~\cite{FM}.
The numerical values are displayed in Table~\ref{tabgl}, where
we used the same $r_0/a$ values as were used in this reference. It is
clear from the discussion in Sec.~\ref{seclatt} that these
are perfectly sensible numbers if
incorporated into a global scheme with renormalon cancellation, for
instance, with the potentials also defined in the lattice scheme as in
Sec.~\ref{seclatt}.
In doing this we are able to independently determine
$\Lambda_B$ in a
different scheme. Consistency would require that after translating
the lattice into the RS scheme
the results should agree with each other. We will check this in this
section. 

\begin{table}[h]
\begin{center}
\begin{tabular}{|l||c|c|c|c|c|}
\hline
$a^{-1}r_0$&$\Lambda_B^Lr_0/\Lambda_B^{\RS}r_0$(LO)&$\Lambda_B^{\RS}r_0$(NLO)&$\Lambda_B^{\RS}r_0$(NNLO)&$\Lambda_B^{\RS}r_0$(NNNLO*)&
$\Lambda_B^{\RS}r_0$\\\hline\hline
2.94&5.33(10)&1.59(19)&2.82(12)&2.37(15)&2.41(10)\\
5.27&6.99(05)&1.97(17)&3.20(10)&2.88(12)&2.89(13)\\
7.32&8.36(05)&2.21(17)&3.55(10)&3.25(13)&3.16(13)\\\hline
\end{tabular}
\end{center}
\caption{{\it The inverse lattice spacing, the mass of the $1^{+-}$
gluelump $\Lambda_B^L$ in the lattice scheme, as well as its
conversion to the RS scheme to different orders in perturbation
theory. NNNLO* stands for an estimate obtained neglecting $1/N_c^2$ corrections, for
details see the text. In the last column, we state the values of
$\Lambda_B^{\RS}(a^{-1})$ using Eq.\ (\ref{fitLambdaRSB}) and the running
according to the PV prescription, Eq.~(\ref{PV}).
The errors only incorporate the statistical uncertainties
as well as the 8~\% uncertainty in $\Lambda_{\overline{MS}}r_0$, added in
quadrature, but no estimates of ``theoretical'' errors.}}
\label{tabgl}
\end{table}

 The master formula that relates the lattice and the RS scheme reads
(known up to NNLO),
\be
\label{trans}
\Lambda_H^{\RS}(\nu_f)=\Lambda_H^{L}(a)-\left[\delta \Lambda_H^L(a)+\delta
\Lambda_H^{\RS}(\nu_f)\right]
\,.
\ee
Both, $\Lambda_H^{\RS}$ and $\Lambda_H^L$ have a power-like dependency
on $\nu_f$ and $a^{-1}$, respectively, but are renormalon-free,
$\Lambda_H^L$ exactly and $\Lambda_H^{\RS}$ within the precision of our estimation of the renormalon contribution.
This implies that the combination $\delta\Lambda_L+\delta\Lambda_{\RS}$ does not
contain a renormalon either if calculated in a consistent way:
$\delta\Lambda_L(a)$ and $\delta
\Lambda^{\RS}(\nu_f)$ contain one and the very same renormalon contribution
(with negative relative sign).
The sum of both terms, expanded in terms of
$\als$ has good convergence properties
(using the same normalization point to
enforce the renormalon cancellation at each order in perturbation
theory). The explicit expression at NNLO reads,
\bea\label{translate}
&&
\delta\Lambda_L(a)+\delta
\Lambda^{\RS}(\nu_f)=\frac{C_A}{2}v_1a^{-1}\als(\nu)
\\
\nn
&&
\qquad
+\left\{
\frac{C_A}{2}\frac{a^{-1}}{4\pi}\left\{v_2+v_1\left[-b_1+2\beta_0\ln(\nu a)\right] \right\}
+\nu_f\left({\tilde V}_{s,1}-{\tilde
V}_{o,1}\right)\right\}\als^2(\nu)+\cdots
\,, 
\eea
where the $v_i$ can be found in Eqs.~(\ref{v1}) and (\ref{v2}),
$b_1$ in Eq.~(\ref{eqb1}) and
${\tilde V}_{o,1}$ and $\tilde{V}_{s,1}$ in Table~\ref{tabv}.
An estimate of the $O(\als^3)$ term can be obtained
from Eq.~(\ref{eqmstat}) below, under the replacements,
$C_f\mapsto C_A$ and
$\tilde{V}_{s,i}\mapsto 2(\tilde{V}_{s,i}-\tilde{V}_{o,i})$.
This estimate will be subject to $O(1/N_c^2)$ corrections to the coefficient
$v_3$.

In principle, $\nu_f$ and $a^{-1}$ need not be
equal but we will take them similar to avoid large logs.
The large numerical values of
$v_2$ and $b_1$ are mainly due to contributions
from lattice-specific tadpole diagrams that arise because
the breaking of Lorentz symmetry becomes particularly evident
at UV scales $\simeq a^{-1}$. This often results in badly
convergent perturbative series when expanded in terms of
$\alpha_L(a)$. However, the convergence is vastly improved,
once the series is re-expressed in terms of a more ``physical''
coupling like $\alpha_s(a^{-1})=\alpha_L[1-b_1\alpha_L^2/(4\pi)+\cdots]$
(see e.g.\ Refs.~\cite{Bali:2002wf,Lepage:1992xa}).
This is also evident from Eq.~(\ref{translate}) above as
$v_1\approx 3.17$, $(v_2-b_1v_1)/(4\pi)\approx -1.97$
[and $[v_3-2b_1v_2-(b_2-2b_1^2)v_1]/(4\pi)^2\approx 14.5$].


\begin{figure}[h]
\hspace{-0.1in}
\epsfxsize=5in
\centerline{
\epsffile{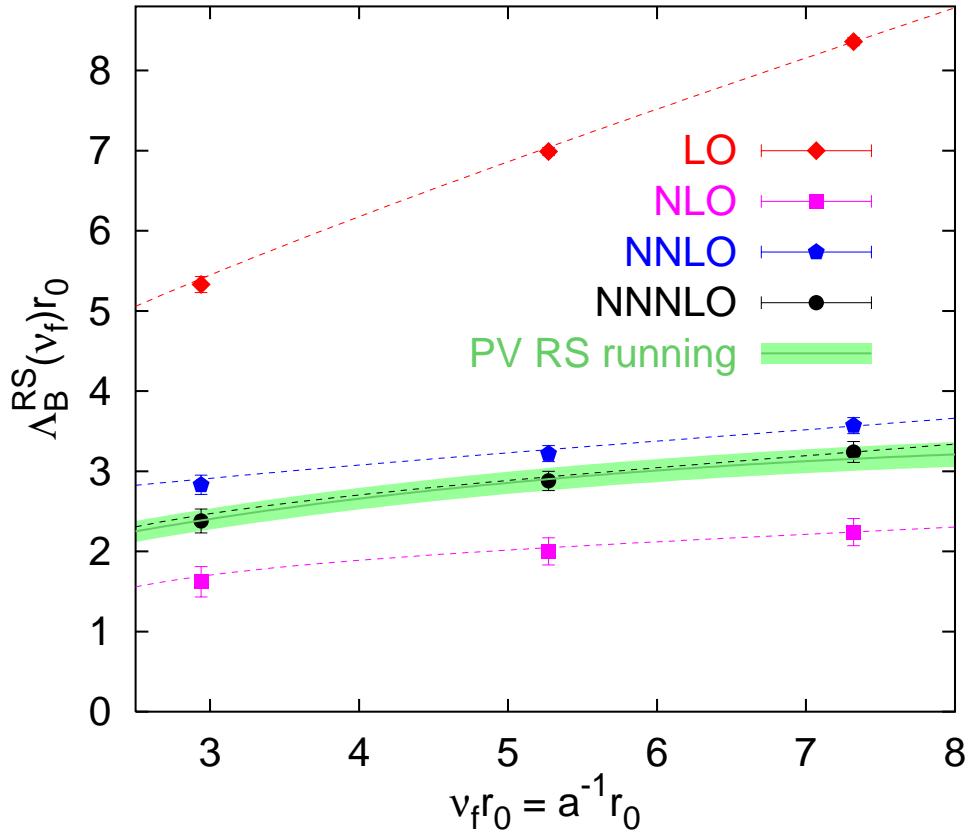}}
\caption {{\it The lowest gluelump mass $\Lambda_B^L$ as obtained on
the lattice (diamonds), as well as converted into the RS scheme
at NLO (squares), NNLO (pentagons) and NNNLO* (NNNLO estimate, circles).
The error band corresponds to the result for
$\Lambda_B^{\RS}$ of Eq.~(\ref{fitLambdaRSB}), without the ``theoretical''
error, run to different scales,
according to the PV prescription Eq.~(\ref{PV}).
The dashed lines, drawn to guide the eye, are explained in the text.}}
\label{figgl}
\end{figure}

We can now translate the $\Lambda_B^L$ values obtained by Foster and
Michael~\cite{FM} into the RS scheme. The results are shown in
Table~\ref{tabgl} and are also displayed in Fig.~\ref{figgl}.  ``NLO''
and ``NNLO'' refer to translating from the lattice scheme to the RS
scheme via Eq.~(\ref{translate}) to $O(\als)$ and $O(\als^2)$,
respectively\footnote{Note that the counting here differs from that used
in Fig.~\ref{PVvsFO}, in the RS scheme,
where we labelled $O(\als^2)$ as ``LO''.}. 
Obviously, to leading order, $\Lambda_B$ is scheme
independent. ``NNNLO*'' stand for an estimate obtained assuming
that the NNNLO contribution to $\delta\Lambda_L$ is equal to the
NNNLO contribution to $\delta m^L$ with the replacement of the overall
factor $C_f \mapsto C_A$. This is correct up to $O(1/N_c^2)$
effects. Finally, the conversion from the lattice to the RS scheme has
been performed using the 4-loop running of $\als$ at $\nu = a^{-1}=\nu_f$.
This accelerates the
convergence to the RS results. If, instead, we use the $n$-loop
running of $\als$ that is consistent with the order of the calculation,
we still see convergence but with, in the NLO and NNLO cases,
larger corrections. This is mainly due to
the fact that within the present window of energies
the values obtained for $\als(\nu)$ from $\Lambda_{\MS}$
{}from a one- or two-loop running are significantly different from those
from the three-loop running (which is close to four-loop).
The lattice prediction of $\Lambda_{\MS}$ that we use as an
input applies to very high energies, such that it is important to
run $\als$ down to $\nu\geq 2.5\,r_0^{-1}$ as
precisely as possible.

Within present errors we can fit the data with straight lines but
there will be logarithmic corrections and, in the gluelump data
$\Lambda_B^Lr_0$, additional $O(a^2)=O(\nu_f^{-2})$ lattice artifacts.
The figure reveals that at the lattice spacings investigated these are
tiny, relative to the linear slope.  Except for these lattice
corrections the running of $\Lambda_B^L$ is non-perturbatively
accurate.  Needless to say that the power dependence on $a^{-1}$ is
universal for all gluelumps, such that gluelump mass splittings have a
well defined continuum limit, which is also confirmed in
Ref.~\cite{FM}.

In lattice perturbation theory we can calculate the ``running'' of the
gluelump data to $O(\als^2)$ [and up to $O(\als^3)$ if we neglect
$O(1/N_c^2)$ effects]. There is a renormalon ambiguity in the
absolute value. However the slope is not affected by this. If we take
the value $\Lambda_{H}^L(7.32\,r_0^{-1})\approx 8.36\,r_0^{-1}$
from Table~\ref{tabgl} and
perform the running with NNNLO* accuracy, we obtain the dashed line that
joins the ``LO'' $\RS(=L)$ points.
We can see that this parametrization
is quite close to the
non-perturbatively evaluated data.
Moreover, there is overall convergence, with
higher order terms being numerically smaller in the lattice scheme. We
will discuss this in more detail in Sec.~\ref{statlight}
below, in the context of the
static-light binding energy $\overline{\Lambda}^L$, which has a similar
perturbative expansion, up to an overall factor $C_f/C_A$,
see Fig.~\ref{runbind}.

In Fig.~\ref{figgl}
we also compare the value obtained in Sec.~\ref{RSglump}
above [Eq.~(\ref{fitLambdaRSB})], with running according to the
PV prescription Eq.~(\ref{PV}), with the results obtained
directly from the  lattice determination of the gluelump mass
via Eq.~(\ref{translate}). We see clear convergence with alternating signs
{}from LO (diamonds), NLO (squares), NNLO (pentagons) and NNNLO* (circles) 
towards the result calculated from the
$\Pi_u$ and $\Sigma_g^+$ potentials in the
previous section and its running (error band).
Our NNNLO* estimates already agree with this error band.
The dashed lines connecting the NLO,
NNLO and NNNLO* points
are the corresponding transformations of the curve
through the LO points and just drawn to guide the eye.
All errors displayed in Fig.~\ref{figgl} are statistical only,
plus the uncertainty on $\Lambda_{\MS}$. Within the theoretical errors
of Eq.~(\ref{fitLambdaRSB}) ($\pm 0.21\,r_0^{-1}$), in fact
we already find agreement at the NNLO level.
In Secs.~\ref{ssssrs} and in particular \ref{v02} below we will
analyse the running of the
binding energy of static-light mesons in more detail,
see also Fig.~\ref{potrun2}.

We obtain an independent second prediction for $\Lambda_B^{\RS}$ from the  
gluelump data. The statistical errors are
smaller in the gluelump case than those we encountered from the continuum
potentials. In a first step we obtain
the fit parameter,
\be
\label{lamb}
\Lambda_B^{\RS}(7.32\,r_0^{-1})
=[3.21\pm 0.04 (\mbox{latt.}) \pm 0.42 (\mbox{th.}) \pm 0.10 (\Lambda_{\MS})]\,
r_0^{-1}\,,
\ee
from a global NNNLO* fit,
\be
\Lambda_H^L(a)=\Lambda_H^{\RS}(\nu_f)+
\left[\delta\Lambda_H^L(a)+\delta\Lambda^{RS}_H(\nu_f)\right]\,,
\ee
where we have chosen $\nu=\nu_f=7.32\,r_0^{-1}$.
We can then convert this result into,
\be
\Lambda_B^{\RS}(2.5\,r_0^{-1})=\left[2.31\pm 0.04 (\mbox{latt.}) \pm 0.33 (\mbox{th.})
{}^{+0.18}_{-0.19} (\Lambda_{\MS})\right]\,
r_0^{-1},
\ee
using the PV running in the RS scheme. This compares well with 
the result from the potentials, Eq.~(\ref{fitLambdaRSB}).

The errors displayed in Eq.~(\ref{lamb}) above are due to the following
sources:

{\bf 1)} ``latt.'' is the sum of the statistical error ($\pm 0.03$)
and the error encountered when varying the fit range (i.e.\ excluding
the left-most data point): $\pm 0.01$.

{\bf 2)} ``th.'' is the sum of perturbative and non-perturbative
errors.  As perturbative errors we take the difference between NNLO and
NNNLO* results ($\pm 0.20$) as well as a 10~\% uncertainty
in $N_{V_s}-N_{V_o}$ ($\pm 0.18$). To investigate possible
non-perturbative effects we include an $a^2$ term into the fit. We
estimate an additional $\pm 0.04$ uncertainty
from this source. Adding these three errors
linearly results in $\pm 0.42$.

{\bf 3)} ``$\Lambda_{\MS}$'' stands for the uncertainty due to the
error of $\Lambda_{\MS}r_0$~\cite{Lambda}: $\pm 0.10$.

Whereas the statistical error is smaller in this determination
than the one of Eq.~(\ref{fitLambdaRSB})
and the uncertainty due to the error of $\Lambda_{\MS}$ is comparable
in size, the systematics are less well under control, which
is reflected in the large theoretical error.
First of all, for the lattice gluelumps we only have the perturbative
result to $O(\als^2)$ with an estimate of the $O(\als^3)$ term while
in Sec.~\ref{RSglump} above we knew the $O(\als^3)$ results and
have an estimate of the $O(\als^4)$ terms. Furthermore,
as the previous analysis was based on
observables with a well defined continuum limit, we circumvented the
problem of disentangling the $a^{-1}$ ``running'' of $\Lambda_Br_0$ from
$O(\lQ^2a^2)$ lattice artifacts. With gluelump data on more, and in particular
finer, lattice spacings the latter disadvantage (which at present is however
not the dominant one) can in principle
be overcome.
In conclusion, it is nice to observe perfect agreement between the
two predictions, which enhances our confidence in the methods applied
and adds further credibility to our error estimates.

\subsection{Higher gluelump excitations}
\label{higherg}
Now that we have fixed the energy of the lightest gluelump, we can
quote absolute values for the remaining gluelump spectrum using
the results of Foster and Michael~\cite{FM}. We display our
predictions in Table~\ref{tablegluelumps} where the errors correspond
to the sum of the individual uncertainties, added linearly.  The
dominant uncertainty is that of $\Lambda_{B}$, as the mass differences
between the different gluelumps have been determined with very good
accuracy. Needless to say that these results are scheme and scale
dependent. The quoted numbers refer to the RS scheme with
$\nu_f=2.5\,r_0^{-1}\approx 1$~GeV. With the information presented in this paper
they can be run to different scales. For ease of reference we also
converted these values into GeV units (using $r_0^{-1}=394$~MeV). 
However, we note that one
should add a scale uncertainty of about 10~\% to them to account for
the fact that all results have only been obtained in the quenched
approximation.

\begin{table}[h]
\begin{center}
\begin{tabular}{|c||c|c|c|}\hline
$J^{PC}$&$H$&$\Lambda_H^{\RS}r_0$&$\Lambda_H^{\RS}$/GeV\\\hline\hline
$1^{+-}$&$B_i$                           &2.25(39)&0.87(15)\\
$1^{--}$&$E_i$                           &3.18(41)&1.25(16)\\
$2^{--}$&$D_{\{i}B_{j\}}$                &3.69(42)&1.45(17)\\
$2^{+-}$&$D_{\{i}E_{j\}}$                &4.72(48)&1.86(19)\\
$3^{+-}$&$D_{\{i}D_{j}B_{k\}}$           &4.72(45)&1.86(18)\\
$0^{++}$&${\mathbf B}^2$                 &5.02(46)&1.98(18)\\
$4^{--}$&$D_{\{i}D_jD_kB_{l\}}$          &5.41(46)&2.13(18)\\
$1^{-+}$&$({\mathbf B}\wedge{\mathbf E})_i$&5.45(51)&2.15(20)\\\hline
\end{tabular}
\end{center}
\caption{{\it Absolute values for the gluelump masses in the continuum limit
in the RS scheme at $\nu_f=2.5\,r_0^{-1}\approx 1$~GeV,
in $r_0$ units and in GeV. Note that an additional
uncertainty of about 10~\% should be added to the last column to account for
the quenched approximation. We also display examples of creation operators
$H$ for these states. The curly braces denote complete
symmetrization of the indices.}}
\label{tablegluelumps}
\end{table}

Note that the gluelump operators can be represented
in terms of gluonic
fields~\cite{pNRQCD,Simonov}.
In general one and the same gluelump
can be created by infinitely many different adjoint operators $H$.
Within each channel we display (one of) the lowest dimensional
such choice(s)
in the table.
The basic building blocks are the covariant derivative $D_i$
(with $J^{PC}=1^{-+}$, dimension 1), the chromomagnetic field
$B_i$ ($1^{+-}$, dimension 2) and the chromoelectric field
$E_i$ ($1^{--}$, dimension 2). The curl of the electric field
has the quantum numbers of the magnetic field,
such that on the lattice all states can be created by
operators
that are local in time. Furthermore, ${\bf D}\cdot {\bf B}$ and
${\bf D}\cdot {\bf E}$ can be eliminated, the first because it is
identically zero, using the Jacobi identity, the second by applying
the equations of motion.
One example: the lowest dimensional operator that creates
the $3^{+-}$ state is
$D_{\{i}D_jB_{k\}}$, where the curly braces denote the sum over all
10 symmetric
permutations of the indices. This includes three terms
$D_i\sum_j D_jB_j=0$ such that indeed there remain only seven
independent operators
to create this seven dimensional representation.
Also note that $D_{\{i}B_{j\}}$ and $D_{\{i}E_{j\}}$ each
only contain five independent operators, consistent with $J=2$ etc..

It is interesting to see that the level ordering roughly corresponds
to the lowest dimension of the creation operator, once the equations
of motion are used to eliminate the $E$
field~\cite{pNRQCD}. This makes the $E$ field ``heavier'' than
a $B$ field, increasing its dimension by one. The $3^{--}$
gluelump (two derivatives and one $E$ which corresponds to
dimension five, after substituting $E$) is not included into the table
as no controlled continuum limit extrapolation was possible. However,
its mass at fixed finite lattice spacing is in the same ball park as
that of the other dimension five states, $4^{--}$ and $1^{-+}$,
in support of this na\"{\i}ve operator counting
picture.

\subsection{Comparison with previous results}
We shall relate our results to previous determinations of the gluelump
masses.  All these suffer from the problem of obtaining the global
constant and, in none of these, the scheme was clearly defined, such
that they need not yield the same results that we obtain.

In Ref.\ \cite{Simonov} the gluelumps were studied within a string
model. One general feature of this approach is the excess of predicted
states. This seems to be a problem of this model since it does not
appear to be compatible with QCD, or more precisely with its
realization for this kinematical regime: pNRQCD \cite{pNRQCD} (see
also the discussion in Ref. \cite{KK}).  The prediction of this model,
$\Lambda_B(n_f=0)=1.87$ GeV, is by a factor of two larger than our
result.

In Ref.~\cite{DiGiacomo}, the same value for the electric and magnetic
correlation length is obtained:
$\Lambda_E(n_f=0)=\Lambda_B(n_f=0)=0.90(5)(10)$~GeV, from lattice
simulations using the cooling method. The number for $\Lambda_B$
coincides with ours.  However, the splitting between chromoelectric
and -magnetic correlators is unaccounted for.  From the results of
Foster and Michael one would then assign a systematic error of the
order of this splitting $\approx 400$~MeV: clearly a better
conceptional understanding of how ``cooling'' removes short distance
fluctuations, without destroying essential infrared physics would be
useful. On the other hand it is comforting that numbers similar to
our results are obtained in this approach, that is also meant to
subtract the perturbative contributions from the low energy matrix element.

In Ref.\ \cite{DEJ} a sum rule analysis of the electric and magnetic
correlator was made. The main result was $\Lambda_E(n_f=0)=(1.9\pm 0.5)$
GeV. It should be noted that the value of $\Lambda_{\MS}$
on the lattice is now smaller by 5\%, compared to the value used in
this analysis.  Taking this into account we find this result
compatible with ours [1.25(16)~GeV], within errors.  Moreover, in this
analysis, evidence for $\Lambda_E > \Lambda_B$ was
reported.

In Ref.\ \cite{bag}, an MIT bag model calculation
was used to obtain the gluelump spectrum. 
No errors were assigned to this evaluation. The value of $\Lambda_B$ is by 
about 500 MeV larger than ours and quite consistent with the
sum rules evaluation. The same holds true for $\Lambda_E$, however, for
the higher excitations the agreement with the results of Foster
and Michael is less convincing.
  
In Ref.\ \cite{BBV}, lattice correlation functions that are needed
to calculate relativistic corrections to the static potential were 
used in order to check the validity of the stochastic vacuum model
in the Gaussian approximation. 
Under this assumption, which was to some extent tested in this reference,
these correlation functions
could be related to gluonic field strength correlators and upper limits
for the gluelump masses were obtained:
$\Lambda_B(n_f=0)\leq 1.64(16)$ GeV and  $\Lambda_E(n_f=0)\leq 1.04(15)$ GeV,
respectively: the ordering of the gluelumps is wrong, however,
the upper limits quoted are in no
contradiction to our results (or indeed to a different ordering).

In Ref.\ \cite{SS}, a constituent quark model was used. The results
roughly agree (within a 200 -- 300 MeV error) with the splittings predicted by
Michael and Foster and the hybrid spectrum at short distances (see
Ref. \cite{KK} for some criticism of this evaluation). For the
lightest gluelump they obtain $\Lambda_B \approx 1.4$ GeV. 

We have seen how different determinations of $\Lambda_{B}$ 
result in values ranging from less than one GeV up to nearly two GeV. 
These numbers are all scheme dependent. This may explain the huge differences
between different results. Our result provides 
strong constraints on vacuum models. Furthermore,
the RS scheme provides a unified framework to study 
the non-perturbative effects in an unambiguous and model independent way.

\section{Static-light systems}
\label{statlight}
The situation discussed above very much resembles the one that one
encounters in heavy-light mesons in the static limit. In this case,
the adjoint source is replaced by a fundamental source which is
not screened by gluonic fields but by a light Dirac quark instead.
(A light
Higgs scalar  in the fundamental representation would be an alternative
possibility.) In these systems
the binding energy $\overline{\Lambda}$ of the $\frac{1}{2}^-$
state (which will correspond to pseudoscalar and vector heavy-light mesons,
once $1/m_b$ corrections and the spin of the heavy quark are taken
into account)
plays a r\^ole similar to that of the $\Lambda_B$ discussed above.
The experimental mass of the $B$ meson $M_B$ can be factorized into,
\begin{equation}
\label{MB2}
M_B=\overline{\Lambda}+m_b+O(1/m_b),
\end{equation}
where both $\overline{\Lambda}$ and $m_b$ depend on scheme and scale.
In the literature~(see e.g.\ Ref.~\cite{Martinelli:1998vt}) the binding
energy in the lattice scheme is referred to as
${\mathcal E}(a)=\overline{\Lambda}^L(a)$, which is renormalon-free but has
an $a^{-1}$ power divergence. For the Wilson action and $n_f=0$ this
$\delta m_{\rm stat}^L(a)$ power term
is known to $O(\als^3)$ in perturbation theory [Eqs.~(\ref{v0}) -- (\ref{v3})].
Subtracting this perturbative result introduces renormalons.

It is also possible to define the binding energy in an entirely
non-perturbative renormalon-free and power-term free way, for instance
by subtracting the energy of a temporal Schwinger line in Coulomb or
Landau gauge~\cite{Gimenez:1996nw}.
In fact the same
can be achieved in the case of the
lowest gluelump mass, either by subtracting the energy of an adjoint
Schwinger-line in a fixed gauge (see also Ref.~\cite{Philipsen:2002az})
or by subtracting the on-shell mass of
an adjoint Polyakov-Wilson line, encircling a compactified lattice
dimension.  From an EFT point of view however one would like to
combine a non-perturbative low energy result with a perturbative
calculation at high energies. For instance to quote a value for the $b$
quark mass in the $\MS$ scheme, the UV renormalon of the binding
energy is required to cancel the IR renormalon of the OS mass
and hence a perturbative subtraction is essential: the renormalon of
the expansion of the power divergence is the same as the one that is
encountered in the conversion {}from the OS mass into the $\MS$ mass.
This procedure has been implemented in the past in calculations of the
$b$ quark mass from lattice simulations in the static
limit~\cite{Martinelli:1998vt}.

The $b$ quark mass has also been obtained in perturbative
QCD in the $\RS$ scheme at $\nu_f=2$ GeV from the $\Upsilon(1S)$ system
using EFTs~\cite{RS}.
Subtracting this value from the spin-averaged mass of the $B$ meson
yields,
\begin{equation}
\label{Lambdabarnf4}
\overline{\Lambda}^{\RS}(\nu_f = 1\,\mbox{GeV})=[0.365 \pm 0.085(\mbox{th.})
{}^{+0.045}_{-0.061} (\Lambda_{\MS})]\,\mbox{GeV}\,.
\end{equation}
This number is different from the value quoted in
Ref.~\cite{RS}\footnote{Again, note that what we call RS scheme here
corresponds to the $\RS'$ scheme of Ref.~\cite{RS}.}, since here we have 
performed the running to $\nu_f=1$ GeV using the PV prescription and
not included $O(1/m_b)$ corrections into the fit (these two effects
partially compensate each other). Using the PV prescription allows 
us to perform the log resummation for the renormalon related terms.
However, the result strongly depends on the value of $\Lambda_{\MS}$. 

Eq.~(\ref{Lambdabarnf4}) has been obtained from the 
physical $\Upsilon$ and $B$ systems,
not in the quenched approximation. The scale
$r_0^{-1}=394\pm 20$~MeV~\cite{Sommer:1993ce,Bali:1997am,boyle} is also
obtained from $\Upsilon$ phenomenology. Re-expressed in terms
of $r_0$ we get,
\begin{equation}
\label{rsval}
\overline{\Lambda}^{\RS}(\nu_f = 2.5\,r_0^{-1})=[0.92 \pm 0.22(\mbox{th.}) 
{}^{+0.15}_{-0.11}(\Lambda_{\MS})]\,r_0^{-1}\,.
\end{equation}

In what follows we will extract $\overline{\Lambda}^{\RS}$ from
lattice data of static-light mesons. After addressing
the $b$ quark mass we will conclude with a more detailed
study of the running in the lattice and RS schemes, using precision data
from the static potential
within an energy range, $2\siml r_0\nu_f=r_0/a\siml 15$.

\subsection{Determination of $\overline{\Lambda}^{\RS}$}
\label{ssssrs}
We will use Eq.~(\ref{rsval}) as
our starting point for the $n_f=0$ situation. In order
to compare with lattice results in the quenched approximation we
will employ the $n_f=0$ running of $\overline{\Lambda}^{\RS}(\nu_f)$
and keep in mind that on top of the errors stated above one might expect
an additional 10~\% quenching error.

\begin{table}[h]
\begin{center}
\begin{tabular}{|l||c|c|c|c|c|c|}
\hline
Ref&$a^{-1}r_0$&$\overline{\Lambda}^Lr_0$
&$\overline{\Lambda}^{\RS}r_0$(NLO)
&$\overline{\Lambda}^{\RS}r_0$(NNLO)
&$\overline{\Lambda}^{\RS}r_0$(NNNLO)
&$\overline{\Lambda}^{\RS}r_0$\\\hline\hline
\cite{Michael:1998sg}&2.93&2.45( 6)&0.79(10)&1.34( 7)&1.16( 8)&0.99(1)\\
\cite{Duncan:1994uq} &2.93&2.22( 4)&0.56( 9)&1.11( 5)&0.93( 6)&0.99(1)\\
\cite{Duncan:1994uq} &4.48&2.86( 4)&0.83( 8)&1.37( 6)&1.23( 6)&1.16(3)\\
\cite{Allton:1993ix} &5.37&3.28( 6)&1.03( 9)&1.59( 7)&1.45( 8)&1.22(4)\\
\cite{Duncan:1994uq} &6.32&3.44( 8)&0.96(11)&1.53( 9)&1.40( 9)&1.28(4)\\
\cite{Allton:1993ix} &7.36&3.83( 8)&1.10(11)&1.70( 9)&1.57(10)&1.34(4)\\
\cite{Ewing:1995ih}  &7.36&3.87(11)&1.14(13)&1.74(12)&1.61(12)&1.34(4)\\
\cite{Duncan:1994uq} &8.49&4.24( 8)&1.24(11)&1.87( 9)&1.74(10)&1.40(4)\\
\cite{Allton:1993ix} &9.76&4.49(10)&1.20(13)&1.85(11)&1.72(12)&1.45(5)\\\hline
\end{tabular}
\end{center}
\caption{{\it The inverse lattice spacing~\cite{Guagnelli:1998ud},
the static-light binding energy $\overline{\Lambda}^L={\mathcal E}$
\cite{Allton:1993ix,Duncan:1994uq,Ewing:1995ih,Michael:1998sg} in the
lattice scheme, as well as its
conversion to the RS scheme to different orders in perturbation
theory. In the last column, we state the values of
$\Lambda_B^{\RS}(a^{-1})$ using the PV running, Eq.~(\ref{PV}),
of the result Eq.~(\ref{rsval}) in the RS scheme.
The errors only incorporate the statistical uncertainties
as well as the 8~\% uncertainty in
$\Lambda_{\overline{MS}}r_0$~\cite{Lambda}, added in
quadrature. The values in the last column, which have been obtained
from the physical $\Upsilon(1S)$ and $B$ meson masses, have additional
errors inherited from Eq.\ (\ref{rsval}), which, however, will only
result in an overall upward or downward shift and 
will not affect their differences.}}
\label{tablam}
\end{table}

\begin{figure}[h]
\hspace{-0.1in}
\epsfxsize=4.8in
\centerline{\epsffile{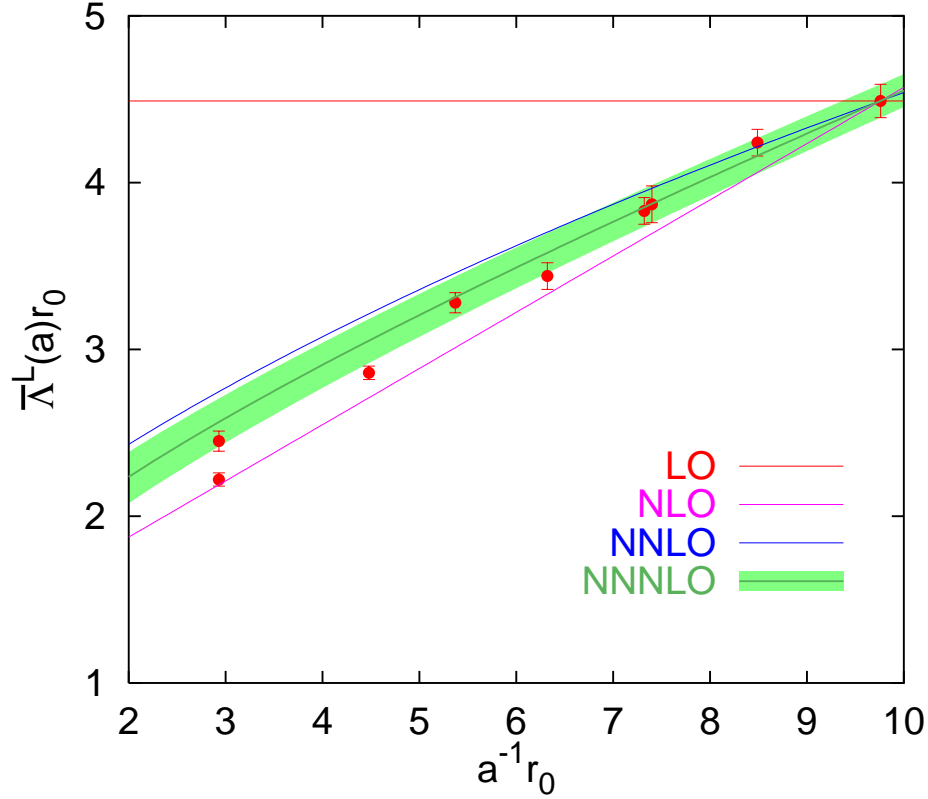}}
\caption {{\it Perturbative running of the binding energy
$\overline{\Lambda}$ in the lattice
scheme, in comparison with lattice data, starting at the smallest available
lattice spacing. The NNNLO error band incorporates the error due to the
uncertainty in $\Lambda_{\MS}$~\cite{Lambda}, and the statistical error.}}
\label{runbind}
\end{figure}

\begin{figure}[h]
\hspace{-0.1in}
\epsfxsize=4.8in
\centerline{\epsffile{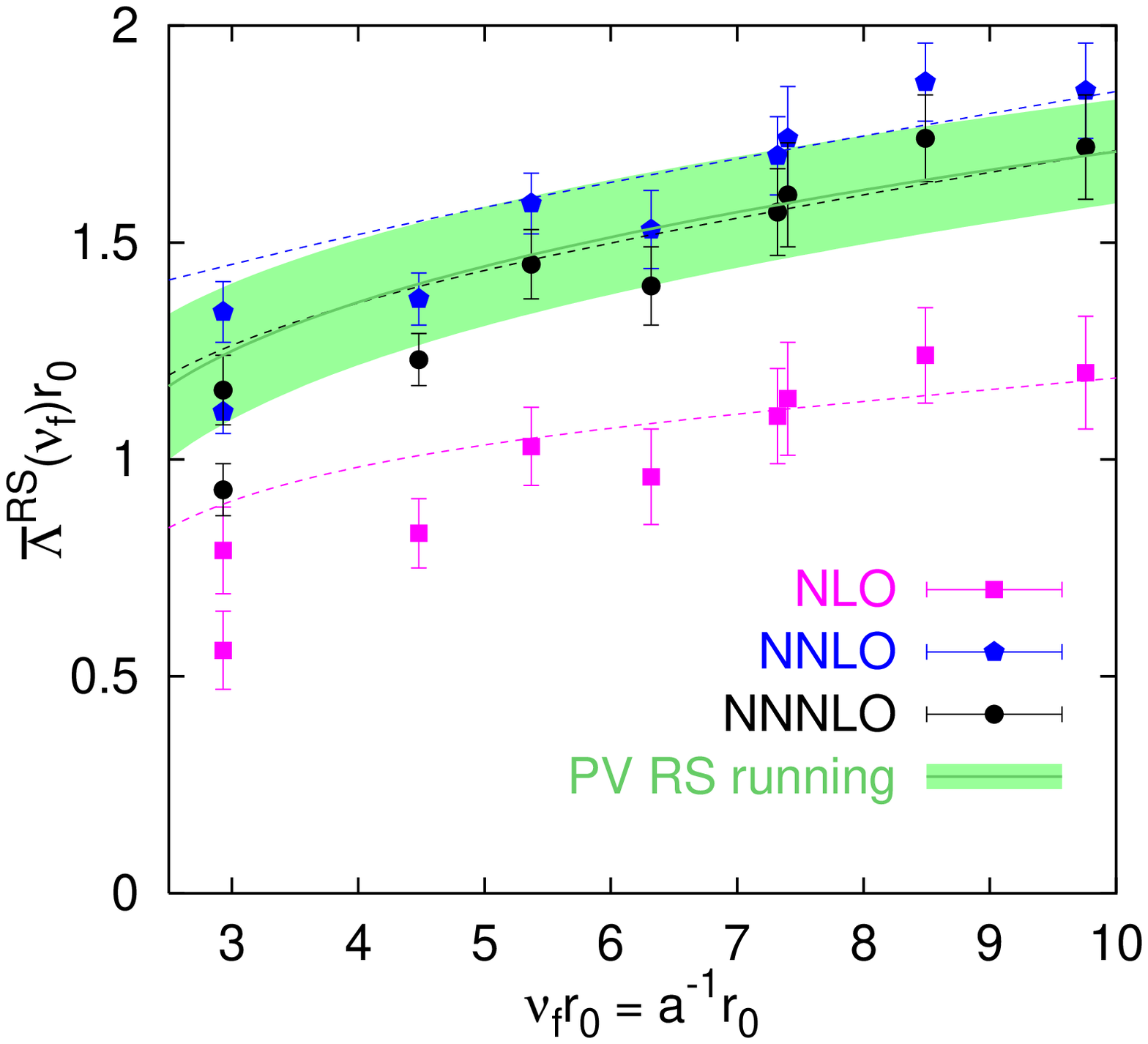}}
\caption {{\it The binding energy $\overline{\Lambda}$,
obtained on the lattice and converted
into the RS scheme at NLO (squares), NNLO (pentagons), NNNLO (circles)
and the result of Eq.~(\ref{lambdalarge}), run to different
scales using the PV prescription (neglecting the ``theoretical'' errors).
The dashed lines are explained in the text.}}
\label{bindrun}
\end{figure}

$\overline{\Lambda}^L(a)$ has been calculated on a variety of lattice spacings
by different collaborations~\cite{Allton:1993ix,Duncan:1994uq,Ewing:1995ih,Michael:1998sg}. The main source of uncertainty in these determinations is the
extrapolation to zero light quark mass.
We used the $r_0/a$ values from the interpolation of
Ref.~\cite{Guagnelli:1998ud} to assign the scale\footnote{These values slightly
differ from those quoted in Ref.~\cite{FM} used in Table~\ref{tabgl} above,
which cover a smaller window of lattice resolutions.}. The results
are displayed
in Table~\ref{tablam} and are roughly
consistent with each other, with the exception
of the coarsest lattice point $r_0\approx 2.93\,a$ that corresponds to
$\beta=5.7$. Here the raw data of Ref.~\cite{Michael:1998sg} are more accurate
but the chiral extrapolation of Ref.~\cite{Duncan:1994uq} should
be better controlled.

We multiply the values obtained for $\Lambda_B^Lr_0$ of Ref.~\cite{FM}
(that are displayed in Table~\ref{tabgl}) by the colour factor
$C_f/C_A$.  At $\beta=5.7, 6.0$ and 6.2, respectively, we obtain the
numerical values $2.37(4)$, $3.11(2)$ and $3.72(2)$. The corresponding
values in Table~\ref{tablam} read $2.45(6)|2.22(4)$, $3.28(6)$ and
$3.83(8)|3.87(11)$ where for both, $\beta=5.7$ and $\beta=6.2$, two
independent determinations exist.
The qualitative agreement is remarkable: not only the
perturbative expansions of $\delta\Lambda$ and $\delta m_{\rm stat}$
are dominated by terms that are proportional to the respective
Casimirs of the gauge group representation of the static
source but also the non-perturbative values themselves.  In fact also
in the $\RS$ scheme the result Eq.~(\ref{fitLambdaRSB}) is close to
the value displayed in Eq.~(\ref{rsval}), multiplied by $C_A/C_f=9/4$.

Similar to the discussion in Sec.~\ref{seclaa} above, we can translate
the results from the lattice scheme into the $\RS$ scheme.
The master formula in this case is very similar to Eq.~(\ref{trans})
and reads (known to NNNLO),
\be
\label{eqmshift}
\overline{\Lambda}^{\RS}(\nu_f)=\overline{\Lambda}^{L}(a)
-\left[\delta m_{\rm stat}^L(a)-\delta m_{\RS}(\nu_f)\right]
\,,
\ee
with
\bea
\nn
\delta m_{\rm stat}^L(a)-\delta m_{\RS}(\nu_f)
&=&\frac{C_f}{2}v_1a^{-1}\als(\nu)
\\
\label{eqmstat}
&+&\left\{
\frac{C_f}{2}\frac{a^{-1}}{4\pi}\left[v_2+v_1B_1(\nu a)\right]
+\frac{\nu_f}{2}{\tilde V}_{s,1}\right\}\als^2(\nu)\\\nn
&+&\left\{\frac{C_f}{2}\frac{a^{-1}}{(4\pi)^2}\left\{v_3+2v_2B_1(\nu a)+v_1\left[
B_2(\nu a)+B_1^2(\nu a)+b_1^2\right]\right\}
\right.
\\\nn
&&
\left.
\quad
+\frac{\nu_f}{2}\left[{\tilde V}_{s,2}-\tilde{V}_{s,1}\frac{\beta_0}{\pi}
\ln\left(\frac{\nu_f}{\nu}\right)\right]
\right\}\als^3(\nu)
+ \cdots
\,, 
\eea
where,
\begin{equation}
B_i(x)=-b_i+2\beta_{i-1}\ln(x),\quad i=1,2\,,
\ee
and the coefficients $\tilde{V}_{s,1}$ and $\tilde{V}_{s,2}$ can be
found in Table \ref{tabv}.
The coefficients $v_i$ and $b_i$  can be found in Eqs.~(\ref{v1})
-- (\ref{v3}) and Eqs.~(\ref{eqb1}) and (\ref{eqb2}), respectively.

Eqs.~(\ref{eqmshift}) and (\ref{eqmstat}) also relate
results obtained at different lattice
spacings to each other,
\be
\label{aapL}
\overline{\Lambda}^L(a')=\overline{\Lambda}^L(a)-\left[\delta m^L_{\rm stat}(a)
-\delta m^L_{\rm stat}(a')\right]\,.
\ee
To illustrate this we display the
$\Lambda^L(a)$ values of Table~\ref{tablam}
in Fig.~\ref{runbind}, together with
the expected running, starting at the finest, i.e. right-most,
lattice point at LO, NLO, NNLO and NNNLO. The NNNLO error band
contains both, the statistical error and that due to the uncertainty
in $\alpha_s(a)$. The running is done in each order in a self-consistent way
to the given order in $\als$,
according to Eq.~(\ref{eqmstat}) (without the $\tilde{V}_{s,i}$ terms).
We used $\nu=9.76\,r_0^{-1}$ and the initial value,
$\als(\nu)$, was calculated from
$\Lambda_{\MS}r_0$ using the four loop running.
We observe convergence and moreover the series is sign alternating.
To NNNLO, except for the lower lying of the two
$r_0/a\approx 2.93$ data points, there
is no contradiction between data and the expectation.
However, the points of Ref.~\cite{Duncan:1994uq}
have a slightly more pronounced slope such that the
$r_0/a\approx 2.9, 4.5$ and 6.3 points ($\beta=5.7, 5.9$ and 6.1)
lie below the curve while the $r_0/a\approx 8.5$ point ($\beta=6.3$)
lies somewhat above.

We also display the data of the table in Fig.~\ref{bindrun}, in analogy
to Fig.~\ref{figgl} but disregard the LO result that is already displayed
in Fig.~\ref{runbind}.
The size of
$\delta m^L_{\rm stat}(a)-\delta m_{\RS}(a^{-1})$ increases
linearly in $a^{-1}$, with logarithmic corrections: at coarse
lattice spacings there might be significant perturbative $O(\alpha_s^4)$
and non-perturbative $O(a^2/r_0^2)$ corrections affecting the
slope of this function while at fine lattice spacings the
slope can be determined accurately but the $\delta m$-difference itself
becomes large. An accurate conversion between the two schemes
can therefore neither be obtained at extremely fine nor at very
coarse lattice spacings. Setting $\nu=\nu_f$,
the difference between NLO and
NNLO translation is minimised for $3\siml r_0\nu_f\siml 4$ while
that between NNLO and NNNLO is minimal for $7.5\siml r_0\nu_f\siml 9$, where
the widths of these bands are determined by our uncertainty
in the value of $\Lambda_{\MS}r_0$.

We choose to translate the lattice scheme results
into the $\RS$ scheme by means of a global NNNLO fit to the
$r_0/a > 5$, i.e. $\beta\geq 6.0$ data, expanded in terms of
$\alpha_s(\nu = 9.76\,r_0^{-1})$, where we set $\nu_f=\nu$.
The result reads,
\be
\label{lambdalarge}
\overline{\Lambda}^{\RS}(\nu_f=9.76\,r_0^{-1})=
\left[1.70\pm 0.08({\rm latt.})\pm 0.18
({\rm th.})\pm 0.04 (\Lambda_{\MS}) \right]\,r_0^{-1}\,.
\ee
Note that $\overline{\Lambda}^{\RS}$ is the only fit parameter.

The dashed curves in Fig.~\ref{bindrun} correspond to such an
NNNLO fit to the LO results, subsequently transformed
in the same way as the data points to NLO, NNLO and NNNLO.
The error band corresponds to the result
Eq.~(\ref{lambdalarge}) above, without the theoretical error,
run to different energies,
using the PV prescription, Eq.~(\ref{PV}):
unlike the band displayed in Fig.~\ref{figgl} above, this is
not the result of an independent determination.
We would also have found agreement with the result Eq.~(\ref{rsval}),
but only within the large theoretical errors of this
un-quenched determination.

At high order in the perturbative expansion
and at high energies
one would expect the slope of the non-perturbative
running in the lattice scheme, translated into the RS scheme,
to approach that of the running within the RS scheme.
Discarding the four data points of Ref.~\cite{Duncan:1994uq},
Fig.~\ref{bindrun} nicely confirms this expectation.
We will investigate
this running with higher accuracy in Sec.~\ref{v02} below.

The errors of the determination Eq.~(\ref{lambdalarge})
above stem from the following sources:

{\bf 1)} ``latt.'' is the sum of the statistical error ($\pm 0.03$)
and the error encountered when varying the fit range
$a^{-1}_{\min}r_0= 4.48,5.37,6.32$ ($\pm 0.05$): $\pm 0.08$.

{\bf 2)} ``th.'' is the sum of perturbative and non-perturbative
errors.  As perturbative error we take the difference between NNLO and
NNNLO results. Varying the fit range as above this difference never
exceeds $\pm 0.04$. We also study the error due to the uncertainty of
$N_{V_s}$ obtaining $\pm 0.06$. To investigate possible
non-perturbative effects we include an $a^2$ term into the fit. We
estimate an additional $\pm 0.08$ uncertainty. Adding these three errors
linearly results in $\pm 0.18$.

{\bf 3)} ``$\Lambda_{\MS}$'' stands for the uncertainty in the
determination of $\Lambda_{\MS}r_0$~\cite{Lambda}: $\pm 0.04$.

Using the running in the RS scheme (we note that the error
due to the uncertainty of $N_{V_s}$ almost cancels in the running) we obtain,
\be
\label{LambdaRS1GEV}
\overline{\Lambda}^{\RS}(\nu_f=2.5\,r_0)=
\left[1.17\pm 0.08({\rm latt.})\pm 0.13
({\rm th.})\pm 0.09 (\Lambda_{\MS}) \right]\,r_0^{-1}\,.\,
\ee
from the value Eq.~(\ref{lambdalarge}). 
This $n_f=0$ result compares reasonably well with the phenomenological
$n_f=4$ value of Eq.~(\ref{rsval}) above and its error is of a comparable size.

\subsection{Comment on the $b$ quark mass}
We cannot resist the temptation to obtain a value for the
$\RS$ scheme bottom quark mass, using Eq.~(\ref{MB2}) and our quenched result
Eq.~(\ref{LambdaRS1GEV}). We obtain,
\be
\label{mRS1GEV}
m_{b,\RS}(\nu_f=1\,{\rm GeV})=
\left[4849\pm 32({\rm latt.})\pm 60 
({\rm th.})\pm 35 (\Lambda_{\MS}) \right]\,{\rm MeV}\,,\,
\ee
where we have translated Eq.~(\ref{LambdaRS1GEV}) into physical units
for $\nu_f=1$ GeV and also added an extra theoretical error of $\pm 30$ MeV,
due to $1/m_b$ corrections,
combined quadratically with the theoretical error inherited
from the lattice determination of $\overline{\Lambda}^{\RS}$. From this
number we can compute the $\MS$ scheme result,
\be
\label{mMSPV}
m_{b,\MS}(m_{b,\MS})=\left[4191\pm 29({\rm latt.})\pm 47
({\rm th.})\pm 1 (\Lambda_{\MS}) \right]\,{\rm
MeV}\,,
\ee
where we have performed any running and manipulation with
$n_f=4$ and used the PV prescription to run $m_{b,\RS}$ from
1 GeV up to the bottom $\MS$ mass\footnote{We ignore the charm mass threshold.
Since the charm quark mass is not much heavier than 1~GeV this is a
small effect anyhow, completely paled by our dominant source of uncertainty,
the $n_f=0$ approximation.}.
In this way higher order terms in the
relation between the $\MS$ and the RS mass are minimized. If instead
one determines $m_{b,\MS}(m_{b,\MS})$ directly
from its perturbative relation with
$m_{b,\RS}(1\,\mbox{GeV})$ one obtains
a somewhat larger result, but with sizeable higher order terms.
Note that some of the theoretical errors, such as 
the uncertainty of $N_{V_s}$, are correlated with the running
of $\als$.

Obviously one has to allow for quenching errors. Na\"\i vely one might assume
an $O(10~\%)$ effect on the binding energy which amounts to
50~MeV in Eq.~(\ref{mRS1GEV}).
However, this might be an underestimate since the running of the mass
with the scale
in the $n_f=0$ case is very different from that for $n_f=4$ and the relative
effect on $\overline{\Lambda}\ll m_b$, due to a different running, is larger
than that on the quark mass.
To illustrate this we also work consistently with $n_f=0$ and obtain,
\be
\label{mMSnf0}
m_{b,\MS}(m_{b,\MS})=\left[4339\pm 29({\rm latt.})\pm 49
({\rm th.})\pm 9 (\Lambda_{\MS}) \right]\,{\rm
MeV}\,:
\ee
this differs from the value Eq.~(\ref{mMSPV}) by almost 150~MeV.
Note that we have used the $n_f=0$ value
$\Lambda_{\MS}=0.602\,r_0^{-1}$ 
to obtain the above results.
Using the $n_f=5$ QCD world average $\alpha_s(M_z) = 
0.118$ instead (running it across the bottom flavour threshold
down to 1~GeV), the central value of Eq.~(\ref{mMSPV}) would read,
$m_{b,\MS}(m_{b,\MS})=4113$~MeV. The difference between these two values
may also be indicative of the typical size of 
the error due to quenching.

We feel that 1 GeV might be a more
natural scale to obtain an $n_f=4$ prediction from the quenched model
than 4 GeV and hence prefer the central value of Eq.~(\ref{mMSPV}).
After all, the quenched model has been adjusted to reproduce
low energy QCD phenomenology and indeed Eqs.~(\ref{rsval}) and
(\ref{LambdaRS1GEV})
agree with each other within errors. However, as discussed above and as
indicated by the 150~MeV difference from using a different perturbative
running, such predictions have to be consumed with some caution.
Eq.~(\ref{mMSPV}) demonstrates the precision that can be achieved in
lattice simulations of static-light mesons with sea quarks to
NNNLO. Obviously, the ``latt.'' error
can systematically be reduced. Note that, with NNNLO perturbative results,
the dominant theoretical uncertainty (apart from the sea quark content)
is due to $1/m_b$ corrections.

\subsection{The running of $\overline{\Lambda}$ from the static lattice
potential}
\label{v02}
To leading order, the singlet static energy $E_s$ is the sum of
twice the heavy quark mass and the singlet potential, Eq.~(\ref{Es}),
while $M_B$ is the sum of the quark mass and the binding energy
$\overline{\Lambda}$, Eq.~(\ref{MB}). Consequently, in the $\OS$ ($\RS$)
schemes $V_s$ contains twice the leading
renormalon (power term) of $\overline{\Lambda}$.
In QCD with sea quarks this is also evident
from the large distance behaviour, where $E_s(r)$ will approach $2M_B$.

In the lattice scheme, the
non-perturbative energy $E^L_{\Sigma_g^+}$
differs from $E_s$ by twice the quark mass, Eq.~(\ref{mLL}), 
and contains the same power term as the static-light energy
$\overline{\Lambda}^L$ (times two). One can explicitly verify
this in perturbation theory. In QCD with sea quarks
$E^L_{\Sigma_g^+}(r)$ will approach $2\overline{\Lambda}^L$
for $r\simg r_c$, where $r_c$ denotes the distance associated
with ``string breaking'' and is implicitly defined by,
$E_s(r_c)=2M_B$. We find the static
potential~\cite{statpot,Bali:1997am}
$E_{\Sigma_g^+}$ to exceed the values of
$2\overline{\Lambda}^L$ of Ref.~\cite{Allton:1993ix}
at $r>r_c=(2.25\pm 0.15)\,r_0$, a distance that is statistically
indistinguishable
from the value $r_c\approx 2.3\,r_0$,
obtained in simulations with $n_f=2$ light sea
quarks~\cite{Bali:2000vr,Bali:2000gf}.

\begin{table}[h]
\begin{center}
\begin{tabular}{|c|c|c|c|c|}
\hline
$a^{-1}r_0$&$\overline{\Lambda}^L_{\rm pot}r_0$
&$\overline{\Lambda}^{\RS}r_0$(NLO)
&$\overline{\Lambda}^{\RS}r_0$(NNLO)
&$\overline{\Lambda}^{\RS}r_0$(NNNLO)\\\hline\hline
 1.95&2.11(10)&0.64(14)&1.30(10)&0.99(12)\\ 
 2.42&2.35(10)&0.81(12)&1.39(10)&1.16(11)\\ 
 2.94&2.51 (4)&0.86 (8)&1.40 (5)&1.22 (6)\\ 
 3.80&2.81 (5)&0.95 (8)&1.49 (5)&1.33 (6)\\ 
 4.47&3.02 (3)&1.00 (7)&1.54 (4)&1.39 (5)\\ 
 5.35&3.29 (2)&1.05 (7)&1.61 (4)&1.46 (5)\\ 
 7.30&3.84 (2)&1.13 (7)&1.73 (4)&1.59 (5)\\ 
 9.89&4.53 (2)&1.21 (8)&1.87 (5)&1.73 (6)\\ 
12.74&5.23 (3)&1.26 (9)&1.99 (6)&1.85 (7)\\ 
14.36&5.47(10)&1.16(13)&1.92(12)&1.78(12)\\\hline
\end{tabular}
\end{center}
\caption{{\it The inverse lattice spacing,
the estimate of the
static-light binding energy in the lattice scheme,
$\overline{\Lambda}^L_{\rm pot}=
E_{\Sigma_g^+}^L(r_0)/2+\Delta$, Eqs.~(\ref{eql1})
and (\ref{eql2}), as well as its
conversion to the RS scheme to different orders in perturbation
theory. The errors only incorporate the statistical uncertainties
of the $E_{\Sigma_g^+}(r_0)$ data,
as well as the 8~\% uncertainty in
$\Lambda_{\overline{MS}}r_0$~\cite{Lambda}, added in
quadrature. The overall error due to the uncertainty in $\Delta$,
which does not affect the running of $\overline{\Lambda}^L$,
is not displayed.}}
\label{tablamm}
\end{table}

The difference $2\overline{\Lambda}^L-E_{\Sigma_g^+}^L(r_0)
=E_{\Sigma_g^+}^L(r_c)-E_{\Sigma_g^+}^L(r_0)$
is a constant, up to $O(a^2)$ lattice artifacts.
In what follows we will investigate the running of,
\begin{equation}
\label{eql1}
\overline{\Lambda}^L_{\rm pot}(a)=\frac{1}{2}E_{\Sigma_g^+}^L(r_0;a)+\Delta,
\end{equation}
as a function of $a^{-1}$. The static lattice potential $E_{\Sigma_g^+}(r_0)/2$
can be determined more precisely than $\overline{\Lambda}^L$:
in terms of computer time it is cheaper to obtain with the
same statistical error and, since no chiral extrapolation is 
involved, with virtually no systematic uncertainties.
However, we do not know the absolute normalization
$\Delta$. We re-analyse the lattice potentials
of Refs.~\cite{statpot,Bali:1997am},
to correctly account for the propagation of the uncertainty
of $r_0$ into the combination
$r_0E_{\Sigma_g^+}(r_0)$.
By matching the lattice potential
$E_{\Sigma_g^+}^L(r_0)/2=(2.856\pm 0.014)\,r_0^{-1}$
to $\overline{\Lambda}^L=(3.844\pm 0.065)\,r_0^{-1}$
at $\beta =6.2$ ($r_0/a\approx 7.3$),
where we have two independent results for the latter
quantity~\cite{Allton:1993ix,Ewing:1995ih}, we obtain
\begin{equation}
\label{eql2}
\Delta=(0.988\pm 0.067)\,r_0^{-1}\,.
\end{equation}
For ease of comparison with Sec.~\ref{ssssrs},
we display the resulting $\overline{\Lambda}^L_{\rm pot}(a)$
in Table~\ref{tablamm} as well as in the figures.
The additional uncertainty due to the error in $\Delta$
should be kept in mind.

\begin{figure}[h]
\hspace{-0.1in}
\epsfxsize=4.8in
\centerline{\epsffile{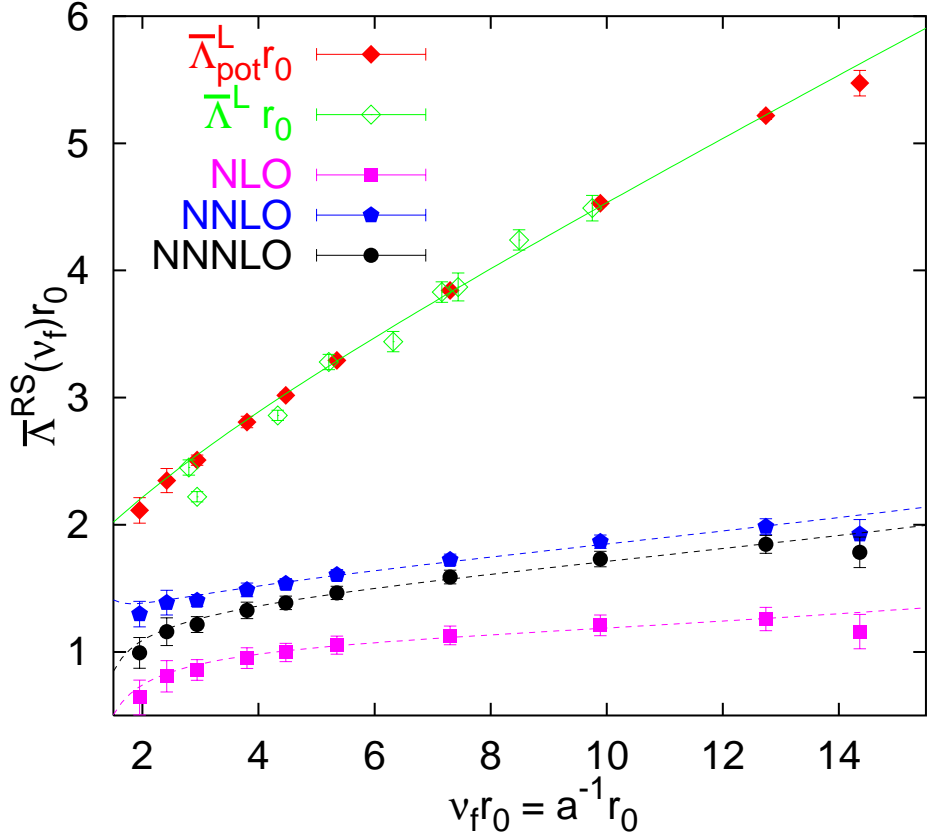}}
\caption {{\it The binding energy
$\overline{\Lambda}^L_{\rm pot}$, Eq.~(\ref{eql1}), in the lattice
scheme (full diamonds), in comparison with 
$\overline{\Lambda}^L$ of Sec.~\ref{ssssrs} (open diamonds).
The constant $\Delta$ has been adjusted by 
requiring agreement between the two data sets
at $r_0\approx 7.3\, a$.
The uncertainty of $\Delta=(0.988\pm 0.067)\,r_0^{-1}$ is
not included into the errors.
NLO, NNLO and NNNLO refer to transformations of
$\overline{\Lambda}^L_{\rm pot}$ into the RS scheme
to different orders in perturbation theory. The solid
line corresponds to the NNNLO expectation
with $\Lambda_{\MS}\approx 0.602\,r_0^{-1}$,
and the central value of Eq.~(\ref{lambdalarge}),
$\overline{\Lambda}^{\RS}(\nu_f=9.76\,r_0^{-1})=
1.70\,r_0^{-1}$.}}
\label{potrun}
\end{figure}

We display $\overline{\Lambda}^L_{\rm pot}$ in Table~\ref{tablamm},
together with conversions into the RS scheme, according to Eqs.\
(\ref{eqmshift}) and (\ref{eqmstat}). The data are also depicted in
Fig.~\ref{potrun} (full diamonds), together with the results
{}from the static-light energies $\overline{\Lambda}^L$ (open diamonds).
Except for the four data points of Ref.~\cite{Duncan:1994uq}
at $r_0/a\approx 2.9, 4.5,6.3$ and 8.5,
whose slope 
is somewhat incompatible with the results from the other references
as well as with perturbation theory (as we already noticed in
Sec.~\ref{ssssrs} above), we find agreement between
the non-perturbative running of $E_{\Sigma_g^+}(r_0)/2$
and that of $\overline{\Lambda}^L$,
down to the lowest scales.
This need not be so since in principle the results may differ
by $O(a^2/r_0^2)$ lattice terms. We also compare this running with
the expectation from the value $\Lambda_{\MS}\approx 0.602\,r_0^{-1}$
(solid line),
where we use the normalization suggested by
the central value of Eq.~(\ref{lambdalarge}),
$\overline{\Lambda}^{\RS}(\nu_f=9.76\,r_0^{-1})\approx
1.70\,r_0^{-1}$.

As can be seen there is no contradiction between the lattice data
and NNNLO perturbation theory down to scales as low as $2\,r_0^{-1}$
and as high as $15\,r_0^{-1}$.
This agreement is quantifiable:
a one-parameter NNNLO fit to the $a^{-1}>5\,r_0^{-1}$ data (setting
$\Lambda_{\MS}=0.602\,r_0^{-1}$)  yields
$\chi^2/N_{DF}=3.53/4$, with the value (translated into the RS scheme
for ease of comparison),
\begin{equation}
\label{lambdalarge2}
\overline{\Lambda}^{\RS}(9.76\,r_0^{-1})=(1.71\pm 0.01)\,r_0^{-1}\,.
\end{equation}
Including all available data results in
$\chi^2/N_{DF}=6.91/9$ with
$\overline{\Lambda}^{\RS}(9.76\,r_0^{-1})=(1.70\pm 0.01)\,r_0^{-1}$.
The errors of the above examples are purely statistical. The uncertainties
in $\Delta$ and $\Lambda_{\MS}$ as well as theoretical errors are unaccounted
for. If we go to NNLO we obtain the $\chi^2/N_{DF}$ values of
16.3 (all data points), 23.0 ($a^{-1}>5\,r_0^{-1}$) and 6.7
($a^{-1}>9\,r_0^{-1}$). Also, the predicted value of
$\overline{\Lambda}^{\RS}(9.76\,r_0^{-1})r_0$ becomes somewhat
unstable, ranging from 1.76 (all data points), 1.79 ($a^{-1}>5\,r_0^{-1}$),
up to 1.91 ($a^{-1}>9\,r_0^{-1}$): within the accuracy of the data it is
essential to go to at least NNNLO in perturbation theory.

\begin{figure}[h]
\hspace{-0.1in}
\epsfxsize=4.8in
\centerline{\epsffile{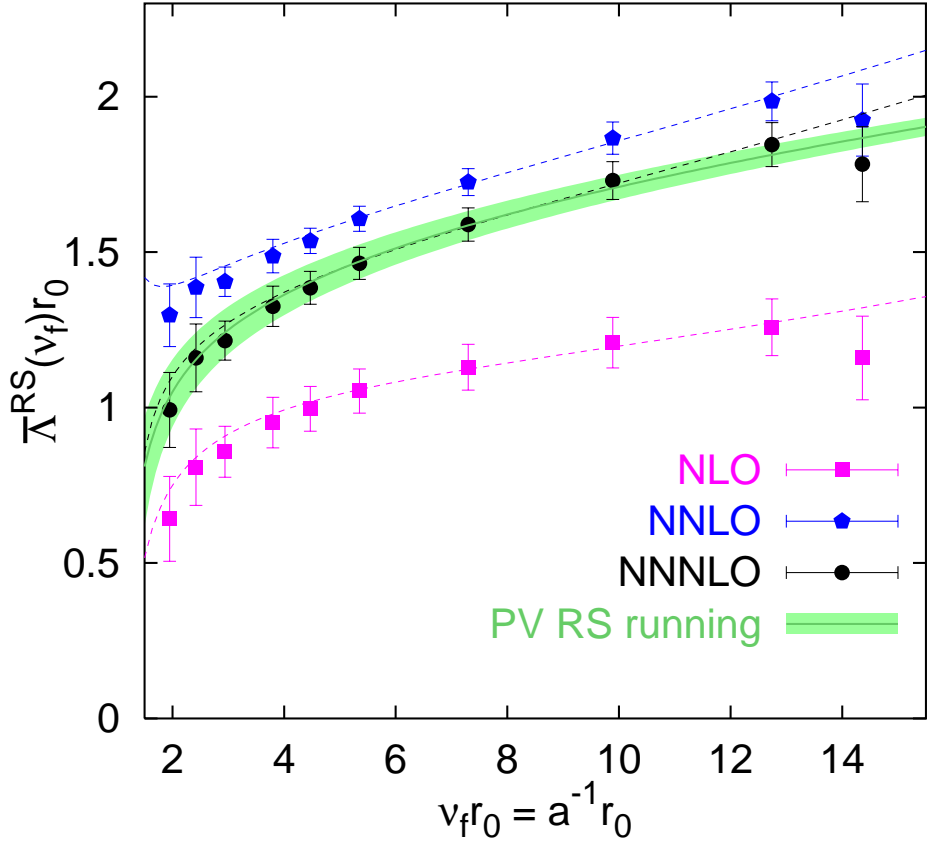}}
\caption {{\it The binding energy
$\overline{\Lambda}^L_{\rm pot}$, Eq.~(\ref{eql1}), translated into
the RS scheme at NLO (squares), NNLO (pentagons) and NNNLO (circles).
We have neglected an overall error in the vertical scale
of $\pm 0.067\,r_0^{-1}$,
due to the uncertainty of $\Delta$, that does not affect the
running. The dashed lines correspond to the NNNLO running in the
lattice scheme with $\Lambda_{\MS}=0.602\,r_0^{-1}$,
where we used the fit result Eq.~(\ref{lambdalarge2}),
$\overline{\Lambda}^{\RS}(9.76\,r_0^{-1})\approx 1.71\,r_0^{-1}$,
as normalization. The error band corresponds to the
prediction Eq.~(\ref{lambdalarge}),
$\overline{\Lambda}^{\RS}(9.76\,r_0^{-1})=(1.70\pm 0.04)\,r_0^{-1}$,
and includes the uncertainty due to
$\Lambda_{\MS}=(0.602\pm 0.048)\,r_0^{-1}$ (but no other errors).}}
\label{potrun2}
\end{figure}

In Fig.~\ref{potrun}  and Table~\ref{tablamm}
we have also displayed the results, translated into
the RS scheme to different orders in perturbation theory.
In Fig.~\ref{potrun2}
we focus on this comparison.
This figure very much resembles Fig.~\ref{bindrun}, only
that now the error bars are smaller as we discard the error of
$\Delta$, which will only affect the overall value of
$\overline{\Lambda}$ but not the running with the scale.
The dashed lines correspond to
NNNLO perturbation theory in the lattice scheme with
$\Lambda_{\MS}=0.602\,r_0^{-1}$, and the central value of
Eq.~(\ref{lambdalarge2}) as normalization point.
This running perfectly agrees with
the data down to very low energies.
As already observed in Sec.~\ref{ssssrs}
above, we also find nice convergence
for $a^{-1}\simg 3\,r_0^{-1}$,
as the order of the perturbation theory
is increased. 
The error band corresponds
to the PV prescription of the running in the RS scheme\footnote{At
$\nu_f\gg\nu=9.76\,r_0^{-1}$ we find some differences between the NNNLO
running in the lattice scheme (dashed black line) and the PV
prediction (error band), due to large logs in the difference
Eq.~(\ref{aapL}), where we have not attempted a log resummation.} with
$\overline{\Lambda}^{\RS}(9.76\,r_0^{-1})=(1.70\pm 0.04)\,r_0^{-1}$,
run to different scales, using
$\Lambda_{\MS}=(0.602\pm 0.048)\,r_0^{-1}$.
Note that the errors that we display in this case are
only due to the uncertainty in $\als$, with
all other error sources of 
Eq.~(\ref{lambdalarge}) (as well as the uncertainty of $\Delta$)
ignored.

We find excellent agreement between data and
the predicted running. In fact, one can in principle determine
$\als$ from the logarithmic corrections to the $a^{-1}$ running of the 
binding energy: in dedicated lattice
simulations of the short distance static potential
tremendous statistical accuracy can be achieved and tiny
lattice spacings are accessible~\cite{NS}. Even using our 
static singlet potentials~\cite{statpot,Bali:1997am} that are less accurate
than those of this recent
reference, a two-parameter NNNLO fit to the
$a^{-1}>5\,r_0^{-1}$ data yields, $\Lambda_{\MS}=(0.590\pm 0.036)\,r_0^{-1}$
and $\overline{\Lambda}^{\RS}(9.76\,r_0^{-1})=(1.73\pm 0.04)\,r_0^{-1}$ with
$\chi^2/N_{DF}=3.35/3$. Including the whole energy range, down to
$a^{-1}\approx 2\,r_0^{-1}$, results in,
$\Lambda_{\MS}=(0.627\pm 0.026)\,r_0^{-1}$
and $\overline{\Lambda}^{\RS}(9.76\,r_0^{-1})=(1.68\pm 0.02)\,r_0^{-1}$, still
with very acceptable $\chi^2/N_{DF}=6.08/8$. The results for
$\overline{\Lambda}^{\RS}$ are in perfect agreement with those obtained
in Eqs.~(\ref{lambdalarge}) and (\ref{lambdalarge2}) above.
Moreover, the fits are consistent with the value of Ref.~\cite{Lambda},
$\Lambda_{\MS}=(0.602\pm 0.048)\,r_0^{-1}$, within statistical errors
smaller than the uncertainty of this reference value.

In conclusion, we have demonstrated that the running
of the binding energy in the lattice scheme can be reproduced
with incredible accuracy in NNNLO perturbation theory, in terms
of $\als$. 
This accuracy is possible since, unlike in the case of the
binding energy itself, there is no leading renormalon contribution
to its running. Down to energies of about 1~GeV
we do not see any sign of a break-down of perturbation theory
or evidence of significant non-perturbative contributions to the
running. We have also confirmed that the
theoretical errors estimated
in Eqs.~(\ref{lambdalarge}) and (\ref{LambdaRS1GEV}) are
indeed conservative.

\section{Gluinonium and other related issues}
\label{gluino}
We already mentioned that gluelumps are interesting
in the context of bound states including heavy adjoint particles,
such as gluinos of SUSY models (even if it is quite likely that 
they will decay before any kind of hadronization takes place). 
In this case, to leading order in HGET (Heavy Gluino Effective Theory), the
gluino mass can be obtained from the relation,
\be
M_{\tilde{G}}=\Lambda_B^{\OS}+m_{\tilde{g},\OS}=\Lambda_B^{\RS}(\nu)+
m_{\tilde{g},\RS}(\nu),
\ee
in a scheme of choice that can then be converted
into the mass in say the $\MS$ scheme
$\overline{m}_{\tilde{g}}(\overline{m}_{\tilde{g}})$,
analogously to the discussion of Sec.~\ref{statlight} above.
We will limit most of the discussion below to the $\RS$ and $\OS$ schemes but
translation into lattice schemes is straight forward.

$M_{\tilde{G}}$ denotes the mass of the lightest (spin-averaged)
glueballino. Note that in this context the gluelump energy
$\Lambda_B$ plays the
same r\^ole as the binding energy $\overline{\Lambda}$ did for
heavy-light mesons.
We have $\Lambda_H^{\RS}(\nu_f)=\Lambda_H^{\OS}-\delta\Lambda_{\RS}(\nu_f)$
and hence,
\be
m_{\tilde{g},\RS}(\nu_f)=m_{\tilde{g},\OS}+\delta\Lambda_{\RS}(\nu_f)\,:
\ee
$\delta m_{\tilde{g},\RS}= - \delta\Lambda_{\RS}$
in the glueballino case corresponds to the
$\delta m_{\RS}$ of heavy-light
mesons. We can also write down the above equations in
the lattice scheme in which case, using the same conventions
as in other parts of this paper, $\delta m_{\tilde{g},L}=
\delta\Lambda_{L}$.

In addition to glueballinos one can imagine gluino-gluino
bound states: gluinonia, $\Gamma$. Their dynamics is dictated by the following
Lagrangian,
\be
L_{\rm pNRQCD,\Gamma} =
\int d^3{\bf r}\,d^3{\bf R}\,{\rm Tr} \,\Biggl\{ {\rm S_{\Gamma}}^\dagger \left( i\partial_0 - V_{A,s}  \right) {\rm S_{\Gamma}} 
+ {\rm O_{\Gamma,1}}^\dagger \left( iD_0 - V_{A,o} \right) {\rm O_{\Gamma,1}} +\cdots \Biggr\}\,,
\label{pppp}
\ee
at
leading order in $1/m_{\tilde{g}}$ and in the multipole expansion. This
is analogous to Eq. (\ref{pnrqcd0}), replacing the static sources in
the fundamental by static sources in the adjoint representation. This
means that there will be further multiplets in Eq.\ (\ref{pppp}) that
we will not consider in this paper.

The singlet potential between two adjoint sources $V_{A,s}(r)$
has been calculated
in perturbation theory to $O(\als^2)$ and the corresponding
energy $E_A^L(r;a)$ was determined in lattice
simulations (see e.g.\ Ref.~\cite{Bali:2000un}).
Up to lattice artifacts $\propto a^{2}$ we can write,
\bea
E_A(r)&=&2m_{\tilde{g},L}(a)+E_A^L(r;a)\\
&=&2m_{\tilde{g},L}(a)+V_{A,s,L}(r;a)+O(r^2)\\
&=&2m_{\tilde{g},\RS}(\nu_f)+V_{A,s,\RS}(r;\nu_f)+O(r^2),
\label{EARS}
\eea
where the normalization of $E_A^L(r)$ with respect to $E_A(r)$
can be obtained from the gluinonium spectrum.
Obviously,
\be
\label{unimp}
\lim_{r\rightarrow\infty}E_A(r)=2M_{\tilde{G}},
\ee
while for the bottomonium energy in QCD with sea quarks one obtains
(up to $1/m_b$ corrections and neglecting radial and gluonic excitations),
\be
\label{unimp2}
\lim_{r\rightarrow\infty}E_s(r)=\lim_{r\rightarrow\infty}E_H(r)=2M_{B}.
\ee

In combining Eq.~(\ref{EARS}) with Eqs.~(\ref{EsRS}) and (\ref{EHRS})
we obtain the important equality,
\be
E_A(r)+2[E_B(r)-E_s(r)]=
V_{A,s,\RS}(r;\nu_f)+2[V_{o,\RS}(r;\nu_f)-V_{s,\RS}(r;\nu_f)]
+2M_{\tilde{G}}+O(r^2)\,,
\label{important}
\ee
where we have used the fact that
$M_{\tilde{G}}=m_{\tilde{g},\RS}+\Lambda_B^{\RS}$ and $E_B
\in\{E_{\Pi_u},E_{\Sigma_u^-}\}$.
The effect of $\delta m_{\RS}$ cancels from the combination $E_B-E_s$
and $\delta\Lambda_{\RS}$ from $E_A+2E_H$. Since the glueballino mass
is a physical observable this implies that, up to $O(r^2)$ corrections,
the combination $V_{A,s}(r)+2[V_o(r)-V_s(r)]$ is scale
independent and free of renormalon and power contributions:
the UV renormalon of $V_o$ is cancelled by the UV renormalon of
$V_s$ while the leading IR renormalon of $V_o$, which we studied 
in this paper, is cancelled by one half of the UV renormalon
of $V_{A,s}$. In fact, to $O(\als^2)$ this combination explicitly vanishes
and the $O(\als^3)$ term is suppressed by a colour factor $1/N_c^2$.

In the above equation $E_B(r)$ corresponds to the
$\Pi_u$ or $\Sigma_u^-$ hybrid levels. For a general
$E_H(r)$ we would have to replace the $M_{\tilde{G}}$ on the {\em rhs}
by the mass of the excited glueballino in the respective channel.
At $r\rightarrow\infty$ the {\em rhs} of Eq.~(\ref{important}) will
approach $2M_{\tilde{G}}$, see Eqs.~(\ref{unimp}) and (\ref{unimp2}).

We wish to compare our expectation with lattice data. This can either
be done after an extrapolation of these to the continuum limit
or at finite lattice spacing in the lattice scheme.
Re-expressing Eq.~(\ref{important}) in terms of the energy
levels as determined in the
lattice scheme [$E_A(r)=E_A^L(r;a)+2m_{\tilde{g},L}(a)$ etc.],
and using the conventions of Sec.~\ref{seclatt} above,
this amounts to,
\bea
\label{im2}
&&E_A^L(r;a)+2[E_{\Pi_u}^L(r;a)-E_{\Sigma_g^+}^L(r;a)-\Lambda_B^L(a)]\\\nn
&&\qquad\qquad
=V_{A,s,L}(r;a)+2[V_{o,L}(r;a)-\delta\Lambda_L(a)-V_{s,L}(r;a)]+O(r^2)+\cdots\\\nn
&&\qquad\qquad
=O(r^2)+O(\als^3/N_c^2)+O(\lQ^2a^2)+O(a^2/r^2)\,.
\eea
Both {\em lhs} and {\em rhs} of the above equation
are explicitly free of $a^{-1}$ power terms (and of leading renormalons).
In fact the {\em rhs} vanishes in
perturbation theory, to at least $O(\als^3/N_c^2)$. As indicated in the
equation, in general there will
be non-perturbative $O(a^2/r^2)$ as well as $O(\lQ^2 a^2)$ lattice
artifacts, in addition to the $O(r^2)$ corrections from higher
terms in the multipole expansion.

The above combination is extremely interesting
as for small $r$ there should only be a quadratic but no linear term.
At $r\simg 2\,r_0$ the adjoint string will break and the {\em lhs}
of the equation will approach zero like $1/r$. In the intermediate region
$0.5\,r_0<r<2\,r_0$ one would expect two non-perturbative contributions,
a linear term from the slope of
$E_A(r)$, with an effective string tension~\cite{Bali:2000un},
$\sigma_{\rm eff}=[3.09\pm 0.10]r_0^{-2}$, as well as
a $1/r$ term that dominantly originates from the difference between
static hybrid and singlet potentials and whose coefficient will
approach $2\pi$ as $r\rightarrow\infty$, in an effective
string model expectation. In fact for $r\approx r_0$ one would
expect this $1/r$ term still to dominate over the linear term.

\begin{figure}[h]
\hspace{-0.1in}
\epsfxsize=4.8in
\centerline{\epsffile{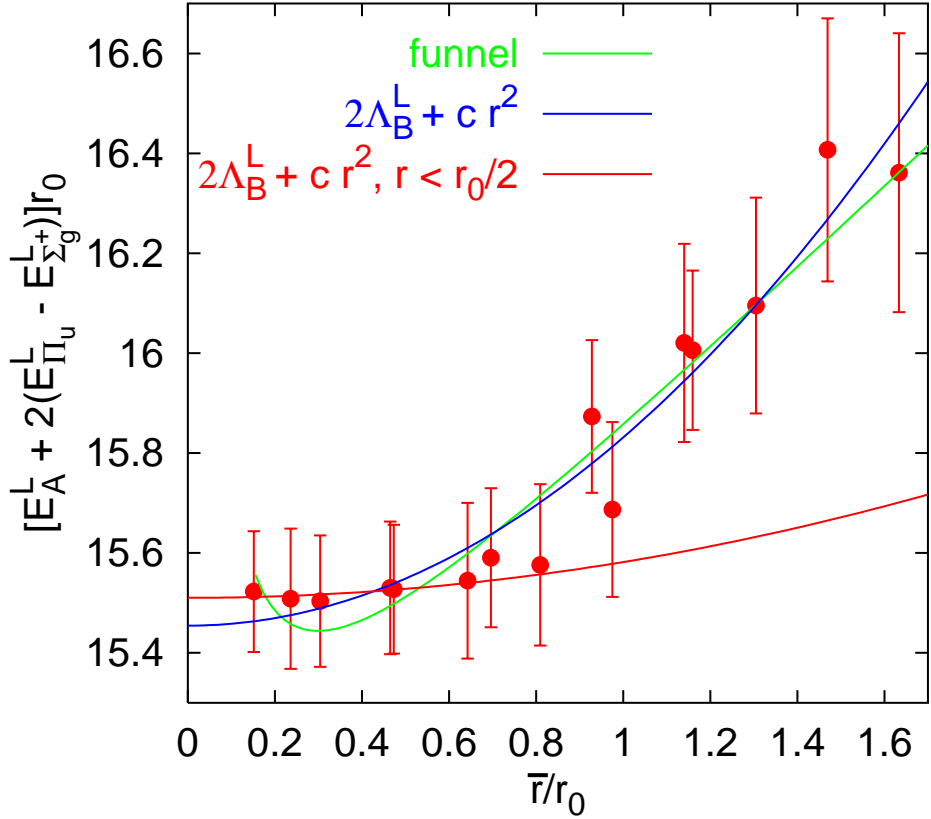}}
\caption {{\it The combination
$E^L_A(r)+2[E^L_{\Pi_u}(r)-E^L_{\Sigma_g^+}(r)]$,
Eq.~(\ref{im2}),
as a function of $\overline{r}=r[1+O(a^2/r^2)]$, Eq.~(\ref{overl}),
together with various fits, on an anisotropic lattice with
resolution $a_{\sigma}\approx 4\,a_{\tau}\approx 0.16\,r_0$.}}
\label{adjoint}
\end{figure}

We wish to compare this expectation to numerical data.
Unfortunately, on isotropic lattices where we know the gluelump
mass in the lattice scheme we did not compute the adjoint potential
while on our anisotropic data sets all potentials, singlet, adjoint
and hybrid are available but the gluelump
mass is unknown. In Fig.~\ref{adjoint} we display the combination
$E^L_A(r)+2[E^L_{\Pi_u}(r)-E^L_{\Sigma_g^+}(r)]$ as
a function of $\overline{r}=r[1+O(a^2/r^2)]$, 
Eq.~(\ref{overl}), at
our finest anisotropic lattice spacing, $a_{\sigma}\approx
0.16\,r_0\approx 4\,a_{\tau}$ which, within errors, is compatible
with the continuum limit, see Sec.~\ref{lattice} and Ref.~\cite{Bali:2000un}.
Note that there is an additional 1~\% overall error on ordinate
and abscissa due to the conversion from lattice units into units
of $r_0$ that we do not display.

{}From Eq.~(\ref{im2}) we would expect the combination shown to approach
the gluelump energy in the lattice scheme, $\Lambda_B^L(a)$, as
$r\rightarrow 0$.
We see that the approach towards this limit is remarkably flat.
In fact, excluding the $r>0.9\,r_0$ data, which are clearly in the
non-perturbative regime anyway, we are unable to resolve deviations
of the data from a constant.
Note that the units on the ordinate, $0.2\,r_0^{-1}
\approx 80$~MeV, are quite small. A linear plus quadratic fit,
\be
E_A^L(r)+2[E_{\Pi_u}^L(r)-E_{\Sigma_g^+}(r)]=2\Lambda_B^L+c\,r^2\,,
\ee
to $r<0.5\,r_0$ data yields,
\be
2\Lambda_B^L=(15.51\pm 0.10)\,r_0^{-1},\qquad c=(0.07\pm 0.70)\,r_0^{-3}\,.
\ee
A purely phenomenological fit to the same functional form for all distances
results in,
\be
2\Lambda_B^L=(15.45\pm 0.06)\,r_0^{-1},\qquad c=(0.38\pm 0.07)\,r_0^{-3}\,,
\ee
while in a physically completely unmotivated funnel parametrisation,
$2\Lambda_B^L+e/r+k\,r$,
we obtain,
\be
2\Lambda_B^L=(14.95\pm 0.20)\,r_0^{-1},\qquad e=(0.08\pm 0.04),\qquad
k = (0.84\pm 0.18)\,r_0^{-2}\,:
\ee
the $r$ dependence is so weak that on the 1~\% error level of the lattice
data we are unable to discriminate between different parametrizations.
However,
we can determine the gluelump mass rather precisely,
$\Lambda_B^L(a_{\sigma},a_{\tau})=(7.75\pm 0.05\pm 0.07)\,r_0^{-1}$,
where the second error reflects the uncertainty in the lattice
determination of $r_0/a_{\tau}$. In fact the same can be done
for the $a_{\sigma}\approx 0.23\,r_0$ and $a_{\sigma}\approx 0.33\,r_0$
data sets. The respective results read,
$\Lambda_B^L=(6.71\pm 0.04\pm 0.09)\,r_0^{-1}$ and
$\Lambda_B^L=(5.75\pm 0.10\pm 0.05)\,r_0^{-1}$, respectively.
The data are in agreement with a linear slope in $a^{-1}$ but,
unfortunately, at present we only know the NLO perturbation theory
for the anisotropic case. After subtracting twice these gluelump
energies, we find scaling of the coarse lattice data with the
results depicted in the figure, within error bars of comparable size.

In particular {}from the fit to the funnel type parametrisation we
see that the data leave little room for perturbation theory style
short-distance Coulomb terms. This is in agreement with our expectation.
However, miraculously there is also no evidence for a quadratic term
in the $r<0.9\,r_0$ data and in fact we can set the limits
$0.46>c\,r_0^3>-0.18$ for such a contribution, from the $r<0.7\,r_0$ data.
We believe that the smallness of this term is accidental as had
we replaced the $\Pi_u$ by the $\Sigma_u^-$ potential, it would
certainly be present, see Sec.~\ref{pisisplit}. One can however
speculate that there might be a cancellation of $r^2$ effects
and that $\Pi_u$ does not receive a significant $r^2$ contribution
in the multipole expansion. This issue should be addressed in
future theoretical and numerical studies with enhanced accuracy.

The observed slope at larger distances
($\approx 0.84\,r_0^{-2}$) is much smaller than that of the
adjoint potential in this region ($\approx 3.09\,r_0^{-2}$),
in agreement with our expectation that the $1/r$ contribution from
the difference $2(E_{\Pi_u}-E_{\Sigma_u^-})$ cannot be neglected.

There is no evidence of a linear non-standard
short-distance term
for $r<0.9\,r_0$, at least not of the size expected in
various models~\cite{linear}. 
A possible explanation of
the absence of such a term
from our calculation of a quantity that vanishes to low
orders in perturbation theory would be that $\alpha_s(q)$ itself receives
$O(1/q^2)$ corrections (see Refs.~\cite{linear}). We remark to
this end that $\alpha_s$ is not a physical observable.
In the $\overline{MS}$ scheme it is perturbatively defined.
The difference between $\als$ and any non-perturbative generalisation
of this coupling, that
would allow inclusion of such singularities, will necessarily
not be universal but
depend on the prescription used. However, we are investigating 
a physical observable here that is scheme independent.

Combinations of different potentials that lead to renormalon
and low order perturbation theory cancellations are certainly
an arena worthwhile to study for a determination of higher
order terms in the multipole expansion and for testing the
validity of the standard operator product expansion picture.
As we shall detail below many such combinations exist.

There are also hybrid excitations in the adjoint channel.
The perturbation theory in this case is richer than for potentials
between fundamental sources as
${\mathbf 8}\otimes{\mathbf 8}={\mathbf 1}\oplus{\mathbf 8}
\oplus{\mathbf 8}\oplus{\mathbf 1}{\mathbf 0}\oplus{\mathbf 1}{\mathbf 0^*}\oplus{\mathbf 2}{\mathbf 7}$:
in addition to singlet and octet, we have another octet, two decuplet
fields and a 27-let which have to be included into Eq.~(\ref{pppp}).
Consequently, adjoint hybrid potentials cannot only have the
octet perturbative expansion but some will correspond to decuplets
and others to 27-lets.
Note that the decuplet representation is not self-adjoint but
has vanishing triality as it should be.

The renormalon of the octet potential between adjoint sources is the
same as in the fundamental case but the decuplet and 27-plet
adjoint potentials contain new renormalons, which are related to
those of the singlet potentials between colour charges in these respective
representations. This exactly resembles the situation discussed
above where the adjoint singlet potential contains the
same renormalon as the fundamental octet potential.
In fact one can define an infinite tower of states with different
renormalons following this construction, a theoretically
interesting enterprise but not likely to be of much direct
phenomenological use.

The inclusion of the octet states of
Eq.~(\ref{pppp}) is necessary for any
consistent pNRQCD calculation of gluino pair production
near threshold
at NLO~\cite{Beenakker:1996ch}. At NNLO the
decuplet and 27-plet fields will also play a r\^ole.
In fact such contributions, depending on the mass of the gluino
(and on its existence), might be of bigger importance than
in the case of $t\bar{t}$ production because there are more of them.
This is an exciting and very clean-cut
situation since $v^2$ and $r^{-1}$ are bigger
by a factor $\sim C_A/C_f$ than
for quarkonia, such that all ``soft'' physics is clearly
and extremely safely within the perturbative domain.

Let us finally mention that $\Lambda_B$, the binding energy
of the lightest glueballino, determines the size of the splitting
between the adjoint singlet potential and the lowest adjoint hybrid potential
at short distances, the latter of which, unfortunately, has never
been determined in lattice simulations.
This is very different from the case of fundamental
sources where binding energies of heavy-light systems,
$\overline{\Lambda}$, are much smaller than the gaps between ground state
and hybrid excitations. In ``hadrinos'', that contain stable adjoint
sources, gluonic excitations
would hence play a very prominent r\^ole and simple constituent-gluino
models might fail terribly.
Unfortunately, in nature we do not encounter such particles.
It would however be most entertaining to confirm this
expectation in lattice simulations.

\section{Conclusions}
\label{conclude}
We report compelling evidence that for distances around 1~GeV${}^{-1}$ 
the gluonic excitations of the static potential are in the
short distance regime.

We are able to obtain a value for the lowest
lying mass $\Lambda_B$ of the bilocal
gluonic correlation functions with well controlled uncertainties, by fitting 
to the difference between the $\Pi_u$ and $\Sigma_g^+$ potentials.
The RS scheme result reads,
\be
\label{eqxxx}
\Lambda_{B}^{\RS}(n_f=0)=\left[2.25\pm 0.10({\rm latt.})\pm 0.21 ({\rm th.})\pm 0.08 
(\Lambda_{\MS})\right]\,r_0^{-1} 
\,,
\ee
for $\nu_f=2.5\,r_0^{-1}\approx 1$~GeV.
Translated into physical units this reads,
\begin{equation}
\Lambda_B^{\RS}(1\,\mbox{GeV})=[0.887\pm 0.039({\rm latt.})\pm 0.083({\rm th.})\pm 0.032({\rm
\Lambda_{\MS}})]\,\mbox{GeV}\,.
\end{equation}
Note that one should also add an extra error of order 10 \%  due to
quenching to these numbers.
With the information presented
in this paper $\nu_f$ can be run to different scales (see
Fig.~\ref{figgl}).
We also obtain values for the masses of other gluelumps,
listed in Table~\ref{tablegluelumps}, as well as for the non-perturbative
slope
$A_{\Pi_u-\Sigma_u^-}=0.92^{+0.53}_{-0.52}\,r_0^{-3}$ of the
quadratic difference between the lowest two hybrid potentials.

In order to state sensible numbers for $\Lambda_{B}$, the scheme for
the renormalon cancellation has to be specified. Otherwise, very
different numbers can be obtained, as we can see from a comparison
of the result in the lattice and the RS schemes. One can translate
from one scheme into the other in a renormalon-free way, order by
order in perturbation theory and check whether both results are
consistent with each other.  We have been able to confirm this.
If we use the gluelump results from Foster and Michael~\cite{FM}
at finite lattice spacing we obtain,
$
\Lambda_B^{\RS}(2.5\,r_0^{-1})=\left[2.31\pm 0.04 (\mbox{latt.}) \pm 0.33 (\mbox{th.})
{}^{+0.18}_{-0.19} (\Lambda_{\MS})\right]\,
r_0^{-1},
$
which is perfectly compatible with the result Eq.~(\ref{eqxxx})
above, albeit with slightly
larger errors.

We also investigate the binding energy of heavy-light
mesons in the static limit and to NNNLO in the conversion. We
arrive at similar conclusions. For the binding energy we
obtain the $n_f=0$ value, 
\be
\overline{\Lambda}^{\RS}(\nu_f=2.5\,r_0)= \left[1.17\pm 0.08({\rm
latt.})\pm 0.13 ({\rm th.})\pm 0.09 (\Lambda_{\MS})
\right]\,r_0^{-1}\,,\, 
\ee 
which is in good agreement with the
phenomenological value, obtained from the
experimental $\Upsilon(1S)$ and $B$ meson systems~\cite{RS},
$\overline{\Lambda}^{\RS}=[0.92 \pm 
0.22 (\mbox{th.}){}^{+0.15}_{-0.11} (\Lambda_{\MS})]\,r_0^{-1}$.

We have demonstrated the internal consistency of our approach. Lattice
predictions for $\Lambda_B^L$ and
$\overline{\Lambda}^L$ at different lattice spacings have been studied.
We have shown that the perturbative series,
Eq.\ (\ref{aapL}), relating
$\Lambda_B^L(a)$ and $\overline{\Lambda}^L(a)$ with $\Lambda_B^L(a')$ and
$\overline{\Lambda}^L(a')$, respectively, in the lattice scheme is
free of renormalon singularities and has nice convergence properties, as
indicated by the consistency with the
non-perturbatively obtained values.
In particular this means that
{}from the knowledge of $\Lambda_B^L$ and $\overline{\Lambda}^L$ at
a given lattice spacing values at different
lattice spacings can accurately be predicted.  We
have studied the conversion of lattice predictions for
$\Lambda_B^L$ and $\overline{\Lambda}^L$ into the RS scheme. This conversion
is also dictated by a perturbative series which is free of
renormalons. We have verified that the values in the lattice scheme
indeed approach the results in the RS scheme with a convergent pattern
and, remarkably, the $\nu_f$-scale dependence predicted by the RS
scheme is reproduced, within errors. We remark
that for the $\nu_f$-scale running it is possible to obtain a resummed
non-perturbative expression in which the renormalon is cancelled and
at the same time the log resummation is performed.

We stress that the RS scheme used here is designed to smoothly
converge to ($\MS$-style dimensional regularization) perturbation
theory at low orders in $\als$; after all, the renormalon effect only
sets in asymptotically at large orders in perturbation
theory. Different values for $\Lambda_{B}$ and for
$\overline{\Lambda}$ can be obtained in other schemes but only at the
inconvenience of having large corrections to ``standard'' perturbation
theory at low orders. In this sense we consider our approach
``natural''; the RS scheme incorporates salient features of both,
dimensional and lattice regularization.  The approach readily benefits
from results computed in the $\MS$ scheme, the scheme in which
perturbative quantities are usually known to the highest order.  On
the other hand, by subtracting renormalons we encounter explicit power
divergencies, which is exactly what one obtains with a hard lattice
cut-off too.

Our model independent non-perturbative predictions can directly be
incorporated into perturbative calculations, within effective field
theories, or exploited in the context of QCD vacuum models or
calculations based on non-local condensates.  Obvious phenomenological
applications in the context of EFTs are pNRQCD in the kinematic domain
$mv^2\siml\lQ<mv$, translating glueballino masses into RS or $\MS$
gluino masses within HGET (Heavy Gluino Effective Theory), or
converting heavy-light meson masses into quark masses within HQET.

We observe that $\Lambda_B\approx (C_A/C_f)\overline{\Lambda}\approx
m_G/2$, where $m_G$ denotes the mass of the lightest glueball.  The
first similarity is not necessarily surprising since there are
technical parallels between $\Lambda_B$, which corresponds to the
binding energy of an adjoint source, and $\overline{\Lambda}$, the
energy of a fundamental source.  We do not intend to advocate a
constituent gluon picture. Nevertheless, it may seem reasonable that
the binding energy of the glue to an adjoint source has about half the
size of the energy of an entirely gluonic state. It should however be
noted that the latter is an unambiguously defined state in the
physical spectrum while for the binding energy $\Lambda_B$ we
necessarily encounter the scheme and scale dependence that we
discussed.

We have also investigated the scenario of gluinonia and other excitations
in non-fun\-da\-men\-tal channels. While gluinos might not exist in nature and
certainly do not form light bound states, such that phenomenological
applications are limited, from a theoretical and conceptual point
of view the existence of this part of the spectrum is very interesting.
The inclusion of such potentials allows one to identify many combinations
in which renormalons and other un-wanted
contributions vanish, opening up a
window to the study of non-perturbative short distance physics.
\vskip 1cm

\noindent {\bf Acknowledgments}\\[.5cm] 
We thank C.\ DeTar and in particular J.\ Juge for sharing their
lattice data with us. A.P. thanks J. Soto and J. Sola for discussions. 
G.B.\ is supported by a PPARC Advanced
Fellowship (grant PPA/A/S/2000/00271) as well as by PPARC grant
PPA/G/0/2002/0463. A.P.\ is supported by MCyT and Feder (Spain),
FPA2001-3598, by CIRIT (Catalonia), 2001SGR-00065 and by the EU
network EURIDICE, HPRN-CT2002-00311.


\end{document}